\documentclass[11pt]{article}
\usepackage{txfonts}
\usepackage{color,epsfig}
\usepackage{mathrsfs}
\usepackage{amssymb}
\usepackage{amsfonts,graphicx,psfrag}

\newtheorem{Th}{Theorem}[section]

\newtheorem{lemma}{Lemma}[section]
\newtheorem{theorem}[lemma]{Theorem}
\newtheorem{corollary}[lemma]{Corollary}

\newtheorem{proposition}[lemma]{Proposition}
\newtheorem{definition}[lemma]{Definition}

\newtheorem{remark}[lemma]{Remark}
\let\lutzremark=\remark
\def\remark{\lutzremark\normalfont}

\newtheorem{notation}[lemma]{Notation}

\def\be{\begin{equation}}
\def\ee{\end{equation}}
\def\bea{\begin{eqnarray}}
\def\eea{\end{eqnarray}}
\def\bes{\begin{eqnarray*}}
\def\ees{\end{eqnarray*}}

\def\nn{\nonumber}
\def\<{\langle}
\def\>{\rangle}
\def\lb{\label}
\def\bs{\setminus}
\def\pt{\partial}

\def\R{{\bf R}}
\def\C{{\bf C}}
\def\Z{{\bf Z}}
\def\N{{\bf N}}
\def\U{{\bf U}}

\def\aa{{\alpha}}
\def\bb{{\beta}}
\def\ga{{\gamma}}
\def\Ga{{\Gamma}}

\def\th{{\theta}}
\def\Th{{\Theta}}
\def\om{{\omega}}

\def\ep{{\epsilon}}
\def\lm{{\lambda}}

\def\sg{{\sigma}}

\def\vf{{\varphi}}

\def\P{{\cal P}}

\def\diag{{\rm diag}}
\def\constant{{\rm constant}}

\def\span{{\rm span}}
\def\Sp{{\rm Sp}}

\def\dm{{\rm \diamond}}

\def\hb{\vrule height0.18cm width0.14cm $\,$}
\def\ol#1{\overline{#1}}

\title{Linear stability of elliptic relative equilibria of
four-body problem with two infinitesimal masses}
\author{Qinglong Zhou$^{1} $\thanks{Partially supported by the Natural Science Foundation of Zhejiang Province (No.Y19A010072) and the Fundamental Research Funds for the Central Universities (No.2019QNA3002). E-mail: zhouqinglong@zju.edu.cn}\quad\\
	$^{1}$ Department of Mathematics\\Zhejiang University, Hangzhou 310027, Zhejiang, China}
\date{}

\begin{document}

\maketitle

\begin{abstract}
{In this paper, we consider the elliptic relative equilibria of four-body problem with two infinitesimal masses.
The most interesting case is when the two small masses tend to the same Lagrangian point $L_4$ (or $L_5$).
In \cite{Xia}, Z. Xia showed that there exist four central configurations: two of them are non-convex, and the other two are convex.
We prove that the elliptic relative equilibria raised from the non-convex central configurations are always linearly unstable;
while for the elliptic relative equilibria raised from the convex central configurations,
the conditions of linear stability with respect to the parameters are given.}
\end{abstract}



\renewcommand{\theequation}{\thesection.\arabic{equation}}

\setcounter{equation}{0}
\setcounter{figure}{0}
\section{Introduction and main results}%{Section 1}
\label{sec:1}

For $n$ particles of mass $m_1,m_2,\ldots,m_n>0$,
let $q_1,q_2,\ldots,q_n\in \R^2$ the position vectors
respectively. Then the system of equations for $n$-body problem is
\be   m_i\ddot{q}_i=\frac{\partial U}{\partial q_i}, \qquad {\rm for}\quad i=1, 2, \ldots, n, \lb{1.1}\ee
where $U(q)=U(q_1,q_2,\ldots,q_n)=\sum_{1\leq i<j\leq n}\frac{m_im_j}{\|q_i-q_j\|}$ is the
potential or force function by using the standard norm $\|\cdot\|$ of vector in $\R^2$.

Note that $2\pi$-periodic solutions of this problem correspond to critical points of the action functional
$$ \mathcal{A}(q)=\int_{0}^{2\pi}\left[\sum_{i=1}^n\frac{m_i\|\dot{q}_i(t)\|^2}{2}+U(q(t))\right]dt $$
defined on the loop space $W^{1,2}(\R/2\pi\Z,\hat{\mathcal {X}})$, where
$$  \hat{\mathcal {X}}:=\left\{q=(q_1,q_2,\ldots,q_n)\in (\R^2)^n\,\,\left|\,\,
       \sum_{i=1}^n m_iq_i=0,\,\,q_i\neq q_j,\,\,\forall i\neq j \right. \right\}  $$
is the configuration space of the planar three-body problem.

Letting $p_i=m_i\dot{q}_i\in\R^2$ for $1\le i\le n$, then (\ref{1.1}) is transformed to a Hamiltonian system
\be \dot{p}_i=-\frac{\partial H}{\partial q_i},\,\,\dot{q}_i
  = \frac{\partial H}{\partial p_i},\qquad {\rm for}\quad i=1,2,\ldots, n,  \lb{1.2}\ee
with Hamiltonian function
\be H(p,q)=H(p_1,p_2,\ldots,p_n, q_1,q_2,\ldots,q_n)=\sum_{i=1}^n\frac{\|p_i\|^2}{2m_i}-U(q_1,q_2,\ldots,q_n).  \lb{1.3}\ee
A {\it central configuration} is a solution $(q_1,q_2,\ldots,q_n)=(a_1,a_2,\ldots,a_n)$ of
\begin{equation}
-\lambda m_iq_i={\pt U\over\pt q_i}(q_1,q_2,\ldots,q_n)
\end{equation}
for some constant $\lambda$.
An easy computation show that $\lambda={U(a)\over2I(a)}>0$,
where $I(a)={1\over2}\sum m_i||a_i||^2$ is the moment of inertia.
Please refer \cite{Win1} and \cite{Moe1} for the properties of central configuration.

It is well known that a planar central configuration of the $n$-body problem give rise to solutions
where each particle moves on a specific Keplerian orbit while the totaly of the particles move on a homographic motion.
Following Meyer and Schmidt \cite{MS}, we call these solutions as {\it elliptic relative equilibria}
and in shorthand notation, simply ERE.
Specially when $e=0$, the Keplerian elliptic
motion becomes circular motion and then all the bodies move around the center of masses along circular
orbits with the same frequency, which are called {\it relative equilibria} traditionally.

In the three-body case, the linearly stability of any ERE is clearly studied recently (c.f.\cite{HLS},\cite{ZL1}).
In fact, the stability of ERE depends on the eccentricity $e$ and a mass parameter $\beta$.
For the elliptic Lagrangian solution, the mass parameter is given by
\begin{equation}\lb{L:bb}
\beta=\frac{m_1m_2+m_2m_3+m_3m_1}{(m_1+m_2+m_3)^2},
\end{equation}
and for the elliptic Euler solution, the mass parameter is given by
\begin{equation}
\beta=\frac{m_1(3x^2+3x+1)+m_3x^2(x^2+3x+3)}{x^2+m_2[(x+1)^2(x^2+1)-x^2]},    \lb{E:bb}
\end{equation}
where $x$ is the unique positive solution of the Euler quintic polynomial equation
\begin{equation}\label{quintic.polynomial}
(m_3+m_2)x^5+(3m_3+2m_2)x^4+(3m_3+m_2)x^3-(3m_1+m_2)x^2-(3m_1+2m_2)x-(m_1+m_2)=0,
\end{equation}
and the three bodies form a central configuration of $m_1,m_2,m_3$,
which are denoted by $q_1=(0,0)^T,q_2=(x,0)^T$ and $q_3=(1+x,0)^T$ with $|q_1-q_2|=x,|q_2-q_3|=1$.
In \cite{HLS} and \cite{ZL1},
Long et al. used Maslov-type index and operator theory to study the stability problem,
and gave out a full description of the bifurcation graph.
For the near-collision Euler solutions of 3-body problem, the linear stability  was
studied by Hu and Ou in \cite{HO1}.

To our knowledge,
for the general $n$ bodies, the elliptic Euler-Moulton solutions is the only case which has been well studied in \cite{ZL2}.
It turns out that the stability
of the elliptic Euler-Moulton solutions depends on $(n-1)$ parameters, namely the eccentricity $e\in [0,1)$ and the $n-2$ mass parameters
$\bb_1,\bb_2,\ldots,\bb_{n-2}$ which defined by (1.14) in \cite{ZL2}.
For some special cases of $n$-body problem,
the linear stability of ERE which raised from an $n$-gon or $(1+n)$-gon central configurations with $n$ equal masses
was studied by Hu, Long and Ou in \cite{HLO}.

For the elliptic relative equilibria raised from a general non-collinear central configuration,
even for $n=4$,
the stability problem is quite open.
We will concern a special case of such ERE,
which raised from a central configuration of two primary masses $m_1,m_2$ and two infinitesimal masses $m_3,m_4$.
For example, the ``massless bodies" can be imaged as two space stations
and the two massive bodies are the sun and the earth.
Or  one could think of the ``massless bodies" as the two planets in a
binary star system.

When $m_3$ and $m_4$ are small enough,
each of them must close to one of the five Lagrangian points of $m_1$ and $m_2$.
If $m_3$ and $m_4$ tend to the different Lagrangian point,
%in the limiting case $m_3,m_4\rightarrow0$,
since the effect between the two small masses is disappeared 
as $m_3,m_4\rightarrow0$,
such a stability problem can be decomposed into two stability problem of the restricted three-body problem respectively:
one of them with masses $m_1,m_2,m_3=0$; and another one with masses $m_1,m_2,m_4=0$.
Then we can study the linear stability of such ERE in details
by using the results of \cite{HLS},\cite{ZL1} and \cite{HO1}.

The more interesting cases occur
when $m_3$ and $m_4$ tend to the same Lagrangian point $L$ 
as $m_3,m_4\rightarrow0$.
In such cases, since the effect between the two small masses is not disappeared,
so the above decomposition will be failed.

If the positions of $m_1,m_2$ and the point $L$ form a Euler central configuration,
then the original central configuration of $m_1,m_2,m_3$ and $m_4$ must be collinear, i.e., an Euler-Moulton central configuration
by \cite{Xia} and \cite{ZL2}.
The stability problem of such ERE was studied well in Section 3 of \cite{ZL2}.
As a matter of fact, in the limiting case $m_3,m_4\rightarrow0$, by (3.69) bellow of \cite{ZL2},
the stability problem is reduced to the linear stability problems of two restricted three-body
problems, for which one has mass parameter $\beta$, and the other has mass parameter $3(\beta+1)$
where $\beta$ is given by (\ref{E:bb}).

If the positions of $m_1,m_2$ and the point $L$ form a Lagrangian central configuration, by \cite{Xia},
we have four central configurations: two of them are non-convex, and the other two are convex.
If the central configuration is convex (non-convex),
we call the corresponding ERE as {\it convex ERE} ({\it non-convex ERE}).
Moreover, when $m_3,m_4\rightarrow0$, 
an ERE which is the limit of a family of convex (non-convex) EREs,
is also called as convex (non-convex) ERE.

In the current paper, we will study the linear stability problem of these two classes of ERE.
We first have the following reduction:

\begin{theorem}\label{main.theorem.decomposition}
In the planar $4$-body problem with given masses $m=(m_1,m_2,m_3,m_4)\in (\R^+)^4$, denote the
ERE with eccentricity $e\in [0,1)$ for $m$ by $q_{m,e}(t)=(q_1(t), q_2(t), q_3(t),q_4(t))$.
When $m_3,m_4$ tend to $0$,
the linearized Hamiltonian system at $q_{m,e}$ is reduced into the sum of $3$ independent
Hamiltonian systems, the first one is the linearized system of the Kepler $2$-body problem at the corresponding Kepler orbit,
the second one is the linearized Hamiltonian system of some ERE of a $3$-body problem with the
original eccentricity $e$ and the mass parameter $\bb$ of (\ref{L:bb}) with $m_3=0$,
and the last one is the essential part of the linearized Hamiltonian system
which depends on the convexity of the corresponding central configuration.

Moreover, in the non-convex case, the essential part is
\begin{equation}\label{Nonconvex:LHS}
z'=J\left(\matrix{1&  0&  0& 1\cr
                0&  1& -1& 0\cr
                0& -1& 1-\frac{9+3\sqrt{9-\bb}}{2(1+e\cos(t))}& 0\cr
                1&  0&   0& 1+\frac{\sqrt{9-\bb}}{1+e\cos(t)}
                }\right)z,
\end{equation}
and in the convex case, the essential part is
\begin{equation}\label{Convex:LHS}
z'=J\left(\matrix{1&  0&  0& 1\cr
                0&  1& -1& 0\cr
                0& -1& 1-\frac{9-3\sqrt{9-\bb}}{2(1+e\cos(t))}& 0\cr
                1&  0&   0& 1-\frac{\sqrt{9-\bb}}{1+e\cos(t)}
                }\right)z,
\end{equation}
where $\bb$ is given by (\ref{L:bb}) with $m_3=0$.
\end{theorem}

\begin{remark}
	Here we do not assume $m_3=m_4$.
	In contrast, we suppose $m_3\slash m_4\rightarrow\tau$ for
	some mass ratio $\tau\ge0$.
	It is surprising that the system (\ref{Nonconvex:LHS}) 
	and (\ref{Convex:LHS}) are all independent with respect to 
	the mass ratio $\tau$.
	That is to say, the linear stability of both non-convex ERE
	and convex ERE are independent with respect to 
	the mass ratio of the two infinitesimal masses.
\end{remark}

Noting that the linear stability of the first two parts are studied in \cite{HLS},
so we just need to study the essential part.
For the linear stability of non-convex ERE, 
we have
\begin{theorem}\label{main.theorem.nonconvex}
The non-convex ERE with two infinitesimal masses is always linearly unstable.
\end{theorem}

For describing more precise results of both non-convex ERE
and convex ERE,
we need some notation.
Following \cite{Lon2} and \cite{Lon4}, for any $\omega\in\U=\{z\in\C\;|\;|z|=1\}$ we can define a real function
$D_\om(M)=(-1)^{n-1}\overline{\om}^n det(M-\om I_{2n})$ for any $M$ in the symplectic group $\Sp(2n)$.
Then we can define $\Sp(2n)_{\om}^0 = \{M\in\Sp(2n)\,|\, D_{\om}(M)=0\}$ and
$\Sp(2n)_{\om}^{\ast} = \Sp(2n)\bs \Sp(2n)_{\om}^0$. The orientation of $\Sp(2n)_{\om}^0$ at any of its point
$M$ is defined to be the positive direction $\frac{d}{dt}Me^{t J}|_{t=0}$ of the path $Me^{t J}$ with $t>0$ small
enough. Let $\nu_{\om}(M)=\dim_{\C}\ker_{\C}(M-\om I_{2n})$. Let
$\mathcal{P}_{2\pi}(2n) = \{\ga\in C([0,2\pi],\Sp(2n))\;|\;\ga(0)=I\}$ and
$\xi(t)=\diag(2-\frac{t}{2\pi}, (2-\frac{t}{2\pi})^{-1})$ for $0\le t\le 2\pi$.

As in \cite{Lon4}, for $\lm\in\R\bs\{0\}$, $a\in\R$, $\th\in (0,\pi)\cup (\pi,2\pi)$,
$b=\left(\matrix{b_1 & b_2\cr
	b_3 & b_4\cr}\right)$ with $b_i\in\R$ for $i=1, \ldots, 4$, and $c_j\in\R$
for $j=1, 2$, we denote respectively some normal forms by
\bea
&& D(\lm)=\left(\matrix{\lm & 0\cr
	0  & \lm^{-1}\cr}\right), \qquad
R(\th)=\left(\matrix{\cos\th & -\sin\th\cr
	\sin\th  & \cos\th\cr}\right),  \nn\\
&& N_1(\lm, a)=\left(\matrix{\lm & a\cr
	0   & \lm\cr}\right), \qquad
N_2(e^{\sqrt{-1}\th},b) = \left(\matrix{R(\th) & b\cr
	0      & R(\th)\cr}\right),  \nn\\
&& M_2(\lm,c)=\left(\matrix{\lm &   1 &       c_1 &         0\cr
	0 & \lm &       c_2 & (-\lm)c_2 \cr
	0 &   0 &  \lm^{-1} &         0 \cr
	0 &   0 & -\lm^{-2} &  \lm^{-1} \cr}\right). \nn\eea
Here $N_2(e^{\sqrt{-1}\th},b)$ is {\bf trivial} if $(b_2-b_3)\sin\th>0$, or {\bf non-trivial}
if $(b_2-b_3)\sin\th<0$, in the sense of Definition 1.8.11 on p.41 of \cite{Lon4}. Note that
by Theorem 1.5.1 on pp.24-25 and (1.4.7)-(1.4.8) on p.18 of \cite{Lon4}, when $\lm=-1$ there hold
\bea
c_2 \not= 0 &{\rm if\;and\;only\;if}\;& \dim\ker(M_2(-1,c)+I)=1, \nn\\
c_2 = 0 &{\rm if\;and\;only\;if}\;& \dim\ker(M_2(-1,c)+I)=2. \nn\eea

Given any two $2m_k\times 2m_k$ matrices of square block form
$M_k=\left(\matrix{A_k&B_k\cr
                   C_k&D_k\cr}\right)$ with $k=1, 2$,
the symplectic sum of $M_1$ and $M_2$ is defined (cf. \cite{Lon2} and \cite{Lon4}) by
the following $2(m_1+m_2)\times 2(m_1+m_2)$ matrix $M_1\dm M_2$:
$$
M_1\dm M_2=\left(\matrix{A_1 &   0 & B_1 &   0\cr
                             0   & A_2 &   0 & B_2\cr
                             C_1 &   0 & D_1 &   0\cr
                             0   & C_2 &   0 & D_2\cr}\right),
$$
and $M^{\dm k}$ denotes the $k$ copy $\dm$-sum of $M$. For any two paths $\ga_j\in\P_{\tau}(2n_j)$
with $j=0$ and $1$, let $\ga_0\dm\ga_1(t)= \ga_0(t)\dm\ga_1(t)$ for all $t\in [0,\tau]$.

For any $\ga\in \mathcal{P}_{2\pi}(2n)$ we define $\nu_\om(\ga)=\nu_\om(\ga(2\pi))$ and
$$  i_\om(\ga)=[\Sp(2n)_\om^0:\ga\ast\xi^n], \qquad {\rm if}\;\;\ga(2\pi)\not\in \Sp(2n)_{\om}^0,  $$
i.e., the usual homotopy intersection number, and the orientation of the joint path $\ga\ast\xi_n$ is
its positive time direction under homotopy with fixed end points. When $\ga(2\pi)\in \Sp(2n)_{\om}^0$,
we define $i_{\om}(\ga)$ be the index of the left rotation perturbation path $\ga_{-\ep}$ with $\ep>0$
small enough (cf. Def. 5.4.2 on p.129 of \cite{Lon4}). The pair
$(i_{\om}(\ga), \nu_{\om}(\ga)) \in \Z\times \{0,1,\ldots,2n\}$ is called the index function of $\ga$
at $\om$. When $\nu_{\om}(\ga)=0$ or $\nu_{\om}(\ga)>0$, the path $\ga$ is called
$\om$-{\it non-degenerate} or $\om$-{\it degenerate} respectively. For more details we refer to the
\cite{Lon4}.

Now we denote by $\ga_{N;\bb,e}:[0,2\pi]\rightarrow\Sp(4)$ the fundamental solution
of the essential part (\ref{Nonconvex:LHS}).
Here the subscript $``N"$ indicates the non-convex ERE.
We have

\begin{theorem}\label{main.theorem.separation.curves}
	Letting
	\bea  
	\hat\bb_n&=&\frac{n^2-3+\sqrt{9n^4-14n^2+9}}{4} \qquad \forall\;n\in\N.  \lb{1.5}
	\\   \hat\bb_{n+\frac{1}{2}}&=&\frac{(n+\frac{1}{2})^2-3+\sqrt{9(n+\frac{1}{2})^4-14(n+\frac{1}{2})^2+9}}{4}
	\qquad \forall\;n\in\N,  \lb{1.8}
	\eea
	and $\tilde\bb=\sqrt{9-\bb},\tilde\ga_{N;\tilde\bb,e}=\ga_{N;\bb,e}$,
	the following results on the
	linear stability separation curves of $\tilde\ga_{\tilde\bb,e}$ in the parameter $(\tilde\bb,e)$ domain $\Th=(-1,3]\times[0,1)$ hold.
	%Letting
	For every $i\in\N,i\ge1$, there exist functions $e\mapsto\tilde\beta_i(1,e)$ and $e\mapsto\tilde\beta_i(-1,e)$,
	defined for $e\in[0,1)$, such that
	$\tilde\bb_{2n}(1,e)=\tilde\bb_{2n+1}(1,e)$ for every $e\in[0,1)$,
	and if we set
	\bea
	\Ga_n &=& \{(\tilde\bb_{n}(1,e),e)\;|\;e\in [0,1)\}, \nn\\
	\Xi_n &=& \{(\tilde\bb_{n}(-1,e),e)\;|\;e\in [0,1)\}, \nn\eea
	we then have the following:
	
	\begin{figure}[ht]
		\centering
		%\resizebox{10cm}{6cm}
		%{\includegraphics*[0cm,0cm][16cm,9cm]{charged3body.jpg}}%MSS-diagram
		\includegraphics[height=9.5cm]{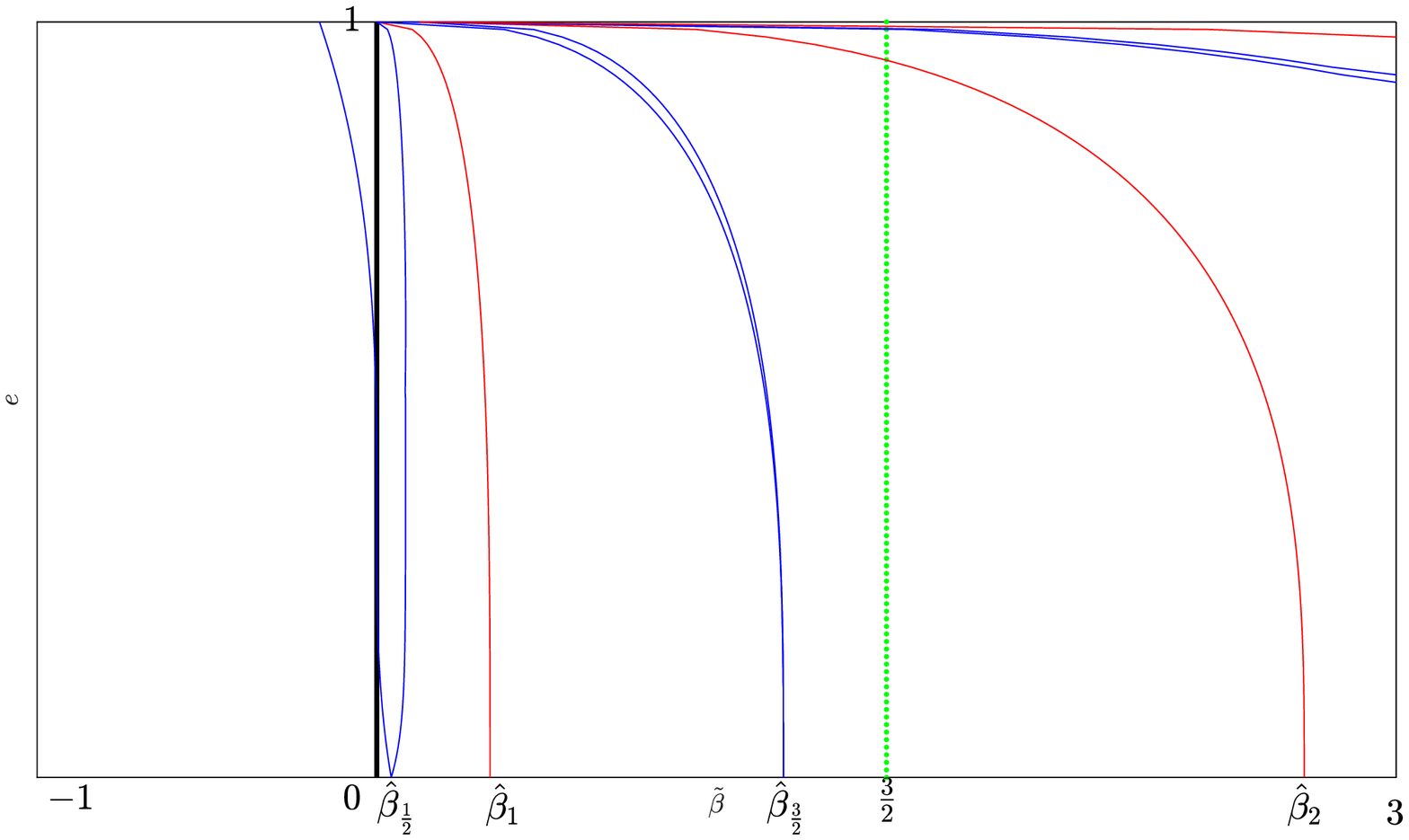}
		\vspace{-1mm}
		\caption{The separation curves with respect to the parameter region in the non-convex case.}
		\label{Nonconvex:bifurcation_curves}
	\end{figure}
	\vspace{2mm}
	
	(i) $\Ga_1$ is the unit segment of the $e$-axis, that is
	$\tilde\bb(1,e)\equiv0$.
	
	(ii) Starting from the point $(\hat\bb_{n},0)$ defined in (\ref{1.5}) for $n\ge1$, there exists
	exactly one $1$-degenerate curve $\Ga_{2n}\;(=\Ga_{2n+1})$
	with multiplicity $2$ of $\tilde\ga_{\tilde\bb,e}(2\pi)$ which is perpendicular to the
	$\tilde\bb$-axis, goes up into the domain $\Th$,
	intersects each horizontal line $e=\constant$ in $\Th$
	precisely once for each $e\in (0,1)$, and satisfies $\nu_1(\tilde\ga_{\tilde\bb_{2n}(1,e),e})=2$ at such an
	intersection point $(\tilde\bb_{2n}(1,e),e)\in\Ga_n$, 
	see Figure \ref{Nonconvex:bifurcation_curves}. 
	Further more, $\tilde\bb_{2n}(1,e)$ is a real analytic function in $e\in [0,1)$.
	
	(iii) Starting from the point $(\hat\bb_{1/2},0)$ defined in (\ref{1.8}), there exist
	exactly two $-1$-degenerate curves $\Xi_{1},\Xi_{2}$ of $\tilde\ga_{\tilde\bb,e}(2\pi)$ goes up into the domain $\Th$. 
	Moreover, for each $e\in (0,1)$,
	if $\tilde\bb_{1}(-1,e)\ne\tilde\bb_{2}(-1,e)$, the two curves intersect each horizontal line $e=\constant$  in $\Th$
	precisely once and satisfy $\nu_1(\tilde\ga_{\tilde\bb_{1}(-1,e),e})=\nu_1(\tilde\ga_{\tilde\bb_{2}(-1,e),e})=1$ at such an
	intersection point $(\tilde\bb_{1}(-1,e),e)\in\Xi_1$ and $(\tilde\bb_{2}(-1,e),e)\in\Xi_2$;
	if $\tilde\bb_{1}(-1,e)=\tilde\bb_{2}(-1,e)$, the two curves intersect each horizontal line $e=\constant$  in $\Th$
	at the same point and satisfy
	$\nu_1(\tilde\ga_{\tilde\bb_{1}(-1,e),e})=2$ at such an intersection point
	$(\tilde\bb_{1}(-1,e),e)\in\Xi_1\cap\Xi_2$. Further more, both $\tilde\bb_{1}(-1,e)$ and
	$\tilde\bb_{2}(-1,e)$ are real piecewise analytic functions in $e\in [0,1)$.
	The tangent directions of $\Xi_1$ and $\Xi_2$ with respect to $e$
	are given by
	\bea
	\pm{41+5\sqrt{1297}\over48\sqrt{1297}}
	\eea
	respectively.
	
	(iv) Starting from the point $(\hat\bb_{n+1/2},0)$ defined in (\ref{1.8}) for $n\ge1$, there exist
	exactly two $-1$-degenerate curves $\Xi_{2n+1},\Xi_{2n+2}$ of $\tilde\ga_{\tilde\bb,e}(2\pi)$ which are perpendicular
	to the $\tilde\bb$-axis goes up into the domain $\Th$. 
	Moreover, for each $e\in (0,1)$,
	if $\tilde\bb_{2n+1}(-1,e)\ne\tilde\bb_{2n+2}(-1,e)$, the two curves intersect each horizontal line $e=\constant$  in $\Th$
	precisely once and satisfy $\nu_1(\tilde\ga_{\tilde\bb_{2n+1}(-1,e),e})=\nu_1(\tilde\ga_{\tilde\bb_{2n+2}(-1,e),e})=1$ at such an
	intersection point $(\tilde\bb_{2n+1}(-1,e),e)\in\Xi_{2n+1}$ and $(\tilde\bb_{2n+2}(-1,e),e)\in\Xi_{2n+2}$;
	if $\tilde\bb_{2n+1}(-1,e)=\tilde\bb_{2n+2}(-1,e)$, the two curves intersect each horizontal line $e=\constant$  in $\Th$
	at the same point and satisfy $\nu_1(\tilde\ga_{\tilde\bb_{2n+1}(-1,e),e})=2$ 
	at such an intersection point
	$(\tilde\bb_{2n+1}(-1,e),e)\in\Xi_{2n+1}\cap\Xi_{2n+2}$. 
	Further more, both $\tilde\bb_{2n+1}(-1,e)$ and
	$\tilde\bb_{2n+2}(-1,e)$ are real piecewise analytic functions in $e\in [0,1)$. Note that in Figure \ref{Nonconvex:bifurcation_curves} the two curves
	which start from the point $(\hat\bb_{n+1/2},0)$ where $n\ge1$ are close enough.
	
	(v) The first $\pm 1$-degenerate curves are intersect each other.
	
	(vi) The $1$-degenerate curves except the first one are all right to 
	the vertical line $\tilde\bb=0$;
	the second $1$-degenerate curve is left to the vertical line
	$\tilde\bb=1$ in the region $\Th$.
	
	(vii) The $1$-degenerate curves except the first one and $-1$-degenerate curves of the
	ERE in Figure \ref{Nonconvex:bifurcation_curves} 
	can be ordered from left to right by
	\begin{equation}
	\Xi_1,\; \Xi_2,\; \Ga_2\;(=\Ga_3),\; \Xi_3,\; \Xi_4,\; \Ga_4(=\Ga_5),\; \ldots,\; \Xi_{2n-1},\; \Xi_{2n},\; \Ga_{2n}(=\Ga_{2n+1}),\;\ldots .
	\end{equation}
	Moreover, for $n_1,n_2\in\N,n_1\ge2$, $\Ga_{n_1}$ and $\Xi_{n_2}$ cannot intersect each other;
	if $n_1\ne n_2,n_1\ge2$, $\Ga_{n_1}$ and $\Ga_{n_2}$ cannot intersect each other.
	More precisely, for each fixed $e\in [0,1)$, we have
	\begin{eqnarray}
	&&\bb_1(-1,e)\le \bb_2(-1,e)<\bb_2(1,e)=\bb_3(1,e)<\bb_3(-1,e)\le \bb_4(-1,e)<\bb_4(1,e)=\bb_5(1,e)<\cdots
	\nonumber
	\\
	&&<\bb_{2n-1}(-1,e)\le \bb_{2n}(-1,e)<\bb_{2n}(1,e)=\bb_{2n+1}(1,e)<\cdots
	\end{eqnarray}
\end{theorem}

\begin{remark}
	Comparing our Figure \ref{Nonconvex:bifurcation_curves} 
	to Figure 1 of \cite{ZL1},
	the most different pattern is that,
	in our Figure \ref{Nonconvex:bifurcation_curves}, 
	the first $-1$-degenerate curve
	crosses the $e$-axis from the right to the left as $e$ increases on $[0,1)$
\end{remark}

Since the first $1$-degenerate curve 
and the first $-1$-degenerate curve intersects each other, 
it is more complicated to consider the region left to the  second $1$-degenerate curve.
In fact, the physical range of $\tilde\bb$ is $[{3\over2},3]$ 
as $\bb\in[0,{27\over4}]$,
and by Theorem \ref{main.theorem.separation.curves}(vi), 
the second $1$-degenerate curve is left to the vertical line
$\tilde\bb={3/2}$.
Therefore, it is reasonable to consider the region right to 
the second $1$-degenerate curve $\tilde\bb_2(1,e)$.
For the normal forms of $\tilde\ga_{\tilde\bb,e}(2\pi)$, 
we have the following theorem.

\begin{theorem}\label{Th:normal.forms.decomposition}
	For the normal forms of $\tilde\ga_{\tilde\bb,e}(2\pi)$ when
	$\tilde\bb\ge\tilde\bb_2(1,e)\;(=\tilde\bb_3(1,e)),0\le e<1,n\in\N$,
	we have the following results:
	
	(i) If $\tilde\bb=\tilde\bb_{2n+1}(1,e)$,
	we have
	$i_1(\tilde\ga_{\tilde\bb,e})=2n-1,\nu_1(\tilde\ga_{\tilde\bb,e})=2$,
	$i_{-1}(\tilde\ga_{\tilde\bb,e})=2n,\nu_{-1}(\tilde\ga_{\tilde\bb,e})=0$
	and $\tilde\ga_{\tilde\bb,e}(2\pi)\approx I_2\diamond D(2)$
	for some $\th\in(0,\pi)$;
	
	(ii) If $\tilde\bb_{2n+1}(1,e)<\tilde\bb<\tilde\bb_{2n+1}(-1,e)$,
	we have
	$i_1(\tilde\ga_{\tilde\bb,e})=2n+1,\nu_1(\tilde\ga_{\tilde\bb,e})=0$,
	$i_{-1}(\tilde\ga_{\tilde\bb,e})=2n,\nu_{-1}(\tilde\ga_{\tilde\bb,e})=0$
	and $\tilde\ga_{\tilde\bb,e}(2\pi)\approx R(\th)\diamond D(2)$;
	
	(iii) If $\tilde\bb=\tilde\bb_{2n+1}(-1,e)=\tilde\bb_{2n+2}(-1,e)$,
	we have
	$i_1(\tilde\ga_{\tilde\bb,e})=2n+1,\nu_1(\tilde\ga_{\tilde\bb,e})=0$,
	$i_{-1}(\tilde\ga_{\tilde\bb,e})=2n,\nu_{-1}(\tilde\ga_{\tilde\bb,e})=2$
	and $\tilde\ga_{\tilde\bb,e}(2\pi)\approx-I_2\diamond D(2)$;
	
	(iv) If $\tilde\bb_{2n+1}(-1,e)\ne\tilde\bb_{2n+2}(-1,e)$ 
	and $\tilde\bb=\tilde\bb_{2n+1}(-1,e)$,
	we have
	$i_1(\tilde\ga_{\tilde\bb,e})=2n+1,\nu_1(\tilde\ga_{\tilde\bb,e})=0$,
	$i_{-1}(\tilde\ga_{\tilde\bb,e})=2n,\nu_{-1}(\tilde\ga_{\tilde\bb,e})=1$
	and $\tilde\ga_{\tilde\bb,e}(2\pi)\approx N_1(-1,-1)\diamond D(2)$;
	
	(v) If $\tilde\bb_{2n+1}(-1,e)\ne\tilde\bb_{2n+2}(-1,e)$ 
	and $\tilde\bb_{2n+1}(-1,e)<\tilde\bb<\tilde\bb_{2n+1}(-1,e)$,
	we have
	$i_1(\tilde\ga_{\tilde\bb,e})=2n+1,\nu_1(\tilde\ga_{\tilde\bb,e})=0$,
	$i_{-1}(\tilde\ga_{\tilde\bb,e})=2n+1,\nu_{-1}(\tilde\ga_{\tilde\bb,e})=0$
	and $\tilde\ga_{\tilde\bb,e}(2\pi)\approx D(-2)\diamond D(2)$;
	
	(vi) If $\tilde\bb_{2n+1}(-1,e)\ne\tilde\bb_{2n+2}(-1,e)$ 
	and $\tilde\bb=\tilde\bb_{2n+2}(-1,e)$,
	we have
	$i_1(\tilde\ga_{\tilde\bb,e})=2n+1,\nu_1(\tilde\ga_{\tilde\bb,e})=0$,
	$i_{-1}(\tilde\ga_{\tilde\bb,e})=2n+1,\nu_{-1}(\tilde\ga_{\tilde\bb,e})=1$
	and $\tilde\ga_{\tilde\bb,e}(2\pi)\approx N_1(-1,1)\diamond D(2)$;
	
	(vii) If $\tilde\bb_{2n+2}(-1,e)<\tilde\bb<\tilde\bb_{2n+2}(1,e)\;(=\tilde\bb_{2n+3}(1,e))$,
	we have
	$i_1(\tilde\ga_{\tilde\bb,e})=2n+1,\nu_1(\tilde\ga_{\tilde\bb,e})=0$,
	$i_{-1}(\tilde\ga_{\tilde\bb,e})=2n+2,\nu_{-1}(\tilde\ga_{\tilde\bb,e})=0$
	and $\tilde\ga_{\tilde\bb,e}(2\pi)\approx R(\th)\diamond D(2)$
	for some $\th\in(\pi,2\pi)$.
\end{theorem}

Here the concept of ``$M\approx N$" for two symplectic matrices $M$ and $N$, i.e., $N\in \Omega^0(M)$,
was first introduced in \cite{Lon4}. This notion is broader than
the symplectic similarity in general as pointed out on p.38 of \cite{Lon4}.

Theorem \ref{main.theorem.nonconvex} is immediately follows 
from Theorem \ref{Th:normal.forms.decomposition}.

For the convex ERE,
we denote by $\ga_{C;\bb,e}:[0,2\pi]\rightarrow\Sp(4)$ the fundamental solution
of the essential part (\ref{Convex:LHS}).
Here the subscript $``C"$ indicates the convex ERE.
We have

\begin{theorem}\label{main.theorem.convex}
For every $e\in[0,1)$, the $-1$ index $i_{-1}(\ga_{C;\bb,e})$ is non-increasing,
and strictly decreasing on two values of $\bb=\bb_1(e,-1)$ and $\bb=\bb_2(e,-1)\in[0,{27\over4}]$
where the second argument $-1$ in $\bb_i,i=1,2$ indicates the $-1$ index.
Define
\begin{equation}
    \beta_l(e)=\min\{\beta_1(e,-1),\beta_2(e,-1)\},\quad
    \beta_m(e)=\max\{\beta_1(e,-1),\beta_2(e,-1)\},
\end{equation}
and
\begin{equation}\label{beta_r}
    \bb_r(e)=\sup\left\{\bb'\in[0,{27\over4}]\;\bigg|\;\sigma(\gamma_{C;\bb,e}(2\pi))\cap{\bf U}\ne\emptyset,\;
    \forall \bb\in[0,\bb']\right\},
\end{equation}
for $e\in[0,1)$.
Let
\begin{equation}
    \Gamma_j=\{(\bb_j(e),e)\in[0,{27\over4}]\times[0,1)\},
\end{equation}
for $j=l,m,r$.
i.e., the curves $\Gamma_l$, $\Gamma_m$ and $\Gamma_r$ are the diagrams of
the functions $\bb_l$, $\bb_m$ and $\bb_r$ with respect to $e\in[0,1)$, respectively.
These three curves separated the parameter rectangle $\Theta=[0,{27\over4}]\times[0,1)$ into four regions,
and we denote them from left to right by I, II, III and IV (see Figure \ref{Convex:bifurcation_curves}), respectively.
Then we have the following:

\begin{figure}[ht]
\centering
%\resizebox{10cm}{6cm}
%{\includegraphics*[0cm,0cm][16cm,9cm]{charged3body.jpg}}%MSS-diagram
\includegraphics[height=9.5cm]{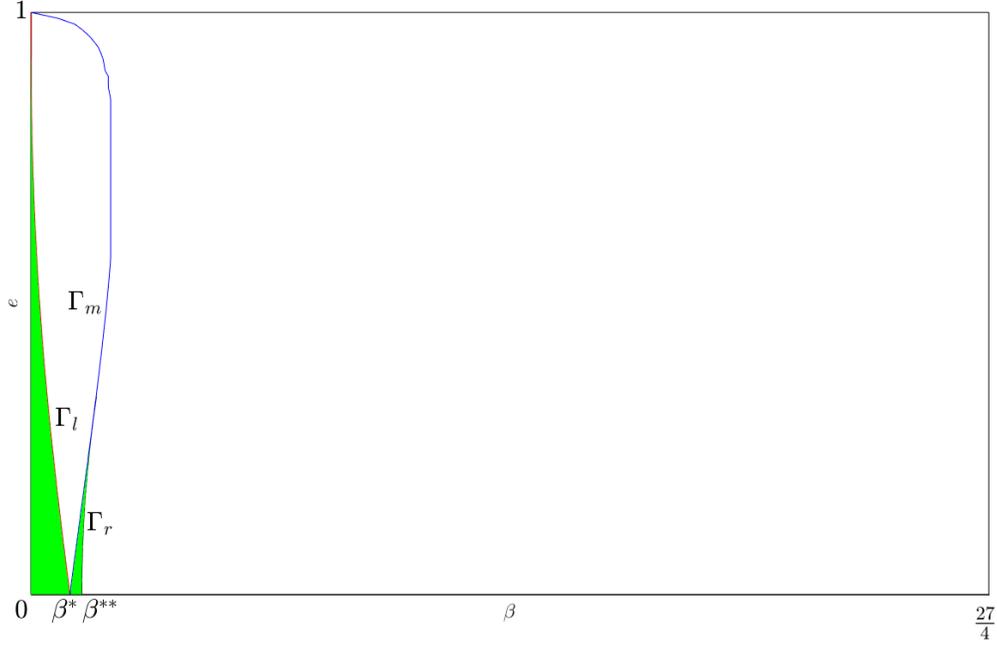}
\vspace{-1mm}
\caption{The three bifurcation curves with respect to the parameter region in the convex case.}
\label{Convex:bifurcation_curves}
\end{figure}
\vspace{2mm}

(i) $0<\bb_i(e,-1)<{27\over4},i=1,2$.
Moreover, $\bb_1(0,-1)=\bb_2(0,-1)=\bb^*$ where $\bb^*$ is given by
\be
\bb^*={1331-35\sqrt{1297}\over288},
\ee
and $\lim_{e\rightarrow1}\bb_1(e,-1)=\lim_{e\rightarrow1}\bb_2(e,-1)=1$;

(ii)
The two functions $\bb_1$ and $\bb_2$ are real analytic in $e$,
and with derivatives $-\frac{2525+67\sqrt{1297}}{288\sqrt{1297}}$,
$\frac{2525+67\sqrt{1297}}{288\sqrt{1297}}$ at $e=0$ with respect to $e$ respectively,
thus they are different and the intersection points of their diagrams must be isolated if there exist when $e\in(0,1)$.
Consequently, $\Gamma_l$ and $\Gamma_m$ are different piecewise real analytic curves;

(iii) We have
\begin{equation}
    i_{-1}(\gamma_{C;\bb,e}) = \left\{\matrix{
   	2, &  {\it if}\;\;\bb\in[0,\bb_l(e)), \cr
   	1, &  {\it if}\;\;\bb\in[\bb_l(e),\mu_m(e)),\cr
   	0, &  {\it if}\;\;\bb\in[\bb_m(e),{27\over4}], }\right.
\end{equation}
and $\Gamma_l$ and $\Gamma_m$ are precisely the $-1$-degenerate curves of the path $\gamma_{C;\bb,e}$
in the $(\bb,e)$-rectangle $\Theta=[0,{27\over4}]\times[0,1)$;

(iv) Every matrix $\gamma_{C;\bb,e}(2\pi)$ is hyperbolic when $\bb\in(\bb_l(e),{27\over4}]$, $e\in[0,1)$, and there holds
\begin{equation}
    \bb_l(e)=\inf\left\{\bb\in[0,{27\over4}]\;\bigg|\;\sigma(\gamma_{C;\bb,e}(2\pi))\cap{\bf U}=\emptyset,\;\forall e\in[0,1)\right\}.
\end{equation}
Consequently, $\Gamma_l$ is the boundary curve of the hyperbolic region of $\gamma_{C;\bb,e}(2\pi)$
in $\Theta$;

(v) $\Gamma_l$ is continuous in $e\in[0,1)$, and $\mu_l(0)=\bb^{**}$
where
\be
\bb^{**}={16\;(182-37\sqrt{21})\over625};
\ee

(vi) $\Gamma_m$ is different from the curve $\Gamma_r$ at least when $e\in[0,\tilde{e})$ for some $\tilde{e}\in(0,1)$;

(vii) In Region I, i.e., when $0<\bb<\bb_l(e)$,
we have $\gamma_{C;\bb,e}(2\pi)\approx R(\theta_1)\diamond R(\theta_2)$ for some $\theta_1,\theta_2\in(\pi,2\pi)$,
and thus it is strongly linear stable;

(viii) In Region II, i.e., when $\bb_l(e)<\bb<\bb_m(e)$,
we have $\gamma_{C;\bb,e}(2\pi)\approx R(\theta)\diamond D(-2)$ for some $\theta\in(\pi,2\pi)$,
and thus it is linearly unstable;

(ix) In Region III, i.e., when $\bb_m(e)<\bb<\bb_r(e)$,
we have $\gamma_{C;\bb,e}(2\pi)\approx R(\theta_1)\diamond R(\theta_2)$ for some $\theta_1\in(\pi,2\pi)$ and $\theta_2\in(0,\pi)$,
and thus it is strongly linear stable;

(x) In Region IV, i.e., when $\bb_r(e)<\bb\le{27\over4}$,
we have $\gamma_{C;\bb,e}(2\pi)$ is hyperbolic, and thus it is linearly unstable.
\end{theorem}

A conjecture of Moeckel \cite{ACS} states that a relative equilibrium is linearly stable only if the corresponding central configuration
is a non-degenerate minimum of the potential function $U$ restricted
to the sphere $I={1\over2}$.
We claim that the conjecture is true
when the two small masses $m_3$ and $m_4$ tend to $0$.
As mentioned before, when $m_3,m_4\rightarrow0$,
the positions of the two small masses must tend to one of the five Lagrangian
points of the primary masses $m_1$ and $m_2$.
When $m_3$ and $m_4$ tend to the different Lagrangian points,
the linear stability of such a relative equilibrium (ERE with $e=0$)
implies they must tend to $L_4, L_5$ (or $L_5, L_4$) respectively.
Then the eigenvalues of the Hessian
$\sg(D^2U|_{I={1\over2}})
=\left\{{3+\sqrt{9-\bb}\over2}\mu_0,{3-\sqrt{9-\bb}\over2}\mu_0,{3+\sqrt{9-\bb}\over2}\mu_0,{3-\sqrt{9-\bb}\over2}\mu_0\right\}$ 
where $\mu_0>0$ is given by (\ref{mu0}) below, 
and hence the corresponding central configuration
is a non-degenerate minimum.
When $m_3$ and $m_4$ tend to the same Lagrangian point, say $L_4$,
then by Theorem \ref{main.theorem.nonconvex} 
and Theorem \ref{main.theorem.convex},
the linear stability of such a relative equilibrium 
implies the corresponding central configuration is convex.
Moreover, the eigenvalues of the Hessian
$\sg(D^2U|_{I={1\over2}})
=\left\{{3+\sqrt{9-\bb}\over2}\mu_0,\;{3-\sqrt{9-\bb}\over2}\mu_0,\;\sqrt{9-\bb}\mu_0\right.$,
$\left.{9-3\sqrt{9-\bb}\over2}\mu_0\right\}$
by (\ref{eig.of.Hessian}) below,
and hence the corresponding central configuration
is a non-degenerate minimum.

This paper is organized as follows.
In Section 2, We reduced the linearized Hamiltonian systems near ERE
for the general $4$-body problem.
In Section 3, we focus on the proof of Theorem \ref{main.theorem.decomposition}.
In Section 4, we study the linear stability of non-convex ERE, 
and also prove Theorem \ref{main.theorem.separation.curves}
and Theorem \ref{Th:normal.forms.decomposition};
In the last section, we study the linear stability of convex ERE,
and Theorem \ref{main.theorem.convex} is also proved there.

\setcounter{equation}{0}%\setcounter{figure}{0}

\section{The symplectic reduction of the linearized Hamiltonian systems near  elliptic relative equilibrium}\label{sec:2}

\subsection{Two Useful Maps}
In this subsection, we introduce two useful maps for our later use.
We define $\Phi,\Psi:\C\rightarrow \R^{2\times2}$ by
\bea
\Phi(z)=\left(\matrix{x & -y\cr
                   y & x}\right) \quad\quad  
               \forall z=x+\sqrt{-1}y\in\C,\;x,y\in\R,   \label{map1}
\\
\Psi(z)=\left(\matrix{x & y\cr
                   y & -x}\right) \quad\quad  
               \forall z=x+\sqrt{-1}y\in\C,\;x,y\in\R.   \label{map2}
\eea
Thus both $\Phi$ and $\Psi$ are real linear maps.
Direct computation shows that:
\begin{lemma}
(i) If $z\in\R$, then
\be
\Phi(z)=zI_2,\quad\quad \Psi(z)=z\left(\matrix{1 & 0\cr 0 & -1}\right);
\ee
(ii) For any $z\in\C$, we have
\bea
\Phi(z)^T=\Phi(\bar{z}),
\\
\Psi(z)^T=\Psi(z);
\eea
(iii) For any $z,w\in\C$, we have
\bea
\Phi(z)\Phi(w)&=&\Phi(zw),
\\
\Psi(z)\Psi(w)&=&\Phi(z\bar{w}),
\\
\Phi(z)\Psi(w)&=&\Psi(zw),
\\
\Psi(z)\Phi(w)&=&\Psi(z\bar{w}).
\eea
Specially, we have
\bea
\Phi(\bar{z})\Phi(z)=\Phi(z)\Phi(\bar{z})=\Phi(|z|^2)=|z|^2I_2,
\\
\Psi(z)\Psi(z)=\Psi(\bar{z})\Psi(\bar{z})=\Phi(|z|^2)=|z|^2I_2.
\eea
\end{lemma}

\begin{remark} \label{map.matrix}
For a $m\times n$  complex matrix $N$, we define $\Phi(N)$ as
\be
\Phi(N)=\left(
\matrix{
\Phi(N_{11})& \Phi(N_{12})& \ldots& \Phi(N_{1n})\cr
\Phi(N_{21})& \Phi(N_{22})& \ldots& \Phi(N_{2n})\cr
\ldots&          \ldots&          \ldots& \ldots\cr
\Phi(N_{m1})& \Phi(N_{m2})& \ldots& \Phi(N_{mn})
}
\right).
\ee
Thus $\Phi(N)$ is a $2m\times 2n$ matrix.
\end{remark}

\subsection{Decomposition of the linearized Hamiltonian Systems for general $4$-body problem}%Subsection 2.2

In \cite{MS} (cf. p.275), Meyer and Schmidt gave the essential part of the fundamental solution of the
elliptic Lagrangian orbit. Their method is explained in \cite{Lon5} too. Our study on ERE is based upon their method.

Suppose the four particles are located at $a_1=(a_{1x},a_{1y}),a_2=(a_{2x},a_{2y}),a_3=(a_{3x},a_{3y}),a_4=(a_{4x},a_{4y})$.
We suppose $a_1$, $a_2,a_3,a_4$ form a collinear central configurations.
For convenience, we define four corresponding complex numbers:
\be
z_{a_i}=a_{ix}+\sqrt{-1}a_{iy},\quad i=1,2,3,4.  \label{complex.rep}
\ee

Without lose of generality, we normalize the three masses by
\begin{equation}\label{nomorlize.the.masses}
\sum_{i=1}^n m_i=1,
\end{equation}
and normalize the positions $a_i,1\le i\le 4$ by
\bea
&&\sum_{i=1}^4 m_ia_i=0,  \label{mass.center}
\\
&&\sum_{i=1}^4 m_i|a_i|^2=2I(a)=1.\label{inertia}
\eea
Using the notations in (\ref{complex.rep}), (\ref{mass.center}) and (\ref{inertia}) are equivalent to
\bea
&&\sum_{i=1}^4 m_iz_{a_i}=0,  \label{mass.center'}
\\
&&\sum_{i=1}^4 m_i|z_{a_i}|^2=2I(a)=1.\label{inertia'}
\eea
Moreover, we define
\begin{equation}\label{mu}
\mu=U(a)=\sum_{1\le i<j\le 4}\frac{m_im_j}{|a_i-a_j|}=\sum_{1\le i<j\le 4}\frac{m_im_j}{|z_{a_i}-z_{a_j}|},
\quad
\sigma=(\mu p)^{1/4},
\end{equation}
and
\be
\tilde{M}=\diag(m_1,m_2,m_3,m_4),\quad M=\diag(m_1,m_1,m_2,m_2,m_3,m_3,m_4,m_4).
\ee
Because $a_1,a_2,a_3,a_4$ form a collinear central configuration, we have
\be\label{eq.of.cc}
\sum_{j=1,j\ne i}^4\frac{m_j(z_{a_{j}}-z_{a_{i}})}{|z_{a_{i}}-z_{a_{j}}|^3}=
\frac{U(a)}{2I(a)}z_{a_{i}}=\mu z_{a_{i}}.
\ee

Let $B$ be a $4\times 4$ symmetric matrix such that
\begin{equation}
B_{ij}=\left\{\begin{array}{c}
               \frac{m_im_j}{|z_{a_i}-z_{a_j}|^3}\quad{if}\;i\ne j,1\le i,j\le 4,\\
               -\sum_{j=1,j\ne i}^4\frac{m_im_j}{|z_{a_i}-z_{a_j}|^3}\quad{if}\;i=j,1\le i\le 4,
              \end{array}\right.
\end{equation}
and
\bea
D&=&\mu I_4+\tilde{M}^{-1}B, \label{D}
\\
\tilde{D}&=&\mu I_4+\tilde{M}^{-1/2}B\tilde{M}^{-1/2}=\tilde{M}^{1/2}D\tilde{M}^{-1/2}. \label{tilde.D}
\eea
where $\mu$ is given by (\ref{mu}).

Firstly, $D$ has two simple eigenvalues: $\lambda_1=\mu$ with $v_1=(1,1,\ldots,1)^T$, and $\lambda_2=0$ with
$v_2=(z_{a_1},z_{a_2},z_{a_3},z_{a_4})^T$.
Exactly, we have
\bea
(Dv_1)_i&=&\mu-\sum_{j=1,j\ne i}^4\frac{m_j}{|z_{a_i}-z_{a_j}|^3}+\sum_{j=1,j\ne i}^4\frac{m_j}{|z_{a_i}-z_{a_j}|^3}=\mu,
\\
(Dv_2)_i&=&(\mu-\sum_{j=1,j\ne i}^4\frac{m_j}{|z_{a_i}-z_{a_j}|^3})z_{a_i}+\sum_{j=1,j\ne i}^4\frac{m_jz_{a_j}}{|z_{a_i}-z_{a_j}|^3}
\nonumber
\\
&=&\mu z_{a_i}+\sum_{j=1,j\ne i}^\frac{m_j(z_{a_j}-z_{a_i})}{|z_{a_i}-z_{a_j}|^3}
\nonumber
\\
&=&\mu z_{a_i}-\mu z_{a_i}
\nonumber
\\
&=&0,
\eea
where in the second last equality, we used (\ref{eq.of.cc}).
Moreover by (\ref{nomorlize.the.masses})-(\ref{inertia}), we have
\bea
\overline{v}_1^T\tilde{M}v_1&=&\sum_{i=1}^n m_i=1,\label{v1.M.v1}
\\
\overline{v}_1^T\tilde{M}v_2&=&\sum_{i=1}^4 m_iz_{a_i}=0,\label{v1.M.v2}
\\
\overline{v}_2^T\tilde{M}v_1&=&\sum_{i=1}^4 m_i\overline{z}_{a_i}=0,\label{v1.M.v2}
\\
\overline{v}_2^T\tilde{M}v_2&=&\sum_{i=1}^n m_i|z_{a_i}|^2=1.\label{v2.M.v2}
\eea
Let $\overline{v}_2=(\overline{z}_{a_1},\overline{z}_{a_2},\overline{z}_{a_3},\overline{z}_{a_4})^T$.
Because, $a_1,a_2,a_3,a_4$ forms a nonlinear central configuration,
$\overline{v}_2$ is independent with $v_2$. Moreover, $\overline{v}_2$ is also independent with $v_1$.
So $\overline{v}_2$ is another eigenvector of $D$ corresponding to eigenvalue $\lambda_3=0$.

Now, we construct $v_3$. We suppose
\be
v_3=k\overline{v}_2+lv_2  \label{v_3}
\ee
with $k\in\R,l\in\C$ will be given later.
If ${v}_2^T\tilde{M}v_2=\sum_{i=1}^n m_iz_{a_i}^2=0$, we set $k=1,l=0$, i.e., $v_3=\overline{v}_2$.
Then we have
\bea
\overline{v}_1^T\tilde{M}v_3&=&\sum_{i=1}^4 m_i\overline{z}_{a_i}=0,\label{v1.M.v3}
\\
\overline{v}_2^T\tilde{M}v_3&=&\sum_{i=1}^4 m_i\overline{z}_{a_i}^2=0,\label{v2.M.v3}
\\
\overline{v}_3^T\tilde{M}v_3&=&\sum_{i=1}^n m_i|z_{a_i}|^2=1.\label{v3.M.v3}
\eea
In the other cases, we also hope (\ref{v1.M.v3})-(\ref{v3.M.v3}) are satisfied.
Thus we have
\bea
0&=&\overline{v}_2^T\tilde{M}v_3=\overline{v}_2^T\tilde{M}(k\overline{v}_2+lv_2)=k\sum_{i=1}^4 m_i\overline{z}_{a_i}^2+l,
\\
1&=&\overline{v}_3^T\tilde{M}v_3=(kv_2+\overline{l}\overline{v}_2)^T\tilde{M}(k\overline{v}_2+lv_2)
=k^2+|l|^2+kl\sum_{i=1}^4 m_iz_{a_i}^2+k\overline{l}\sum_{i=1}^4 m_i\overline{z}_{a_i}^2.
\eea
Therefore, we have
\bea
k=\frac{1}{\sqrt{1-|\sum_{i=1}^4 m_i\overline{z}_{a_i}^2|^2}}, \label{param.k}
\\
l=-\frac{\sum_{i=1}^4 m_i\overline{z}_{a_i}^2}{\sqrt{1-|\sum_{i=1}^4 m_i\overline{z}_{a_i}^2|^2}}. \label{param.l}
\eea

We now construct a unitary matrix $\tilde{A}$ based on $v_1,v_2$ and $v_3$.
That is
\be
\tilde{A}=
\left(
\matrix{
1\quad z_{a_1}\quad b_1\quad c_1\cr
1\quad z_{a_2}\quad b_2\quad c_2\cr
1\quad z_{a_3}\quad b_3\quad c_3\cr
1\quad z_{a_4}\quad b_4\quad c_4
}
\right),
\ee
where $(b_1,b_2,b_3,b_4)=v_3^T$,i.e., $b_i=k\overline{z}_{a_i}+lz_{a_i},1\le i\le4$.
Then $c_i=A_{i4}$, where $A_{i4}$ is the algebraic cofactor of $c_i$.

In the other hand, the signed area of the triangle formed by $a_i,a_j$ and $a_k$ is given by
\be
\Delta_{ijk}=\frac{\sqrt{-1}}{4}\det
\left(
\matrix{
1\quad z_{a_1}\quad \overline{z}_{a_1}\cr
1\quad z_{a_2}\quad \overline{z}_{a_2}\cr
1\quad z_{a_3}\quad \overline{z}_{a_3}
}
\right).
\ee
Then $c_1=\overline{4k\sqrt{-1}\Delta_{234}}=-4k\sqrt{-1}\Delta_{234}$ and so on.
Note that, for any $\om\in\C,|\om|=1$, if $c_i$ are replaced by $\om c_i,i=1,2,3,4$, $\tilde{A}$ is also a unitary matrix.
Thus we can let
\be
(c_1,c_2,c_3,c_4)=({4k\rho\over m_1}\Delta_{234},-{4k\rho\over m_2}\Delta_{134},{4k\rho\over m_3}\Delta_{124},-{4k\rho\over m_4}\Delta_{123}), \label{c}
\ee
where
\be
\rho=\sqrt{m_1m_2m_3m_4}.
\ee
For convenience, we also write $v_4$ as
\be
v_4=(c_1,c_2,c_3,c_4)^T\in\R^4.  \label{v_4}
\ee
Now $v_1,v_2,v_3,v_4$ forms a unitary basis of $\C^n$.
Note that $v_1,v_2,v_3$ are eigenvectors of matrix $D$, then $v_4$ is also an eigenvector of $D$ with the corresponding eigenvalue
\be
\lambda_4=tr(D)-\lambda_1-\lambda_2-\lambda_3=tr(D)-\mu.
\ee
Moreover, we define
\bea
\bb_1&=&-\frac{\lambda_3}{\mu}=0,  \label{bb1}
\\
\bb_2&=&-\frac{\lambda_4}{\mu}=1-\frac{tr(D)}{\mu}.  \label{bb2}
\eea

In the following, if there is no confusion, we will use $a_i$ to represent $z_{a_i},1\le i\le4$.
By the definition of (\ref{v_3}) and (\ref{v_4}),
$Dv_k=\lambda_kv_k,k=3,4$ reads
\bea
\mu b_i-\sum_{j=1,j\ne i}^4\frac{m_j(b_j-b_i)}{|a_i-a_j|^3}=\lambda_3 b_i,\quad 1\le i\le 4,
\\
\mu c_i-\sum_{j=1,j\ne i}^4\frac{m_j(c_j-c_i)}{|a_i-a_j|^3}=\lambda_4 c_i,\quad 1\le i\le 4,
\eea
Let
\be
F_i=\sum_{j=1,j\ne i}^4\frac{m_im_j(b_i-b_j)}{|a_i-a_j|^3},\;
G_i=\sum_{j=1,j\ne i}^4\frac{m_im_j(c_i-c_j)}{|a_i-a_j|^3},\quad 1\le i\le 4, \label{F_ki}
\ee
then we have
\be
F_i=(\mu-\lambda_3)m_ib_i=\mu(1+\bb_1)m_ib_i,\quad G_i=(\mu-\lambda_4)m_ic_i=\mu(1+\bb_2)m_ic_i. \label{Fi.bi}
\ee
%Moreover, we have
%\be
%\sum_{i=1}^nF_{ik}b_{ik}=\sum_{i=1}^n(\mu-\lambda_k)m_ib_{ik}^2=\mu-\lambda_k=\mu(1+\bb_{k-2}), \label{sum.of.Fi.bi}
%\ee
%where in the last equality, we used (\ref{bi}).

%For later convenience, we introduce a new equation about $b_i,c_i,1\le i\le4$:
%\be
%\sum_{1\le i<j\le4}\frac{m_im_j(b_i-b_j)(c_i-c_j)}{|a_i-a_j|^3}=0. \label{eq.BC.8'}
%\ee
%By (\ref{AMA}), $\tilde{M}^{1/2}\tilde{A}$ is an orthogonal matrix, thus we have
%\be
%c_i=\frac{\rho}{m_i}\tilde{A}_{i4},
%\ee
%where $\tilde{A}_{i4}$ is the algebraic complement of $c_i$.
%Then we have
%\bea
%&&\sum_{1\le i<j\le4}\frac{m_im_j(b_i-b_j)(c_i-c_j)}{|a_i-a_j|^3}
%\nonumber
%\\
%&&=\rho\sum_{1\le i<j\le4}\frac{(b_i-b_j)(m_j\tilde{A}_{i4}-m_i\tilde{A}_{j4})}{|a_i-a_j|^3}
%\nonumber\\
%&&=\rho\left(\sum_{1\le i<j\le4}\tilde{A}_{i4}\frac{m_j(b_i-b_j)}{|a_i-a_j|^3}
%-\sum_{1\le i<j\le4}\tilde{A}_{j4}\frac{m_i(b_i-b_j)}{|a_i-a_j|^3}\right)
%\nonumber
%\\
%&&=\rho\left(\sum_{1\le i<j\le4}\tilde{A}_{i4}\frac{m_j(b_i-b_j)}{|a_i-a_j|^3}
%+\sum_{1\le j<i\le4}\tilde{A}_{i4}\frac{m_j(b_i-b_j)}{|a_i-a_j|^3}\right)
%\nonumber
%\\
%&&=\rho\sum_{i=1}^4\tilde{A}_{i4}\sum_{j=1,j\ne i}^4\frac{m_j(b_i-b_j)}{|a_i-a_j|^3}.
%\eea
%Let
%\begin{equation}
%F=
%\left(
%\matrix{
%1\quad a_{1x}\quad b_1\quad F_1/m_1\cr
%1\quad a_{2x}\quad b_2\quad F_2/m_2\cr
%1\quad a_{3x}\quad b_3\quad F_3/m_3\cr
%1\quad a_{4x}\quad b_4\quad F_4/m_4
%}
%\right).
%\end{equation}
%Then we obtain
%\be
%\sum_{1\le i<j\le4}\frac{m_im_j(b_i-b_j)(c_i-c_j)}{|a_i-a_j|^3}=\det F=0,
%\ee
%where the last equality is obtained by (\ref{Fi.bi}).

Now as in p.263 of \cite{MS}, Section 11.2 of \cite{Lon5}, we define
\begin{equation}\label{PQYX}
P=\left(\matrix{p_1\cr p_2\cr p_3\cr p_4}\right),
\quad
Q=\left(\matrix{q_1\cr q_2\cr q_3\cr q_4}\right),
\quad
Y=\left(\matrix{G\cr Z\cr W_1\cr  W_2}\right),
\quad
X=\left(\matrix{g\cr z\cr w_1\cr w_2}\right),
\end{equation}
where $p_i$, $q_i$, $i=1,2,3,4$ and $G$, $Z$, $W_1\;W_2$, $g$, $z$, $w_1,w_2$ are all column vectors in $\R^2$.
We make the symplectic coordinate change
\be\lb{transform1}  P=A^{-T}Y,\quad Q=AX,  \ee
where the matrix $A$ is constructed as in the proof of Proposition 2.1 in \cite{MS}.
Concretely, the matrix $A\in {\bf GL}(\R^{8})$ is given by
\begin{equation}
A=
\left(
\matrix{
I\quad A_1\quad B_{1}\quad C_{1}\cr
I\quad A_2\quad B_{2}\quad C_{2}\cr
I\quad A_3\quad B_3\quad   C_3\cr
I\quad A_4\quad B_{4}\quad C_4
}
\right),
\end{equation}
where each $A_i$ is a $2\times2$ matrix given by
\begin{eqnarray}
A_i &=& (a_i, Ja_i)=\Phi(a_i),  \label{Aa}
\\
B_i &=& (b_i, Jb_i)=\Phi(b_i),  \label{Bb}
\\
C_i &=& (c_i, Jc_i)=\Phi(c_i)=c_iI_2, \label{Cc}
\end{eqnarray}
where $\Phi$ is given by (\ref{map1}).
Moreover, by the definition of $v_i,1\le i\le 4$, we obtain
\bea
\overline{\tilde{A}}^T\tilde{M}\tilde{A}&=&(\overline{v}_1,\overline{v}_2,\overline{v}_3,\overline{v}_4)^T\tilde{M}(v_1,v_2,v_3,v_4)
=I_4
\eea
By (\ref{map.matrix}), we have
$A^TMA=\Phi(\tilde{A})^T\Phi(\tilde{M})\Phi(\tilde{A})=\Phi(\overline{\tilde{A}}^T\tilde{M}\tilde{A})=\Phi(I_4)=I_{8}$
is fulfilled (cf. (13) in p.263 of \cite{MS}).

Now we consider the Hamiltonian function of the four-body problem.
Under the coordinate change (\ref{transform1}), we get the kinetic enrgy
\begin{equation}
K=\frac{1}{2}(|G|^2+|Z|^2+|W_1|^2+|W_2|^2),
\end{equation}
and the potential function
\begin{eqnarray}
U_{ij}(z,w_1,w_2)&=&\frac{m_im_j}{d_{ij}(z,w_1,w_2)}, \label{U_ij}
\\
U(z,w_1,w_2)&=&\sum_{1\le i<j\le 4}U_{ij}(z,w_1,w_2),\label{U}
\end{eqnarray}
with
\begin{eqnarray}
d_{ij}(z,w_1,w_2)&=&|(A_i-A_j)z+(B_i-B_j)w_1+(C_i-C_j)w_2|
\nonumber
\\
&=&|\Phi(a_i-a_j)z+\Phi(b_i-b_j)w_1+\Phi(c_i-c_j)w_2|,
\end{eqnarray}
where we used (\ref{Aa})-(\ref{Cc}).

Let $\theta$ be the true anomaly.
Then under the same steps of symplectic transformation in the proof of Theorem 11.10 (p. 100 of \cite{Lon5}),
the resulting Hamiltonian function of the 3-body problem is given by
\begin{eqnarray}\label{new.H.function}
&&H(\theta,\bar{Z},\bar{W_1},\bar{W}_2,\bar{z},\bar{w_1},\bar{w}_2)=\frac{1}{2}(|\bar{Z}|^2+|\bar{W_1}|^2+|\bar{W_2}|^2)
+(\bar{z}\cdot J\bar{Z}+\bar{w_1}\cdot J\bar{W_1}+\bar{w_2}\cdot J\bar{W_2})
\nonumber
\\
&&\quad\quad\quad\quad\quad+\frac{p-r(\theta)}{2p}(|\bar{z}|^2+|\bar{w_1}|^2+|\bar{w_2}|^2)
-\frac{r(\theta)}{\sigma}U(\bar{z},\bar{w_1},\bar{w}_2),
\end{eqnarray}
where $\mu$ is given by (\ref{mu}) and
\begin{equation}
r(\theta)=\frac{p}{1+e\cos\theta}.
\end{equation}

We now derived the linearized Hamiltonian system at the elliptic relative equilibrium.
\begin{proposition}\label{linearized.Hamiltonian}
Using notations in (\ref{PQYX}), elliptic Euler solution $(P(t),Q(t))^T$ of the system (\ref{1.2}) with
\begin{equation}
Q(t)=(r(t)R(\theta(t))a_1,r(t)R(\theta(t))a_2,r(t)R(\theta(t))a_3,r(t)R(\theta(t))a_4)^T,\quad P(t)=M\dot{Q}(t)
\end{equation}
in time $t$ with the matrix $M=diag(m_1,m_1,m_2,m_2,m_3,m_3,m_4,m_4)$,
is transformed to the new solution $(Y(\theta),X(\theta))^T$ in the variable true anomaly $\theta$
with $G=g=0$ with respect to the original Hamiltonian function $H$ of (\ref{new.H.function}), which is given by
\begin{equation}
Y(\theta)=\left(
\matrix{
\bar{Z}(\theta)\cr
\bar{W}_1(\theta)\cr
\bar{W}_2(\theta)}
\right)
=\left(
\matrix{
0\cr
\sigma\cr
0\cr
0\cr
0\cr
0}
\right),
\quad
X(\theta)=\left(
\matrix{
\bar{z}(\theta)\cr
\bar{w_1}(\theta)\cr
\bar{w_2}(\theta)}
\right)
=\left(
\matrix{
\sigma\cr
0\cr
0\cr
0\cr
0\cr
0}
\right).
\end{equation}

Moreover, the linearized Hamiltonian system at the elliptic Euler solution
${\xi}_0\equiv(Y(\theta),X(\theta))^T =$
\newline
$(0,\sigma,0,0,0,0,\sigma,0,0,0,0,0)^T\in\R^{12}$
depending on the true anomaly $\theta$ with respect to the Hamiltonian function
$H$ of (\ref{new.H.function}) is given by
\begin{equation}
\dot\zeta(\theta)=JB(\theta)\zeta(\theta),  \label{general.linearized.Hamiltonian.system}
\end{equation}
with
\begin{eqnarray}
B(\theta)&=&H''(\theta,\bar{Z},\bar{W_1},\bar{W}_2,\bar{z},\bar{w_1},\bar{w}_2)|_{\bar\xi=\xi_0}
\nonumber
\\
&=&\left(
\matrix{
I& O& O& -J&  O&  O\cr
O& I& O&  O& -J&  O\cr
O& O& I&  O&  O& -J\cr
J& O& O& H_{\bar{z}\bar{z}}(\theta,\xi_0)& O& O\cr
O& J& O& O& H_{\bar{w_1}\bar{w_1}}(\theta,\xi_0)& H_{\bar{w_1}\bar{w_2}}(\theta,\xi_0)\cr
O& O& J& O& H_{\bar{w_2}\bar{w_1}}(\theta,\xi_0)& H_{\bar{w_2}\bar{w_2}}(\theta,\xi_0)
}
\right),
\end{eqnarray}
and
\begin{eqnarray}
H_{\bar{z}\bar{z}}(\theta,\xi_0)&=&\left(
\matrix{
-\frac{2-e\cos\theta}{1+e\cos\theta} & 0\cr
0 & 1
}
\right),
\quad
\\
H_{\bar{w_i}\bar{w_i}}(\theta,\xi_0)&=&I_2-\frac{r}{p}\left[\frac{3+\bb_i}{2}I_2+\Psi(\bb_{ii})\right],\ \ i=1,2, \label{H_12}
\\
H_{\bar{w_1}\bar{w_2}}(\theta,\xi_0)&=&-\frac{r}{p}\Psi(\bb_{12}),
\end{eqnarray}
where $\bb_1=0$ and
$\bb_2$ are given by (\ref{bb2}),
and $\bb_{11},\bb_{12},\bb_{22}$ are given by
\bea
\bb_{11}&=&{3\over2\mu}\sum_{1\le i<j\le 4}\frac{m_im_j(a_i-a_j)^2(\overline{b}_i-\overline{b}_j)^2}{|a_i-a_j|^5},  \label{bb_11}
\\
\bb_{12}&=&{3\over2\mu}\sum_{1\le i<j\le 4}\frac{m_im_j(a_i-a_j)^2(\overline{b}_i-\overline{b}_j)(\overline{c}_i-\overline{c}_j)}{|a_i-a_j|^5},
\label{bb_12}
\\
\bb_{22}&=&{3\over2\mu}\sum_{1\le i<j\le 4}\frac{m_im_j(a_i-a_j)^2(\overline{c}_i-\overline{c}_j)^2}{|a_i-a_j|^5},\label{bb_22}
\eea
and $H''$ is the Hession Matrix of $H$ with respect to its variable $\bar{Z}$,
$\bar{W_1},\bar{W}_2$, $\bar{z}$, $\bar{w_1},\bar{w}_2$.
The corresponding quadratic Hamiltonian function is given by
\begin{eqnarray}
H_2(\theta,\bar{Z},\bar{W_1},\bar{W}_2,\bar{z},\bar{w_1},\bar{w}_2)
&=&\frac{1}{2}|\bar{Z}|^2+\bar{Z}\cdot J\bar{z}+\frac{1}{2}H_{\bar{z}\bar{z}}(\theta,\xi_0)|\bar{z}|^2+H_{\bar{w_1}\bar{w_2}}(\theta,\xi_0)\bar{w_1}\cdot\bar{w_2}
\nonumber\\
&&+\left(\frac{1}{2}|\bar{W_1}|^2+\bar{W_1}\cdot J\bar{w_1}+\frac{1}{2}H_{\bar{w_1}\bar{w_1}}(\theta,\xi_0)|\bar{w_1}|^2\right)
\nonumber\\
&&+\left(\frac{1}{2}|\bar{W_2}|^2+\bar{W_2}\cdot J\bar{w_2}+\frac{1}{2}H_{\bar{w_2}\bar{w_2}}(\theta,\xi_0)|\bar{w_2}|^2\right).
\end{eqnarray}
\end{proposition}

{\bf Proof.} The proof is similar to those of Proposition 11.11 and Proposition 11.13 of \cite{Lon5}.
We just need to compute $H_{\bar{z}\bar{z}}(\theta,\xi_0)$, $H_{\bar{z}\bar{w_i}}(\theta,\xi_0)$
and $H_{\bar{w_i}\bar{w_j}}(\theta,\xi_0)$ for $i,j=1,2$.

For simplicity, we omit all the upper bars on the variables of $H$ in (\ref{new.H.function}) in this proof.
By (\ref{new.H.function}), we have
\bea
H_z&=&JZ+\frac{p-r}{p}z-\frac{r}{\sigma}U_z(z,w_1,w_2),  \nn\\
H_{w_i}&=&JW_i+\frac{p-r}{p}w_i-\frac{r}{\sigma}U_{w_i}(z,w_1,w_2), \quad i=1,2, \nn
\eea
and
\be\lb{Hessian}\left\{
\begin{array}{l}
H_{zz}=\frac{p-r}{p}I-\frac{r}{\sigma}U_{zz}(z,w_1,w_2),
\\
H_{zw_i}=H_{w_lz}=-\frac{r}{\sigma}U_{zw_i}(z,w_1,w_2),\quad i=1,2,
\\
H_{w_iw_i}=\frac{p-r}{p}I-\frac{r}{\sigma}U_{w_iw_i}(z,w_1,w_2),\quad i=1,2,
\\
H_{w_1w_2}=H_{w_2w_1}=-\frac{r}{\sigma}U_{w_1w_2}(z,w_1,w_2),
\end{array}\right. \ee
where we write $H_z$ and $H_{zw_i}$ etc to denote the derivative of $H$ with respect to $z$,
and the second derivative of $H$ with respect to $z$ and then $w_i$ respectively.
Note that all the items above are $2\times2$ matrices.

For $U_{ij}$ defined in (\ref{U_ij}) with $1\le i<j\le n,1\le l\le n-2$,
we have
\bea
\frac{\partial U_{ij}}{\partial z}(z,w_1,w_2) &=& -\frac{m_im_j\Phi(a_i-a_j)^T}{|\Phi(a_i-a_j)z+\Phi(b_i-b_j)w_1+\Phi(c_i-c_j)w_2|^3}
\nn\\
&&\qquad\cdot\left[\Phi(a_i-a_j)z+\Phi(b_i-b_j)w_1+\Phi(c_i-c_j)w_2\right],
\\
\frac{\partial U_{ij}}{\partial w_1}(z,w_1,w_2) &=& -\frac{m_im_j\Phi(b_i-b_j)^T}{|\Phi(a_i-a_j)z+\Phi(b_i-b_j)w_1+\Phi(c_i-c_j)w_2|^3}
\nn\\
&&\qquad\cdot\left[\Phi(a_i-a_j)z+\Phi(b_i-b_j)w_1+\Phi(c_i-c_j)w_2\right],
\\
\frac{\partial U_{ij}}{\partial w_2}(z,w_1,w_2)&=& -\frac{m_im_j\Phi(c_i-c_j)^T}{|\Phi(a_i-a_j)z+\Phi(b_i-b_j)w_1+\Phi(c_i-c_j)w_2|^3}
\nn\\
&&\qquad\cdot\left[\Phi(a_i-a_j)z+\Phi(b_i-b_j)w_1+\Phi(c_i-c_j)w_2\right],
\eea
and
\bea
\frac{\partial^2 U_{ij}}{\partial z^2}(z,w_1,w_2)
&=&-\frac{m_im_j|a_i-a_j|^2I_2}{|\Phi(a_i-a_j)z+\Phi(b_i-b_j)w_1+\Phi(c_i-c_j)w_2|^3}  \nn\\
&&+3\frac{m_im_j}{|\Phi(a_i-a_j)z+\Phi(b_i-b_j)w_1+\Phi(c_i-c_j)w_2|^5}
\nn\\
&&\qquad\cdot\Phi(a_i-a_j)^T\left[\Phi(a_i-a_j)z+\Phi(b_i-b_j)w_1+\Phi(c_i-c_j)w_2\right]  \nn\\
&&\qquad\cdot\left[\Phi(a_i-a_j)z+\Phi(b_i-b_j)w_1+\Phi(c_i-c_j)w_2\right]^T\Phi(a_i-a_j),  \\
\frac{\partial^2 U_{ij}}{\partial z\partial w_1}(z,w_1,w_2)&=&
-\frac{m_im_j\Phi(a_i-a_j)^T\Phi(b_i-b_j)}{|\Phi(a_i-a_j)z+\Phi(b_i-b_j)w_1+\Phi(c_i-c_j)w_2|^3} \nn\\
&&+3\frac{m_im_j}{|\Phi(a_i-a_j)z+\Phi(b_i-b_j)w_1+\Phi(c_i-c_j)w_2|^5}
\nn\\
&&\qquad\cdot\Phi(a_i-a_j)^T\left[\Phi(a_i-a_j)z+\Phi(b_i-b_j)w_1+\Phi(c_i-c_j)w_2\right]  \nn\\
&&\qquad\cdot\left[\Phi(a_i-a_j)z+\Phi(b_i-b_j)w_1+\Phi(c_i-c_j)w_2\right]^T\Phi(b_i-b_j),  \\
\frac{\partial^2 U_{ij}}{\partial {w_1}^2}(z,w_1,w_2)
&=&-\frac{m_im_j|b_i-b_j|^2I_2}{|\Phi(a_i-a_j)z+\Phi(b_i-b_j)w_1+\Phi(c_i-c_j)w_2|^3}  \nn\\
&&+3\frac{m_im_j}{|\Phi(a_i-a_j)z+\Phi(b_i-b_j)w_1+\Phi(c_i-c_j)w_2|^5}
\nn\\
&&\qquad\cdot\Phi(b_i-b_j)^T\left[\Phi(a_i-a_j)z+\Phi(b_i-b_j)w_1+\Phi(c_i-c_j)w_2\right]   \nn\\
&&\qquad\cdot\left[\Phi(a_i-a_j)z+\Phi(b_i-b_j)w_1+\Phi(c_i-c_j)w_2\right]^T\Phi(b_i-b_j).
\eea

Let
$$ K=\left(\matrix{2 & 0\cr
                   0 & -1}\right), \quad
K_1=\left(\matrix{1 & 0\cr
                  0 & 0}\right),\quad
K_2=\left(\matrix{1 & 0\cr
                  0 & -1}\right)=\Psi(1),  $$
where $\Psi$ is given by (\ref{map2}).
Now evaluating these functions at the solution $\bar\xi_0=(0,\sigma,0,0,0,0,\sigma,0,0,0,0,0)^T\in\R^8$
 with $z=(\sigma,0)^T,w_i=(0,0)^T,1\le i\le 2$, and summing them up,
we obtain
\begin{eqnarray}
\frac{\partial^2 U}{\partial z^2}\left|_{\xi_0}\right.&=&
\sum_{1\le i<j\le 4}\frac{\partial^2 U_{ij}}{\partial z^2}\left|_{\xi_0}\right.
\nonumber\\
&=&\sum_{1\le i<j\le 4}\left(-\frac{m_im_j|a_i-a_j|^2}{|(a_i-a_j)\sigma|^3}I
                        +3\frac{m_im_j\sigma^2|a_i-a_j|^2K_1|a_i-a_j|^2}{|(a_i-a_j)\sigma|^5}\right)
\nonumber\\
&=&\frac{1}{\sigma^3}\left(\sum_{1\le i<j\le4}\frac{m_im_j}{|a_i-a_j|}\right)K
\nonumber\\
&=&\frac{\mu}{\sigma^3}K,  \label{U_zz}
\\
\frac{\partial^2 U}{\partial w_1^2}\left|_{\xi_0}\right.&=&
\sum_{1\le i<j\le 4}\frac{\partial^2 U_{ij}}{\partial w_l^2}\left|_{\xi_0}\right.
\nonumber\\
&=&\sum_{1\le i<j\le 4}\left(-\frac{m_im_j|b_i-b_j|^2}{|(a_i-a_j)\sigma|^3}I
                        +3\frac{m_im_j\sigma^2\Phi(b_i-b_j)^T\Phi(a_i-a_j)K_1\Phi(a_i-a_j)^T\Phi(b_i-b_j)}{|(a_i-a_j)\sigma|^5}\right)
\nonumber\\
&=&\sum_{1\le i<j\le 4}\left(-\frac{m_im_j|b_i-b_j|^2}{|(a_i-a_j)\sigma|^3}I
                        +3\frac{m_im_j\sigma^2\Phi(b_i-b_j)^T\Phi(a_i-a_j)\frac{I_2+K_2}{2}\Phi(a_i-a_j)^T\Phi(b_i-b_j)}{|(a_i-a_j)\sigma|^5}\right)
\nonumber\\
&=&\sum_{1\le i<j\le 4}\left(-\frac{m_im_j|b_i-b_j|^2}{|(a_i-a_j)\sigma|^3}I
                        +{3\over2}\frac{m_im_j\sigma^2\Phi(b_i-b_j)^T\Phi(a_i-a_j)\Phi(a_i-a_j)^T\Phi(b_i-b_j)}{|(a_i-a_j)\sigma|^5}\right)
\nonumber\\
&&+\sum_{1\le i<j\le 4}
\left({3\over2}\frac{m_im_j\sigma^2\Phi(b_i-b_j)^T\Phi(a_i-a_j)\Psi(1)\Phi(a_i-a_j)^T\Phi(b_i-b_j)}{|(a_i-a_j)\sigma|^5}\right)
\nonumber\\
&=&\sum_{1\le i<j\le 4}\left(-\frac{m_im_j|b_i-b_j|^2}{|(a_i-a_j)\sigma|^3}I
                        +{3\over2}\frac{m_im_j\sigma^2\Phi(|b_i-b_j|^2|a_i-a_j|^2)}{|(a_i-a_j)\sigma|^5}\right)
\nonumber\\
&&+\sum_{1\le i<j\le 4}
\left({3\over2}\frac{m_im_j\sigma^2\Psi((a_i-a_j)^2(\overline{b}_i-\overline{b}_j)^2)}{|(a_i-a_j)\sigma|^5}\right)
\nonumber\\
&=&{1\over2\sigma^3}\sum_{1\le i<j\le 4}\left(\frac{m_im_j|b_i-b_j|^2}{|a_i-a_j|^3}\right)I_2
+{1\over\sigma^3}\Psi\left({3\over2}\sum_{1\le i<j\le 4}
\frac{m_im_j(a_i-a_j)^2(\overline{b}_i-\overline{b}_j)^2}{|a_i-a_j|^5}\right)
\nonumber\\
&=&\frac{1}{2\sigma^3}\left(\sum_{i=1}^4 \overline{b}_i\sum_{j=1,j\ne i}^4\frac{m_im_j(b_i-b_j)}{|a_i-a_j|^3}\right)I_2
+{1\over\sigma^3}\Psi\left({3\over2}\sum_{1\le i<j\le 4}
\frac{m_im_j(a_i-a_j)^2(\overline{b}_i-\overline{b}_j)^2}{|a_i-a_j|^5}\right)
\nonumber
\\
&=&\frac{1}{2\sigma^3}\left(\sum_{i=1}^4 \overline{b}_iF_i\right)I_2
+{1\over\sigma^3}\Psi\left({3\over2}\sum_{1\le i<j\le 4}
\frac{m_im_j(a_i-a_j)^2(\overline{b}_i-\overline{b}_j)^2}{|a_i-a_j|^5}\right)
\nonumber\\
&=&\frac{\mu(1+\bb_1)}{2\sigma^3}I_2
+{1\over\sigma^3}\Psi\left({3\over2}\sum_{1\le i<j\le 4}
\frac{m_im_j(a_i-a_j)^2(\overline{b}_i-\overline{b}_j)^2}{|a_i-a_j|^5}\right)
\nonumber\\
&=&\frac{\mu(1+\bb_1)}{2\sigma^3}I_2
+{\mu\over\sigma^3}\Psi(\bb_{11}), \label{U_w1w1}
\end{eqnarray}
where in the third equality of the first formula, we used (\ref{F_ki}),
and in the last equality of the second formula, we use the definition (\ref{Fi.bi}) and (\ref{bb_11}).
Similarly, we have
\bea
\frac{\partial^2 U}{\partial w_2^2}\left|_{\xi_0}\right.&=&
\frac{\mu(1+\bb_2)}{2\sigma^3}I_2
+{1\over\sigma^3}\Psi\left({3\over2}\sum_{1\le i<j\le 4}
\frac{m_im_j(a_i-a_j)^2(\overline{c}_i-\overline{c}_j)^2}{|a_i-a_j|^5}\right)
\nonumber
\\
&=&\frac{\mu(1+\bb_2)}{2\sigma^3}I_2
+{\mu\over\sigma^3}\Psi(\bb_{22}), \label{U_w2w2}
\\
\frac{\partial^2 U}{\partial w_1\partial w_2}\left|_{\xi_0}\right.&=&
{1\over\sigma^3}\Psi\left({3\over2}\sum_{1\le i<j\le 4}
\frac{m_im_j(a_i-a_j)^2(\overline{b}_i-\overline{b}_j)(\overline{c}_i-\overline{c}_j)}{|a_i-a_j|^5}\right)
\nonumber\\
&=&{\mu\over\sigma^3}\Psi(\bb_{12}). \label{U_w1w2}
\eea

Moreover, we have
\begin{eqnarray}
\frac{\partial^2 U}{\partial z\partial w_1}\left|_{\xi_0}\right.&=&
\sum_{1\le i<j\le 4}\frac{\partial^2 U_{ij}}{\partial z\partial w_1}\left|_{\xi_0}\right.
\nonumber\\
&=&\sum_{1\le i<j\le4}\left(-\frac{m_im_j\Phi(a_i-a_j)^T\Phi(b_i-b_j)}{|(a_i-a_j)\sigma|^3}
                        +3\frac{m_im_j\sigma^2|a_i-a_j|^2K_1\Phi(a_i-a_j)^T\Phi(b_i-b_j)}{|(a_i-a_j)\sigma|^5}\right)
\nonumber\\
&=&\frac{K}{\sigma^3}\left(\sum_{1\le i<j\le 4}\frac{m_im_j\Phi((\overline{a}_i-\overline{a}_j)(b_i-b_j))}{|a_i-a_j|^3}\right)
\nonumber\\
&=&\frac{K}{\sigma^3}\Phi\left(\sum_{1\le i<j\le 4}\frac{m_im_j(\overline{a}_i-\overline{a}_j)(b_i-b_j)}{|a_i-a_j|^3}\right)
\nonumber\\
&=&\frac{K}{\sigma^3}\Phi\left(\sum_{1\le i<j\le 4}\frac{m_im_j\overline{a}_i(b_i-b_j)}{|a_i-a_j|^3}
-\sum_{1\le i<j\le 4}\frac{m_im_j\overline{a}_j(b_i-b_j)}{|a_i-a_j)|^3}\right)
\nonumber\\
&=&\frac{K}{\sigma^3}\Phi\left(\sum_{1\le i<j\le 4}\frac{m_im_j\overline{a}_i(b_i-b_j)}{|a_i-a_j|^3}
-\sum_{1\le j<i\le 4}\frac{m_jm_i\overline{a}_i(b_j-b_i)}{|a_j-a_i)|^3}\right)
\nonumber\\
&=&\frac{K}{\sigma^3}\Phi\left(\sum_{i=1}^4\overline{a}_i\sum_{j=i+1}^4\frac{m_im_j(b_i-b_j)}{|a_i-a_j|^3}
+\sum_{i=1}^4\overline{a}_i\sum_{j=1}^{i-1}\frac{m_im_j(b_i-b_j)}{|a_i-a_j|^3}\right)
\nonumber\\
&=&\frac{K}{\sigma^3}\Phi\left(\sum_{i=1}^4\overline{a}_i\sum_{j=1,j\ne i}^4\frac{m_im_j(b_i-b_j)}{|a_i-a_j|^3}\right)
\nonumber\\
&=&\frac{K}{\sigma^3}\Phi\left(\sum_{i=1}^4\overline{a}_iF_i\right)
\nonumber\\
&=&\frac{K}{\sigma^3}\Phi\left(\mu(1+\bb_1)\sum_{i=1}^4m_i\overline{a}_ib_i\right)
\nonumber\\
&=&O,\label{U_zw1}
\end{eqnarray}
where in the second last equation, we used (\ref{eq.of.cc}), and in the last equality, we used (\ref{Fi.bi}).
Similarly, we have
\be
\frac{\partial^2 U}{\partial z\partial w_2}\left|_{\xi_0}\right.=0. \label{U_zw2}
\ee

By \label{U_zw1} and (\ref{U_zz})-(\ref{U_zw2}), we have
\begin{eqnarray}
H_{zz}|_{\xi_0}&=&\frac{p-r}{p}I-\frac{r\mu}{\sigma^4}K=I-\frac{r}{p}I-\frac{r\mu}{p\mu}K
=I-\frac{r}{p}(I+K)
=\left(\matrix{-\frac{2-e\cos\theta}{1+e\cos\theta} & 0\cr
               0 & 1}\right),  \nn\\
H_{zw_i}|_{\xi_0}&=&-\frac{r}{\sigma}\frac{\partial^2U}{\partial z\partial w_i}|_{\xi_0}=O,\quad 1\le i\le 2,
\nn\\
H_{w_1w_1}|_{\xi_0}&=&\frac{p-r}{p}I-\frac{r}{\sigma}
   \left[\frac{\mu(1+\bb_1)}{2\sigma^3}I_2+{\mu\over\sigma^3}\Psi(\bb_{11})\right]
=I-\frac{r}{p}I-\frac{r}{p}\left[\frac{1+\bb_1}{2}I_2+\Psi(\bb_{11})\right]
\nonumber
\\
&=&I-\frac{r}{p}\left[\frac{3+\bb_1}{2}I_2+\Psi(\bb_{11})\right],
\nn\\
H_{w_2w_2}|_{\xi_0}&=&\frac{p-r}{p}I-\frac{r}{\sigma}
   \left[\frac{\mu(1+\bb_2)}{2\sigma^3}I_2+{\mu\over\sigma^3}\Psi(\bb_{22})\right]
=I-\frac{r}{p}I-\frac{r}{p}\left[\frac{1+\bb_2}{2}I_2+\Psi(\bb_{22})\right]
\nonumber
\\
&=&I-\frac{r}{p}\left[\frac{3+\bb_2}{2}I_2+\Psi(\bb_{22})\right],
\nn\\
H_{w_1w_2}|_{\xi_0}&=&H_{w_2w_1}|_{\xi_0}=-\frac{r}{\sigma}\frac{\partial^2U}{\partial w_1\partial w_2}|_{\xi_0}=-\frac{r}{p}\Psi(\bb_{12}).
\end{eqnarray}
Thus the proof is complete.\hb

\begin{remark}
If $\beta_{12}=0$, by (\ref{H_12}), we have $H_{w_1w_2}=O$, and hence
the linearized Hamiltonian system (\ref{general.linearized.Hamiltonian.system})
can be separated into three independent Hamiltonian systems,
the first one is the linearized Hamiltonian system of the Kepler two-body problem at Kepler elliptic orbit,
and each of the other two systems can be written as
\be
\dot{\zeta}_i(\th)=JB_{i,0}(\th)\zeta_i(\th), \label{linearized.system.sep_i}
\ee
with
\be
B_{i,0}=\left(\matrix{I_2& -J_2\cr J_2& I_2-{r\over p}\left[{3+\bb_{i,0}\over2}I_2+\Psi(\bb_{ii,0})\right]}\right),
\ee
for $i=1,2$.
Thus the linear stability problem of the elliptic relative equilibrium of the four-body problem
can be reduced to the linear stability problems of system (\ref{linearized.system.sep_i}) with $i=1,2$.
\end{remark}

However in general, $\beta_{12}=0$ does not hold.
But in some special cases, such as the four-body system with two small masses,
we precisely have $\beta_{12}=0$, and we will study such system below.

\setcounter{equation}{0}
\section{Two small masses}

From \cite{Xia}, for an elliptic relative equilibrium of four-body problem,
we know that the two small masses must close to the Lagrangian points of the two-body system of the primaries respectively.
If the two small masses tend to the different Lagrangian points, 
the system is equivalent to the combinations of
two restricted three-body problem systems when their masses are all tend to zero,
and hence the linear stability of the elliptic relative equilibrium of such system can be reduced to
the linear stability of the corresponding elliptic relative equilibria of two three-body problem,
and which are studied well by \cite{HLS}, \cite{ZL1} and \cite{HO1}.

We now consider the linear stability of special central configurations in the four
body problem with two small masses which are closed to each other.
A typical example is the EEM orbit of the
$4$-bodies, the Earth, the Moon and two space stations near the same Lagrangian point of the Earth and the Moon.
We try to give an analytical way 
when the masses of two small bodies tend to zero.
%following which one can numerically find out the best elliptic-hyperbolic
%positions for the two space stations using results 
%in \cite{MSS1} and \cite{MSS2}.
Specially, for the four masses we fix $m_1=m\in (0,1)$,
and let $m_2=1-m-(\tau+1)\ep$, $m_3=\ep$, $m_4=\tau\ep$ with $0<\tau\le1$ and
$0<\ep<\frac{1-m}{\tau+1}$. They satisfy
\be  m_1+m_2+m_3+m_4=1.  \lb{3.1}\ee

Let $q_1=0$ and $q_2=1$.
%When $\epsilon\to0$, we know that $q_3$ and $q_4$ must tend to
%one of the Lagrangian points of $m_1$ and $m_2$ respectively.
%Furthermore, if $q_3$ and $q_4$ tend to the same Lagrangian point,
%the linear stability of such problem is equivalent to the linear stability of the elliptic relavtive equilibria of
%two restricted three-body problem when $\ep\to0$.
Let $L_i,1\le i\le5$ be the five Lagrangian points of $m_1$ and $m_2$.
If $q_3$ and $q_4$ tend to the same point $L_1$ (or $L_2,L_3$) as $\ep\to0$,
the linear stability of the elliptic relative equilibria of such problem is studied in \cite{ZL2}.
Hence, the most interesting case is when $q_3$ and $q_4$ tend to the same point $L_4$ (or $L_5$) as $\ep\to0$ (see Figure 1).

\begin{figure}[ht]
\centering
%\resizebox{10cm}{6cm}
%{\includegraphics*[0cm,0cm][16cm,9cm]{charged3body.jpg}}%MSS-diagram
\includegraphics[height=7.5cm]{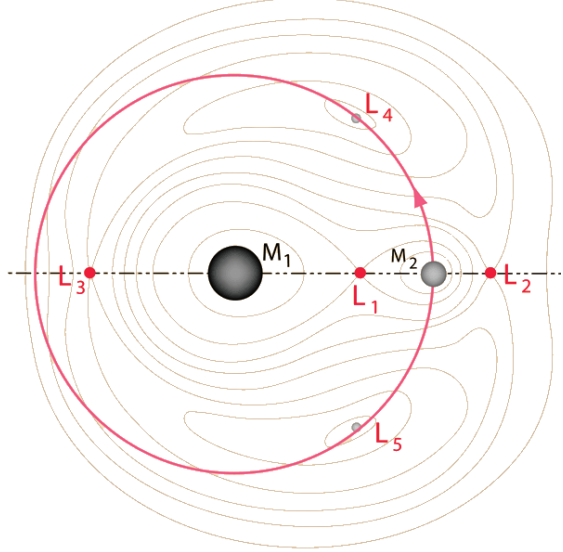}
\caption{The five Lagrangian points of $m_1$ and $m_2$.}
\end{figure}
\vspace{2mm}

Using the complex plane, we have
\be
z_{L_4}={1\over2}+\sqrt{-1}{\sqrt{3}\over2},
\ee
and
\be
\lim_{\ep\to0}q_3=\lim_{\ep\to0}q_4=z_{L_4}.  \label{lim.of.q2.q3}
\ee
The center of mass of the four particles is
\bea
q_c
= m_1q_1+m_2q_2+m_3q_3+m_4q_4
=(1-m)-(1+\tau+q_3+\tau q_4)\epsilon.
\eea

For $i=1$, $2$, $3$ and $4$, let $a_i=(q_i-q_c)\alpha$ for $\alpha>0$, we have
\bea
a_1&=&(q_1-q_c)\alpha=[-(1-m)+(1+\tau+q_3+\tau q_4)\epsilon]\alpha,
\\
a_2&=&(q_2-q_c)\alpha=[m+(1+\tau+q_3+\tau q_4)\epsilon]\alpha,
\\
a_3&=&(q_3-q_c)\alpha=[(q_3+m-1)+(1+\tau+q_3+\tau q_4)\epsilon]\alpha,
\\
a_4&=&(q_4-q_c)\alpha=[(q_4+m-1)+(1+\tau+q_3+\tau q_4)\epsilon]\alpha.
\eea
From $\sum_{i=1}^4m_i|a_i|^2=1$, we have
\bea
{1\over\alpha^2}&=&m|-(1-m)+(1+\tau+q_3+\tau q_4)\epsilon|^2
                 +(1-m-(1+\tau)\epsilon)|m+(1+\tau+q_3+\tau q_4)\epsilon|^2
                 \nonumber\\
                 &&+\epsilon|(q_2+m-1)+(1+\tau+q_3+\tau q_4)\epsilon|^2
                 +\tau\epsilon|(q_3+m-1)+(1+\tau+q_3+\tau q_4)\epsilon|^2
                 \nonumber\\
                 &=&m(1-m)+\epsilon[m^2(1+\tau)+|q_3+m-1|^2+\tau|q_4+m-1|^2]+o(\epsilon)
\eea

Moreover, let
\be  \alpha_0=\lim_{\epsilon\to 0}\alpha=[m(1-m)]^{-{1\over2}},  \lb{3.8}\ee
and
\be  q_{c,0}=\lim_{\epsilon\to 0}q_c=(1-m)\alpha_0,   \lb{3.9}\ee
and hence
\bea
z_{a_1,0} &=& \lim_{\epsilon\to 0}z_{a_1} = -(1-m)\alpha_0,  \lb{3.10}\\ %(a_10)
z_{a_2,0} &=& \lim_{\epsilon\to 0}z_{a_2} = m\alpha_0,  \lb{3.11}\\
z_{a_3,0} &=& \lim_{\epsilon\to 0}z_{a_3} = (z_{L_4}+m-1)\alpha_0,  \\ %(a_10)
z_{a_4,0} &=& \lim_{\epsilon\to 0}z_{a_4} = (z_{L_4}+m-1)\alpha_0.
\eea %(a_40)
The potential $\mu$ is given by
\be
\mu=\mu_{\epsilon,\tau}=\sum_{1\le i<j\le4}\frac{m_im_j}{|a_i-a_j|},
\ee
and by Lemma 3 of \cite{IM}, we have
\be
\mu_0=\lim_{\epsilon\to 0}\mu=\frac{m(1-m)}{\alpha_0}=\alpha_0^{-3}.\label{mu0}
\ee
In the following, we will use the subscript $0$ to denote the limit value of the parameters
when $\epsilon\to 0$.

We now calculate $k$ and $l$ defined by (\ref{param.k})-(\ref{param.l}) for our case.
We first have
\bea
\sum_{i=1}^4m_i\bar{z}_{a_i}^2&=&\alpha^2\bigg\{m[-(1-m)+(1+\tau+\bar{q}_3+\tau\bar{q}_4)\epsilon]^2
                 +(1-m-(1+\tau)\epsilon)[m+(1+\tau+\bar{q}_3+\tau\bar{q}_4)\epsilon]^2
                   \nonumber\\
                 &&+\epsilon[(\bar{q}_3+m-1)+(1+\tau+\bar{q}_3+\tau\bar{q}_4)\epsilon]^2
                 +\tau\epsilon[(\bar{q}_4+m-1)+(1+\tau+\bar{q}_3+\tau\bar{q}_4)\epsilon]^2\bigg\}
                   \nonumber\\
                 &=&\alpha^2\bigg\{m(1-m)+\epsilon[m^2(1+\tau)+(\bar{q}_3+m-1)^2+\tau(\bar{q}_4+m-1)^2]+o(\epsilon)\bigg\}
                   \nonumber\\
                 &=&1+{1\over m(1-m)}\bigg\{\bar{q}_3^2-|q_3|^2+\tau(\bar{q}_4^2-|q_4|^2)-(1-m)[\bar{q}_3-q_3+\tau(\bar{q}_4-q_4)]\bigg\}\epsilon+o(\epsilon)
                   \nonumber\\
                 &=&1+{1\over m(1-m)}\bigg\{\bar{z}_{L_4}^2-|z_{L_4}|^2+\tau(\bar{z}_{L_4}^2-|z_{L_4}|^2)-(1-m)[\bar{z}_{L_4}-z_{L_4}+\tau(\bar{z}_{L_4}-z_{L_4})]\bigg\}\epsilon+o(\epsilon),
                 \nn\\
\eea
where we used (\ref{lim.of.q2.q3}) in the last equality.
Hence by (\ref{v_3}), we have
\be
\sqrt{\epsilon}k={\sqrt{\epsilon}\over\sqrt{1-|\sum_{i=1}^4m_i\bar{z}_{a_i}^2|^2}}
=\sqrt{m(1-m)\over2|z_{L_4}|^2-z_{L_4}^2-\bar{z}_{L_4}^2+\tau(2|z_{L_4}|^2-z_{L_4}^2-\bar{z}_{L_4}^2)}
+o(1)
=\sqrt{m(1-m)\over3(1+\tau)}+o(1)
\ee
and
\bea
k+l&=&\frac{1-\sum_{i=1}^4m_i\bar{z}_{a_i}^2}{\sqrt{1-|\sum_{i=1}^4m_i\bar{z}_{a_i}^2|^2}}
\nonumber\\
&=&\sqrt{\epsilon}k\cdot{1-\sum_{i=1}^4m_i\bar{z}_{a_i}^2\over\sqrt{\ep}}
\nonumber\\
&=&(\sqrt{\epsilon}k){|z_{L_4}|^2-\bar{z}_{L_4}^2+\tau(|z_{L_4}|^2-\bar{z}_{L_4}^2)-(1-m)[z_{L_4}-\bar{z}_{L_4}+\tau(z_{L_4}-\bar{z}_{L_4})]\over m(1-m)}\sqrt{\ep}
+o(\epsilon^{1\over2}).
\eea
Moreover, we have
\bea
\lim_{\ep\to0}\sqrt{\epsilon}k&=&\sqrt{m(1-m)\over3(1+\tau)},
\\
\lim_{\ep\to0}k+l&=&0.
\eea
By (\ref{v_3}), we have
\bea
b_1&=&k\bar{a}_1+la_1=k(\bar{a}_1-a_1)+(k+l)a_1
\nonumber\\
&=&k\epsilon[(\bar{q}_3-q_3)+\tau(\bar{q}_4-q_4)]\alpha+(k+l)a_1
\nonumber\\
&=&(\sqrt{\ep}k)[(\bar{q}_3-q_3)+\tau(\bar{q}_4-q_4)]\alpha\sqrt{\ep}+(k+l)a_1,
\eea
and hence
\be
\lim_{\ep\to0}b_1=0. \label{Case2.lim.b1}
\ee
Similarly, we have
\bea
\lim_{\ep\to0}b_2&=&0,
\\
\lim_{\ep\to0}\sqrt{m_3}b_3&=&\lim_{\ep\to0}\sqrt{\epsilon}b_3=-\sqrt{1\over\tau+1}\sqrt{-1},
\\
\lim_{\ep\to0}\sqrt{m_4}b_4&=&\lim_{\ep\to0}\sqrt{\tau\epsilon}b_4=-\sqrt{\tau\over\tau+1}\sqrt{-1}.
\eea

Also by (\ref{lim.of.q2.q3}), we have
\be
\lim_{\ep\to0}(\Delta_{234},-\Delta_{134},\Delta_{124},\Delta_{123})=(0,0,{\sqrt{3}\over4},-{\sqrt{3}\over4})\alpha_0^2,
\ee
and hence by (\ref{c}), we obtain
\bea
\lim_{\ep\to0}c_1&=&\lim_{\ep\to0}c_2=0,
\\
\lim_{\ep\to0}\sqrt{m_3}c_3&=&\lim_{\ep\to0}\sqrt{m_1m_2m_4}\Delta_{124}=\sqrt{\tau\over\tau+1},\label{Case2.lim.c3}
\\
\lim_{\ep\to0}\sqrt{m_4}c_4&=&\lim_{\ep\to0}-\sqrt{m_1m_2m_3}\Delta_{123}=-\sqrt{1\over\tau+1}. \label{Case2.lim.c4}
\eea
%Moreover,
%\be
%\Delta_{134}=Im(q_3\bar{q}_4)=Im()
%\ee

Following pp.171 in \cite{Xia}, for $q=(q_x,q_y)^T\in\R^2$, we define
\be
V_2(q)=\frac{m}{|a_{1,0}-q|}+\frac{1-m}{|a_{2,0}-q|}+{1\over2}\alpha_0^{-3}|q|^2
\ee
where $\alpha_0^{-3}$ is an extra parameter because Z. Xia fixed $\lambda=1$ of (1) in \cite{Xia},
but here we have $\lambda=\alpha_0^{-3}$.
Then we have
\bea
\frac{\partial^2V_2}{\partial^2q_x}&=&-\frac{m}{|a_{1,0}-q|^3}-\frac{1-m}{|a_{2,0}-q|^3}+\frac{1}{\alpha_0^3}
      +3\left[\frac{m(-(1-m)\alpha_0-q_x)^2}{|a_{1,0}-q|^5}+\frac{(1-m)(m\alpha_0-q_x)^2}{|a_{2,0}-q|^5}\right],
\\
\frac{\partial^2V_2}{\partial q_x\partial q_y}
      &=&-3\left[\frac{m(-(1-m)\alpha_0-q_x)q_y}{|a_{1,0}-q|^5}+\frac{(1-m)(m\alpha_0-q_x)q_y}{|a_{2,0}-q|^5}\right]
\\
\frac{\partial^2V_2}{\partial^2q_x}&=&-\frac{m}{|a_{1,0}-q|^3}-\frac{1-m}{|a_{2,0}-q|^3}+\frac{1}{\alpha_0^3}
      +3\left[\frac{mq_y^2}{|a_{1,0}-q|^5}+\frac{(1-m)q_y^2}{|a_{2,0}-q|^5}\right].
\eea
Therefore, at the critical point $q=a_{3,0}=((m-{1\over2})\alpha_0,{\sqrt{3}\over2}\alpha_0)^T$, we have
\bea
\frac{\partial^2V_2}{\partial^2q_x}\bigg|_{q=((m-{1\over2})\alpha_0,{\sqrt{3}\over2}\alpha_0)^T}
&=&-\frac{m}{\alpha_0^3}-\frac{1-m}{\alpha_0^3}+\frac{1}{\alpha_0^3}  \nn\\
    &&+3\left[\frac{m[-(1-m)-(m-{1\over2})]^2\alpha_0^2}{\alpha_0^5}+\frac{(1-m)[m-(m-{1\over2})]^2\alpha_0^2}{\alpha_0^5}\right]
\nonumber
\\
&=&{3\over4}\alpha_0^{-3},
\\
\frac{\partial^2V_2}{\partial q_x\partial q_y}\bigg|_{q=((m-{1\over2})\alpha_0,{\sqrt{3}\over2}\alpha_0)^T}
&=&-3\left[\frac{m[-(1-m)-(m-{1\over2})]{\sqrt{3}\over2}\alpha_0^2}{\alpha_0^5}+\frac{(1-m)[m-(m-{1\over2})]{\sqrt{3}\over2}\alpha_0^2}{\alpha_0^5}\right]
\nonumber\\
&=&-{3\sqrt{3}(1-2m)\over4}\alpha_0^{-3},
\\
\frac{\partial^2V_2}{\partial^2q_x}\bigg|_{q=((m-{1\over2})\alpha_0,{\sqrt{3}\over2}\alpha_0)^T}
&=&-\frac{m}{\alpha_0^3}-\frac{1-m}{\alpha_0^3}+\frac{1}{\alpha_0^3}
   +3\left[\frac{{3\over4}m\alpha_0^2}{\alpha_0^5}+\frac{{3\over4}(1-m)\alpha_0^2}{\alpha_0^5}\right]
\nonumber
\\
&=&{9\over4}\alpha_0^{-3},
\eea
and hence
\be
D^2V_2(q)|_{q=((m-{1\over2})\alpha_0,{\sqrt{3}\over2}\alpha_0)^T}
 =\left(\matrix{{3\over4}\alpha_0^{-3}& -{3\sqrt{3}(1-2m)\over4}\alpha_0^{-3}\cr
                       -{3\sqrt{3}(1-2m)\over4}\alpha_0^{-3}& {9\over4}\alpha_0^{-3}}\right).
\ee
The two eigenvalues of $D^2V_2(q)|_{q=((m-{1\over2})\alpha_0,{\sqrt{3}\over2}\alpha_0)^T}$ are given by
\bea
\delta_1&=&\frac{3+\sqrt{9-\beta}}{2\alpha_0^3},  \label{delta1}
\\
\delta_2&=&\frac{3-\sqrt{9-\beta}}{2\alpha_0^3},  \label{delta2}
\eea
where we defined
\be\label{bb}
\beta = 27m(1-m).
\ee
Here $\beta$ coincides with the same parameter of (1.4) in \cite{HLS} with our case.
Now since $m_3,m_4\rightarrow0$, the range of $\beta$ is $[0,{27\over4}]$ and hence
\be\label{range.of.tilde.bb}
{3\over2}\le\sqrt{9-\bb}\le3.
\ee
Letting
\be\label{lambda.1&2}
\lambda_{1}={3+\sqrt{9-\bb}\over2},\quad \lambda_{2}={3-\sqrt{9-\bb}\over2},
\ee
by (\ref{delta1}) and (\ref{delta2}), we have
\be
\delta_i=\lambda_i\alpha_0^{-3},\qquad i=1,2.
\ee
Then by the Case (ii) in p.173 of \cite{Xia}, we have
\be
\lim_{\epsilon\rightarrow0}\frac{|a_3-a_4|}{(m_3+m_4)^{1\over3}}=\lim_{\epsilon\rightarrow0}r_2'
=(\delta_i)^{-{1\over3}}=[\lambda_i\alpha_0^{-3}]^{-{1\over3}},
\ee
for $i=1$ or $i=2$,
and hence
\bea
\lim_{\epsilon\rightarrow0}\frac{m_3}{|a_3-a_4|^3}
&=&\frac{1}{1+\tau}\lim_{\epsilon\rightarrow0}\frac{m_3+m_4}{|a_3-a_4|^3}=\frac{\lambda_i\alpha_0^{-3}}{1+\tau}
=\frac{\lambda_i\mu_0}{1+\tau},
\label{m3_a34}
\\
\lim_{\epsilon\rightarrow0}\frac{m_4}{|a_3-a_4|^3}
&=&\frac{\tau}{1+\tau}\lim_{\epsilon\rightarrow0}\frac{m_3+m_4}{|a_3-a_4|^3}=\frac{\tau\lambda_i\alpha_0^{-3}}{1+\tau}
=\frac{\tau\lambda_i\mu_0}{1+\tau}.
\label{m4_a34}
\eea

\begin{figure}[ht]
\centering
%\resizebox{10cm}{6cm}
%{\includegraphics*[0cm,0cm][16cm,9cm]{charged3body.jpg}}%MSS-diagram
\includegraphics[height=9.5cm]{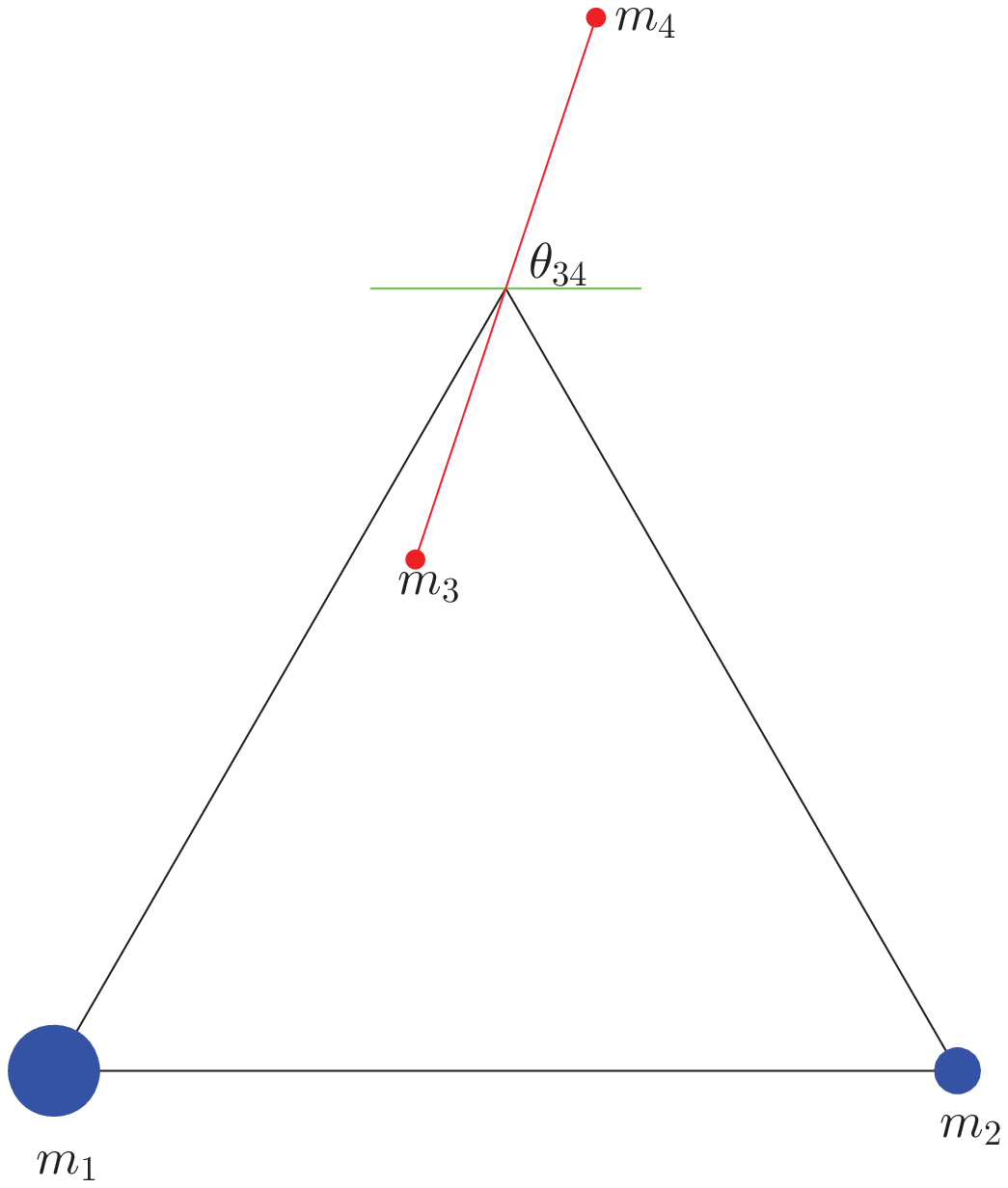}
\vspace{-2mm}
\caption{The non-convex central configuration.}
\label{nonconvex.cc}
\end{figure}
\vspace{2mm}
The direction of $a_4-a_3$ near the Lagrangian point $L_4$ is
the eigenvector of $D^2V_2(q)|_{q=((m-{1\over2})\alpha_0,{\sqrt{3}\over2}\alpha_0)^T}$ with respect
to its eigenvalue $\lambda_i,i=1,2$ respectively.
Thus, we have
\be\label{direction.of.34}
\lim_{\ep\to0}\frac{a_4-a_3}{|a_4-a_3|}=\pm\frac{9-4\lambda_i+3\sqrt{3}(1-2m)\sqrt{-1}}{2\sqrt{27-2\beta-6\lambda_i}},
\ee
and we denote its angle with respect to the horizontal direction by $\th_{34}\in[-\pi,\pi]$.
Then we has two cases.
Concretely, when $i=1$, we have
\bea
\tan^2\th_{34}&=&\left({3\sqrt{3}(1-2m)\over9-4\lambda_1}\right)^2
\nn\\
&=&{27-108m(1-m)\over(3-2\sqrt{9-\bb})^2}
\nn\\
&=&{27-4\bb\over27-4\bb-{9\over2}\cdot{8\over3}(\sqrt{9-\bb}-{3\over2})}
\nn\\
&\ge&{27-4\bb\over27-4\bb-({\sqrt{9-\bb}+{3\over2}})\cdot{8\over3}(\sqrt{9-\bb}-{3\over2})}
\nn\\
&=&{27-4\bb\over27-4\bb-{8\over3}({27\over4}-\bb)}
\nn\\
&=&3,
\eea
where we have used (\ref{bb}) and (\ref{range.of.tilde.bb}).
Thus we have $\th_{34}\ge{\pi\over3}$ or $\th_{34}\le-{\pi\over3}$ (see Figure \ref{nonconvex.cc}).
Therefore, the limit ERE is convergence by a family of non-convex EREs.

\begin{figure}[ht]
\centering
%\resizebox{10cm}{6cm}
%{\includegraphics*[0cm,0cm][16cm,9cm]{charged3body.jpg}}%MSS-diagram
\includegraphics[height=9.5cm]{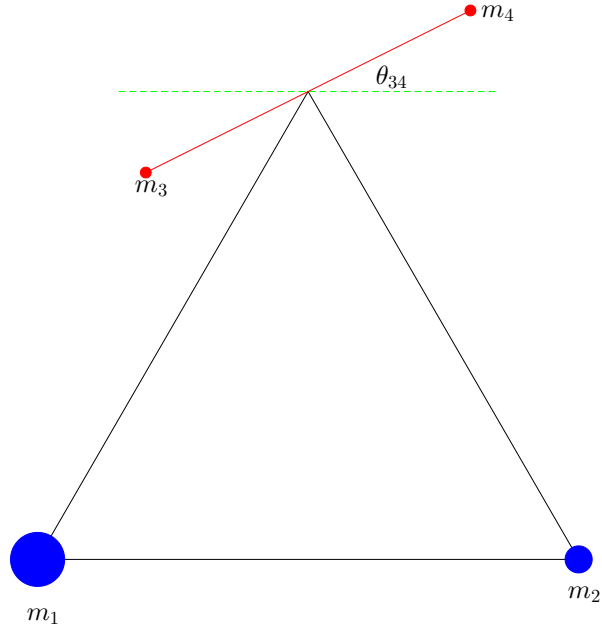}
\vspace{-2mm}
\caption{The convex central configuration.}
\label{convex.cc}
\end{figure}

On the other hand, when when $i=2$, we similarly have $-{\pi\over3}\le\th_{34}\le{\pi\over3}$,
and hence the resulting ERE is a limit of a family of convex EREs (see Figure \ref{convex.cc}).
We will study the non-convex ERE and convex ERE in Section 4 and Section 5 below respectively.

\vspace{2mm}

Form the definition of $D$ by (\ref{D}), we have
\bea
\lim_{\ep\to0}tr(D)&=&\lim_{\ep\to0}\left(4\mu-\sum_{i=1}^4\sum_{j=1,j\ne i}\frac{m_j}{|z_{a_i}-z_{a_j}|^3}\right)
\nonumber\\
&=&4\mu_0-\lim_{\ep\to0}\left(\sum_{i<j}{m_i+m_j\over|z_{a_i}-z_{a_j}|^3}\right)
\nonumber\\
&=&4\mu_0-\left[{1\over\alpha_0^3}+{m\over\alpha_0^3}+{m\over\alpha_0^3}+{1-m\over\alpha_0^3}+{1-m\over\alpha_0^3}
+\lim_{\ep\to0}{m_3+m_4\over|z_{a_3}-z_{a_4}|^3}\right]
\nonumber\\
&=&(1-\lambda_i)\mu_0.
\label{Case2.trD}
\eea
Plugging (\ref{Case2.trD}) into (\ref{bb2}), we have
\be
\bb_{2,0}=\lim_{\ep\to0}\bb_2=1-\lim_{\ep\to0}{tr(D)\over\mu}=\lambda_i.
\ee
Moreover,
by (\ref{bb_11}) and (\ref{Case2.lim.b1})-(\ref{Case2.lim.c4}), we have
\bea
\lim_{\ep\to0}\frac{m_1m_3(a_1-a_3)^2(\bar{b}_1-\bar{b}_3)^2}{|a_1-a_3|^5}
&=&\lim_{\ep\to0}\frac{m_1(a_1-a_3)^2}{|a_1-a_3|^5}\cdot\lim_{\ep\to0}[(\sqrt{m_3}\bar{b}_1-\sqrt{m_3}\bar{b}_3)]^2
\nonumber\\
&=&\frac{m({1\over2}-{\sqrt{3}\over2}\sqrt{-1})}{(\tau+1)\alpha_0^3},
\eea
and
\bea
\lim_{\ep\to0}\frac{m_3m_4(a_3-a_4)^2(\bar{c}_3-\bar{c}_4)^2}{|a_3-a_4|^5}
&=&\left(\lim_{\ep\to0}\frac{(a_3-a_4)}{|a_3-a_4|}\right)^2
   \cdot\lim_{\ep\to0}\frac{m_3}{|a_1-a_3|^3}
   \cdot\lim_{\ep\to0}[(\sqrt{m_4}\bar{c}_3-\sqrt{m_4}\bar{c}_4)]^2
\nonumber\\
&=&\left(\frac{9-4\lambda_i+3\sqrt{3}(1-2m)\sqrt{-1}}{2\sqrt{27-2\beta-6\lambda_i}}\right)^2
   \cdot{\lambda_i\mu_0\over\tau+1}\cdot\left(\sqrt{\tau}\sqrt{\tau\over\tau+1}+\sqrt{1\over\tau+1}\right)^2
\nonumber\\
&=&{\lambda_i\mu_0\over4}\cdot\frac{(9-4\lambda_i)^2-27(1-2m)^2+6\sqrt{3}(1-2m)(9-4\lambda_i)\sqrt{-1}}{27-2\beta-6\lambda_i}
\nonumber\\
&=&{\lambda_i\mu_0\over4}\cdot\frac{6(9-4\lambda_i)}{27-2\beta-6\lambda_i}\left[1+\sqrt{3}(1-2m)\sqrt{-1}\right]
\eea
where we have used (\ref{Case2.lim.c3}),(\ref{Case2.lim.c4}), (\ref{m3_a34}) and (\ref{direction.of.34}) in the second equality,
and (\ref{lambda.1&2}) in the last equality.

Hence
\bea
\bb_{11,0}&=&{3\over2\mu_0}\lim_{\ep\to0}\sum_{1\le i<j\le4}\frac{m_im_j(a_i-a_j)^2(\bar{b}_i-\bar{b}_j)^2}{|a_i-a_j|^5}
\nonumber\\
&=&{3\over2\mu_0}\left[0+\frac{m({1\over2}-{\sqrt{3}\over2}\sqrt{-1})}{(\tau+1)\alpha_0^3}
+\frac{\tau m({1\over2}-{\sqrt{3}\over2}\sqrt{-1})}{(\tau+1)\alpha_0^3}\right.
\nonumber\\
&&\left.+\frac{(1-m)({1\over2}+{\sqrt{3}\over2}\sqrt{-1})}{(\tau+1)\alpha_0^3}
+\frac{\tau(1-m)({1\over2}+{\sqrt{3}\over2}\sqrt{-1})}{(\tau+1)\alpha_0^3}
+0\right]
\nonumber\\
&=&\frac{3[1-\sqrt{3}(1-2m)\sqrt{-1}]}{4}.
\eea
Similarly,
we have
\bea
\bb_{12,0}&=&{3\over2\mu_0}\lim_{\ep\to0}
\sum_{1\le i<j\le4}\frac{m_im_j(a_i-a_j)^2(\bar{b}_i-\bar{b}_j)(\bar{c}_i-\bar{c}_j)}{|a_i-a_j|^5}
=0,  \label{TwoSmallSec:bb_120}
\\
\bb_{22,0}&=&{3\over2\mu_0}\lim_{\ep\to0}
\sum_{1\le i<j\le4}\frac{m_im_j(a_i-a_j)^2(\bar{c}_i-\bar{c}_j)^2}{|a_i-a_j|^5}
\nonumber\\
&=&-\frac{3[1-\sqrt{3}(1-2m)\sqrt{-1}]}{4}+{3\over2\mu_0}\lim_{\ep\to0}\frac{m_3m_4(a_3-a_4)^2(\bar{c}_3-\bar{c}_4)^2}{|a_3-a_4|^5}
\nonumber\\
&=&-\frac{3[1-\sqrt{3}(1-2m)\sqrt{-1}]}{4}
   +\frac{9\lambda_i(9-4\lambda_i)}{27-2\beta-6\lambda_i}\left[{1+\sqrt{3}(1-2m)\sqrt{-1}\over4}\right]
\nonumber\\
&=&\frac{3[1-\sqrt{3}(1-2m)\sqrt{-1}]}{4}\left(-1+\cdot\frac{3\lambda_i(9-4\lambda_i)}{27-2\beta-6\lambda_i}\right).
\eea

When $\epsilon\to0$,
since $\bb_{12,0}=0$ of (\ref{TwoSmallSec:bb_120}),
the linearize Hamiltonian system (\ref{general.linearized.Hamiltonian.system})
can be separated into three independent Hamiltonian systems,
the first one is the linearized Hamiltonian system of the Kepler two-body problem at Kepler elliptic orbit,
and each of the other two systems can be written as
\be
\dot{\zeta}_i(\th)=JB_{i,0}(\th)\zeta_i(\th), \label{TwoSmallSec:linearized.system.sep_i}
\ee
with
\be
B_{i,0}=\left(\matrix{I_2& -J_2\cr J_2& I_2-{r\over p}\left[{3+\bb_{i,0}\over2}I_2+\Psi(\bb_{ii,0})\right]}\right),
\ee
for $i=1,2$.
Thus the linear stability problem of the elliptic relative equilibria of the four-body problem with two small masses
can be reduced to the linear stability problems of system (\ref{TwoSmallSec:linearized.system.sep_i}) with $i=1,2$.

%As a similar analysis in Section 3,
%the system (\ref{TwoSmallSec:linearized.system.sep_i}) with $i=1$ can be transformed to the same system of
%the essential part of the linearized Hamiltonian system near the elliptic Lagrangian
%relative equilibria of the three-body problem with masses $m,1-m$ and $0$.
%Therefore, the system (\ref{TwoSmallSec:linearized.system.sep_i}) for $i=2$ corresponds to
%effect of the two small bodies when $\epsilon\to0$.

Letting
\be
D_i={3+\bb_{i,0}\over2}I_2+\Psi(\bb_{ii,0}), \label{D_i}
\ee
for $i=1,2$,
then
\be
D_1=\left(\matrix{{9\over4}& -{3\sqrt{3}\over4}(1-2m)\cr
          -{3\sqrt{3}\over4}(1-2m)& {3\over4}}\right),
\ee
and hence the two characteristic roots of $D_1$ are $\lambda_1$ and $\lambda_2$ which given by (\ref{lambda.1&2}).

As in the proof of Theorem11.14 of \cite{Lon4},
the system (\ref{TwoSmallSec:linearized.system.sep_i}) for $i=1$ becomes
\be\label{TwoSmallSec:linearized.system.sep_1}
\dot{\zeta}_1(\th)
=J\left(\matrix{1&  0&  0& 1\cr
                0&  1& -1& 0\cr
                0& -1& 1-\frac{3+\sqrt{9-\bb}}{2(1+e\cos\th)}& 0\cr
                1&  0&   0& 1-\frac{3-\sqrt{9-\bb}}{2(1+e\cos\th)}
                }\right)
 \zeta_1(\th),
\ee
thus the system coincides with the essential part of the linearized Hamiltonian system near the elliptic Lagrangian
relative equilibria of the three-body problem with masses $m_1=m,m_2=1-m$ and $m_3=0$.
%Therefore the system (\ref{linearized.system.sep_i}) for $i=2$ corresponds to
%the sub-system formed by the equations associated to the action of the three bodies on the one with zero mass.
Therefore, the linear stability property of such system 
has been studied well in \cite{HLS},
then we just need to study the linear stability of system (\ref{TwoSmallSec:linearized.system.sep_i}) for $i=2$.

By (\ref{D_i}) and $i=2$, the characteristic polynomial is
\be
\lambda^2-(3+\bb_{2,0})\lambda+({3+\bb_{2,0}\over2})^2-|\bb_{22,0}|^2,
\ee
and the characteristic roots of $D_2$ are given by
\be
\lambda_{3,4}={3+\bb_{2,0}\over2}\pm|\bb_{22,0}|,
\ee
and hence, as a similar analogue for $i=1$,
the system (\ref{TwoSmallSec:linearized.system.sep_i}) for $i=2$ becomes
\be\label{TwoSmallSec:linearized.system.sep_2}
\dot{\zeta}_2(\th)
=J\left(\matrix{1&  0&  0& 1\cr
                0&  1& -1& 0\cr
                0& -1& 1-\frac{\lambda_3}{1+e\cos\th}& 0\cr
                1&  0&   0& 1-\frac{\lambda_4}{1+e\cos\th}
                }\right)
 \zeta_2(\th).
\ee
where
\be
\lambda_{3,4}={3+\bb_{2,0}\over2}\pm|\bb_{22,0}|.  \label{lambda_3&4}
\ee
Now, we just need to study the linear stability property of system (\ref{TwoSmallSec:linearized.system.sep_2}).

However, 
our $B(t)$ is the smae matrix as $\mathcal{B}(t)$ of (2.2) in \cite{HO1}.
Since $R=I_{2n-4}+\mathcal{D}$ there can be considered as
the regularized Hessian of the central configurations.
In fact, for $a_0\in\mathscr{E}$ which is a central configurations, 
then $I(a_0) = 1$. With respect to the mass matrix $M$ inner product, 
the Hessian of the restriction of the potential to the inertia ellipsoid, 
is given by
\be
D^2U|_{\mathscr{E}}(a_0)=M^{-1}D^2U(a_0)+U(a_0),
\ee
and thus
\be
R = {1\over U(a_0)}P^{-1}D^2U|_{\mathscr{E}}(a_0)P|_{w\in\R^{2n-4}}.
\ee
On the otherhand, by $\bb_{12,0}=0$ and
(\ref{TwoSmallSec:linearized.system.sep_1}),
(\ref{TwoSmallSec:linearized.system.sep_2}), we have
$\sg(R)=\left\{{3+\sqrt{9-\bb}\over2},{3+\sqrt{9-\bb}\over2},\lambda_3,\lambda_4\right\}$.
Thus the eigenvalues of the Hessian
\be\label{eig.of.Hessian}
\sg(D^2U|_{\mathscr{E}}(a_0))=\left\{{3+\sqrt{9-\bb}\over2}\mu_0,{3-\sqrt{9-\bb}\over2}\mu_0,\lambda_3\mu_0,\lambda_4\mu_0\right\},
\ee
where $\mu_0=U(a_0)$ is given by (\ref{mu0}).

Recall that we have two different possible values of $\beta_{2,0}$ (also $\beta_{22,0}$).
One corresponds to the case when the equilateral triangle central configuration
is the limit of a sequence of non-convex central configurations,
another corresponds to the case when the equilateral triangle central configuration
is the limit of a sequence of convex central configurations.
We call the corresponding ERE as ``the non-convex ERE" and ``the convex ERE" respectively.

\setcounter{equation}{0}
\section{The non-convex ERE}

\subsection{The corresponding second order differential operator}

We first give the relation for the Morse index
and the Maslov-type index which covers the applications to our problem.

For $T>0$, suppose $x$ is a critical point of the functional
$$ F(x)=\int_0^TL(t,x,\dot{x})dt,  \qquad \forall\,\, x\in W^{1,2}(\R/T\Z,\R^2), $$
where $L\in C^2((\R/T\Z)\times \R^{2}\times\R^2,\R)$ and satisfies the
Legendrian convexity condition $L_{p,p}(t,x,p)>0$. It is well known
that $x$ satisfies the corresponding Euler-Lagrangian
equation:
\begin{eqnarray}
&& \frac{d}{dt}L_p(t,x,\dot{x})-L_x(t,x,\dot{x})=0,    \label{A.7}\\
&& x(0)=x(T),  \qquad \dot{x}(0)=\dot{x}(T).    \label{A.8}
\end{eqnarray}

For such an extremal loop, define
\begin{eqnarray}
P(t) &=& L_{p,p}(t,x(t),\dot{x}(t)),  \nonumber\\
Q(t) &=& L_{x,p}(t,x(t),\dot{x}(t)),  \nonumber\\
R(t) &=& L_{x,x}(t,x(t),\dot{x}(t)).  \nonumber
\end{eqnarray}

For $\omega\in\U$, set
\begin{equation}
D(\omega,T)=\{y\in W^{1,2}([0,T],\C^2)\,|\, y(T)=\omega y(0) \}.  
\label{domain.D.om}
\end{equation}
and
\begin{equation}    \label{closure.of.D.om}
\overline{D}(\omega,T)= \{y\in W^{2,2}([0,T],\C^2)\,|\, y(T)=\omega y(0), \dot{y}(T)=\omega\dot{y}(0) \}.
\end{equation}

Suppose $x$ is an extreme of $F$ in $\overline{D}(\omega,T)$.
The index form of $x$ is given by
\begin{equation}
I(y_1,y_2)=\int_0^T\{(P\dot{y}_1+Q{y_2})\cdot\dot{y}_2+Q^T\dot{y}_1\cdot{y_2}+R{y_1}\cdot{y_2}\},\;\;y_1,y_2\in{D}(\omega,T).
\end{equation}
The Hessian of $F$ at $x$ is given by
\begin{equation}
I(y_1,y_2)=\< F''(x)y_1,y_2\>,\quad y_1,y_2\in D(\omega,T),
\end{equation}
where $\<\cdot,\cdot\>$ is the inner product in $L^2$.
Linearization of (\ref{A.7}) at $x$ is given by
\begin{equation}
-{d\over dt}(P(t)\dot{y}+Q(t)y)+Q^T\dot{y}+R(t)y=0,  \label{A.24}
\end{equation}
and $y$ is solution of (\ref{A.24}) if and only if $y\in\ker(I)$.

We define the $\omega$-Morse index $\phi_\omega(x)$ of $x$ to be the dimension of the
largest negative definite subspace of the index form $I$ which was defined on $D(\omega,T)\times D(\omega,T)$.
Moreover, $F''(x)$ is a self-adjoint operator on $L^2([0,T],\R^2)$ with domain $\overline{D}(\omega,T)$.
We also define
$$\nu_\omega(x)=\dim\ker(F''(x)).$$

In general, for a self-adjoint operator $A$ on the Hilbert space $\mathscr{H}$, we set
$\nu(A)=\dim\ker(A)$ and denote by $\phi(A)$ its Morse index which is the maximum dimension
of the negative definite subspace of the symmetric form $\< A\cdot,\cdot\>$. Note
that the Morse index of $A$  is equal to the total multiplicity of the negative eigenvalues
of $A$.

On the other hand, $\tilde{x}(t)=(\partial L/\partial\dot{x}(t),x(t))^T$ is the solution of the
corresponding Hamiltonian system of (\ref{A.7})-(\ref{A.8}), and its fundamental solution
$\gamma(t)$ is given by
\begin{eqnarray}
\dot{\gamma}(t) &=& JB(t)\gamma(t),  \label{2.12}\\
\gamma(0) &=& I_{4},  \label{2.13}
\end{eqnarray}
with
\begin{equation}
B(t)=\left(\matrix{P^{-1}(t)& -P^{-1}(t)Q(t)\cr
	-Q(t)^TP^{-1}(t)& Q(t)^TP^{-1}(t)Q(t)-R(t)}\right). \label{2.14}
\end{equation}

\begin{lemma}\label{Lemma:indices.equalities}
	(Y.~Long, \cite{Lon4}, p.172)\label{LA.3}
	For the $\omega$-Morse index $\phi_\omega(x)$ and nullity $\nu_\omega(x)$ of the solution $x=x(t)$
	and the $\omega$-Maslov-type index $i_\omega(\gamma)$ and nullity $\nu_\omega(\gamma)$ of the symplectic
	path $\gamma$ corresponding to $\tilde{x}$, for any $\omega\in\U$ we have
	\begin{equation}
	\phi_\omega(x) = i_\omega(\gamma), \qquad \nu_\omega(x) = \nu_\omega(\gamma).  
	\end{equation}
\end{lemma}

A generalization of the above lemma to arbitrary  boundary conditions is given in \cite{HS1}.
For more information on these topics, we refer to \cite{Lon4}.
In particular, we have for any $\bb$ and $e$,
the Morse index $\phi_{\om}(A(\bb,e))$ and nullity $\nu_{\om}(A(\bb,e))$
of the operator $A(\bb,e)$ on the domain $\ol{D}(\omega,2\pi)$ satisfy
\begin{equation}
\phi_{\om}(A(\bb,e)) = i_{\om}(\ga_{\bb,e}), \quad
\nu_{\om}(A(\bb,e)) = \nu_{\om}(\ga_{\bb,e}), \qquad
\forall \,\om\in\U.
\end{equation}

\medskip

Now we consider the linear stability of
non-convex case ERE.
By the argument below (\ref{direction.of.34}), we have
\be
\beta_{2,0}=\lambda_1={3+\sqrt{9-\beta}\over2},
\ee
and hence by (\ref{TwoSmallSec:bb_120}), we have
\bea
\beta_{22,0}
&=&\frac{3[1-\sqrt{3}(1-2m)\sqrt{-1}]}{4}\left(-1+\frac{3\lambda_i(9-4\lambda_i)}{27-2\beta-6\lambda_i}\right)
\nonumber\\
&=&\frac{3[1-\sqrt{3}(1-2m)\sqrt{-1}]}{4}\left(-1+{3(3+\sqrt{9-\beta})\over2}\cdot\frac{9-6+2\sqrt{9-\beta}}{27-2\beta-9-3\sqrt{9-\beta}}\right)
\nonumber\\
&=&\frac{3[1-\sqrt{3}(1-2m)\sqrt{-1}]}{4}\left(-1+{3(3+\sqrt{9-\beta})\over2}\cdot\frac{1}{-\sqrt{9-\beta}}\right)
\nonumber\\
&=&-\frac{3[1-\sqrt{3}(1-2m)\sqrt{-1}]}{4}\cdot{9+5\sqrt{9-\beta}\over2\sqrt{9-\beta}}.
\eea
Then the modulus of $\beta_{22,0}$ is:
\bea
|\beta_{22,0}|&=&{\sqrt{9-\beta}\over2}\cdot{9+5\sqrt{9-\beta}\over2\sqrt{9-\beta}}
\nonumber\\
&=&{9+5\sqrt{9-\beta}\over4}.
\eea
By (\ref{lambda_3&4}), we have
\bea
\lambda_{3}&=&{3+\bb_{2,0}\over2}+|\bb_{22,0}|={9+3\sqrt{9-\beta}\over2},
\\
\lambda_{4}&=&{3+\bb_{2,0}\over2}-|\bb_{22,0}|=-\sqrt{9-\beta}.
\eea

Now let $\ga_N=\ga_{N;\beta,e}(t)$ is the fundamental solution of system (\ref{TwoSmallSec:linearized.system.sep_2}), i.e.,
\be  \label{TwoSmallSec:fundamtal.system.nonconvex}
\left\{
\matrix{
\dot\ga(t)&=&JB_N(t)\ga(t),
\cr
\ga(0)&=&I_4,
}
\right.
\ee
with
\be
B_N(t)=\left(\matrix{1&  0&  0& 1\cr
                0&  1& -1& 0\cr
                0& -1& 1-\frac{9+3\sqrt{9-\bb}}{2(1+e\cos(t))}& 0\cr
                1&  0&   0& 1+\frac{\sqrt{9-\bb}}{1+e\cos(t)}
                }\right),
\ee
where $e$ is the eccentricity, and from now on, $t$ is used to instead of $\theta$ as the truly anomaly.
If there is no confusion, we will omit the subscript ``N", which indicates the ``non-convex case".

Let
\be  J_2=\left(\matrix{ 0 & -1 \cr 1 & 0 \cr}\right), \qquad
    K_{N;\bb,e}(t)=\left(\matrix{\frac{9+3\sqrt{9-\bb}}{2(1+e\cos(t))} & 0 \cr
                                     0 & -\frac{\sqrt{9-\bb}}{1+e\cos(t)} \cr}\right),  \lb{Nonconvex:K_tbe}\ee
and set
\be L(t,x,\dot{x})=\frac{1}{2}\|\dot{x}\|^2 + J_2x(t)\cdot\dot{x}(t) + \frac{1}{2}K_{N;\bb,e}(t)x(t)\cdot x(t),
       \qquad\quad  \forall\;x\in W^{1,2}(\R/2\pi\Z,\R^2),  \lb{TwoSmallSec:Lagrangian}\ee
where $a\cdot b$ denotes the inner product in $\R^2$. Obviously the origin in the configuration space is a
solution of the corresponding Euler-Lagrange system. By Legendrian transformation, the corresponding
Hamiltonian function is
$$   H(t,z)=\frac{1}{2}B_N(t)z\cdot z,\qquad \forall\; z\in\R^4.  $$

In order to transform the Lagrangian system (\ref{TwoSmallSec:Lagrangian}) to a simpler linear operator corresponding to
a second order Hamiltonian system with the same linear stability as $\ga_{\bb,e}(2\pi)$, using $R(t)$
and $R_4(t)=\diag(R(t),R(t))$ as in Section 2.4 of \cite{HLS}, we let
\be  \xi_{\bb,e}(t) = R_4(t)\ga_{\bb,e}(t), \qquad
\forall\; t\in [0,2\pi],\;\; (\bb,e)\in[0,{27\over4}]\times [0,1), \lb{2.25}\ee
One can show by direct computations that
\be  \frac{d}{dt}\xi_{\bb,e}(t)
  = J \left(\matrix{I_2 & 0 \cr
                    0 & R(t)(I_2-K_{N;\bb,e}(t))R(t)^T \cr}\right)\xi_{\bb,e}(t). \lb{2.26}\ee
Note that $R_4(0)=R_4(2\pi)=I_4$, so $\ga_{\bb,e}(2\pi)=\xi_{\bb,e}(2\pi)$ holds.
Then the linear stabilities
of the systems (\ref{TwoSmallSec:fundamtal.system.nonconvex}) and (\ref{2.26}) are determined by the same matrix and thus is precisely the same.

By the homotopy invariance of the Maslov-type index (cf. Section 2.4 on p.14 of \cite{HLS} for a detailed analysis) we obtain
\be  i_{\om}(\xi_{\bb,e}) = i_{\om}(\ga_{\bb,e}), \quad \nu_{\om}(\xi_{\bb,e}) = \nu_{\om}(\ga_{\bb,e}),
       \qquad \forall \,\omega\in\U, \; (\bb,e)\in [0,{27\over4}]\times [0,1). \lb{2.27}\ee
Note that the first order linear Hamiltonian system (\ref{2.26}) corresponds to the following second order
linear Hamiltonian system
\be  \ddot{x}(t)=-x(t)+R(t)K_{N;\bb,e}(t)R(t)^Tx(t). \lb{2.28}\ee

For $(\bb,e)\in [0,{27\over4}]\times [0,1)$, the second order differential operator corresponding to (\ref{2.28}) is given by
\bea  A_N(\bb,e)
&=& -\frac{d^2}{dt^2}I_2-I_2+R(t)K_{N;\bb,e}(t)R(t)^T  \nn\\
&=& -\frac{d^2}{dt^2}I_2-I_2+\frac{1}{1+e\cos t}
        \left({9+\sqrt{9-\beta}\over4}I_2+{9+5\sqrt{9-\beta}\over4}S(t)\right),  \lb{A_N}\eea
where $S(t)=\left(\matrix{ \cos 2t & \sin 2t \cr
                           \sin 2t & -\cos 2t \cr}\right)$, 
defined on the domain $\ol{D}(\omega,2\pi)$
in (\ref{closure.of.D.om}) below with $n=2$. Then it is self-adjoint and depends on the parameters $\bb$ and $e$. 
By Lemma
\ref{Lemma:indices.equalities}, we have for any $\bb$ and $e$, the Morse index $\phi_{\om}(A_N(\tilde\tau,\bb,e))$ and nullity $\nu_{\om}(A_N(\bb,e))$
of the operator $A_N(\bb,e)$ on the domain $\ol{D}(\omega,2\pi)$ satisfy
\be  \phi_{\om}(A_N(\bb,e)) = i_{\om}(\xi_{\bb,e}), \quad
      \nu_{\om}(A_N(\bb,e)) = \nu_{\om}(\xi_{\bb,e}), \qquad
           \forall \,\om\in\U. \lb{2.30}\ee

We shall use both of the paths $\ga_{\bb,e}$ and $\xi_{\bb,e}$ to study
the linear stability of $\ga_{\bb,e}(2\pi)=\xi_{\bb,e}(2\pi)$. Because of (\ref{2.27}), in many cases and
proofs below, we shall not distinguish these two paths.
Hence, if there is no confusion,
we will use $i_\om(\bb,e)$ and $\nu_\om(\bb,e)$ to represent $i_{\om}(\ga_{\bb,e})$
and $\nu_{\om}(\ga_{\bb,e})$ respectively.

For convenience, we set $\tilde{\beta}=\sqrt{9-\beta}$ and
\begin{equation}\label{tilde.A_N}
\tilde{A}_N(\tilde\beta,e)=A_N(\beta,e)=-{d^2\over dt^2}I_2-I_2+{1\over1+e\cos t}\left({9+\tilde\beta\over4}I_2+{9+5\tilde\beta\over4}S(t)\right).
\end{equation}
Here the range of $\tilde\beta$ is $[{3\over2},3]$ as $\beta\in[0,{27\over4}]$.
We also let $\tilde\ga_{N;\tilde\beta,e}(t)$ (in short, by $\tilde\ga_{\tilde\beta,e}$) be the fundamental solution of system (\ref{TwoSmallSec:fundamtal.system.nonconvex})
with $\sqrt{9-\beta}$ in $B_N(t)$ replaced by $\tilde\beta$.

However, for our convenience, we will enlarge the range of $\tilde\beta$ to $[-{9\over5},+\infty)$.
When $\tilde\beta\in[{3\over2},3]$, we have
\bea
i_\om(\ga_{\bb,e})=\phi_\omega(A_N(\bb,e)) &=&\phi_\omega(\tilde{A}_N(\tilde\bb,e))=i_\om(\tilde\ga_{\tilde\bb,e}),\\
i_\om(\ga_{\bb,e})=\nu_\omega(A_N(\bb,e)) &=&\nu_\omega(\tilde{A}_N(\tilde\bb,e))=\nu_\om(\tilde\ga_{\tilde\bb,e}).
\eea

\subsection{The monotonicity of the Morse indices $\phi_\om(\tilde{A}_N(\tilde\beta,e))$ with respect to $\tilde\beta$}

When $\tilde\beta=-{9\over5}$, we have
\begin{equation}
\tilde{A}_N(-{9\over5},e)=-{d^2\over dt^2}I_2-I_2+{9\over5(1+e\cos t)}I_2.
\end{equation}
By by Lemma 4.1 in \cite{ZL1}, $\tilde{A}_N(-{9\over5},e)$ is positive definite on $\overline{D}(\om,2\pi)$ for every $\om\in\U$.
Then for $(\tilde\bb,e)\in [1,{27\over4}]\times [0,1)$,
using (\ref{tilde.A_N}) we can rewrite $\tilde{A}_N(\tilde\bb,e)$ as follows
\begin{eqnarray}
\tilde{A}_N(\tilde\bb,e) &=& -{d^2\over dt^2}I_2-I_2+{9\over5(1+e\cos t)}I_2
   +\frac{9+5\tilde\beta}{20(1+e\cos t)}(I_2+5S(t))
\nonumber\\
&=& (9+5\tilde\beta)\bar{A}_N(\tilde\bb,e), \lb{4.11}
\end{eqnarray}
where we define
\be\label{bar.A_N}
\bar{A}_N(\tilde\bb,e)=\frac{-\frac{d^2}{dt^2}I_2-I_2+\frac{9}{5(1+e\cos t)}I_2}{9+5\tilde\beta}
                          +\frac{I_2+5S(t)}{20(1+e\cos t)}
                          =\frac{\tilde{A}_N(-{9\over5},e)}{9+5\tilde\beta}
                          +\frac{I_2+5S(t)}{20(1+e\cos t)}
.
\ee
Therefore when $\tilde\beta>-{9\over5}$, we have
\bea
\phi_\omega(\tilde{A}_N(\tilde\bb,e)) &=& \phi_\omega(\bar{A}_N(\tilde\bb,e)), \lb{4.12}\\
\nu_\omega(\tilde{A}_N(\tilde\bb,e)) &=& \nu_\omega(\bar{A}_N(\tilde\bb,e)).   \lb{4.13}
\eea

Now motivated by Lemma 4.4 in \cite{HLS} or Lemma 4.2 in \cite{ZL1}, and modifying its proof to our case,
we get the following lemma:

\begin{lemma}\label{Nonconvex.Lemma:decreasing.of.index.wrt.beta}
(i) For each fixed $e\in [0,1)$, the operator $\bar{A}_N(\tilde\bb,e)$ is decreasing
with respect to $\beta>-{9\over5}$ for any fixed $\omega\in\U$. Specially
\be  \frac{\pt}{\pt\beta}\bar{A}_N(\tilde\beta,e)|_{\bb=\bb_0}
     = -\frac{5}{(9+5\tilde\beta)^2}\tilde{A}(-{9\over5},e),  \lb{4.14}\ee
is a negative definite operator for every $\bb_0>-{9\over5}$.

(ii) For every eigenvalue $\lm_{\bb_0}=0$ of $\bar{A}_N(\tilde\bb_0,e_0)$ with $\om\in\U$ for some
$(\tilde\bb_0,e_0)\in (-{9\over5},+\infty)\times [0,1)$, there holds
\be \frac{d}{d\bb}\lm_{\bb}|_{\bb=\bb_0} < 0.  \lb{4.15}\ee

(iii)  For every $(\tilde\bb_0,e_0)\in(-{9\over5},+\infty)\times[0,1)$ and $\om\in\U$,
there exist $\epsilon_0=\epsilon_0(\tilde\bb_0,e_0)>0$ small enough such that for all $\epsilon\in(0,\epsilon_0)$ there holds
\be
i_\om(\tilde\ga_{\tilde\bb_0+\epsilon,e_0}) - i_\om(\tilde\ga_{\tilde\bb_0,e_0})=\nu_\om(\tilde\ga_{\tilde\bb_0,e_0}).
\ee
\end{lemma}

\subsection{The $\om$-indices on the boundary segment $\{-1\}\times[0,1)$}

Recall the real range of $\tilde\beta$ is $[{3\over2},3]$ as the range of $\beta$ is $[0,{27\over4}]$.
But when $\tilde\beta={3\over2}$ or $3$, it is hard to compute $i_\om(\tilde\ga_{\tilde\beta,e})$.
Noting that when $\tilde\beta=-1$, we have
\begin{eqnarray}
\tilde{A}_N(-1,e)&=&-{d^2\over dt^2}I_2-I_2+{1\over1+e\cos t}\left(2I_2+S(t)\right)  \nn\\
&=&-{d^2\over dt^2}I_2-I_2+{1\over1+e\cos t}I_2+{1\over1+e\cos t}\left(I_2+S(t)\right) \nn\\
&\ge&-{d^2\over dt^2}I_2-I_2+{1\over1+e\cos t}I_2.
\end{eqnarray}
Then by Lemma 4.1 in \cite{ZL1}, $-{d^2\over dt^2}I_2-I_2+{1\over1+e\cos t}I_2$ is positive definite on $\overline{D}(\om,2\pi)$
for any $\om\in\U\backslash\{1\}$, and semi-positive definite on $\overline{D}(1,2\pi)$.
Moreover, form (4.8) in \cite{ZL1}, if $\left(\matrix{c_1(1+e\cos t)\cr c_2(1+e\cos t)}\right)\in\ker\tilde{A}_N(-1,e)$, we have
\begin{eqnarray}
0&=&\left\<\tilde{A}_N(-1,e)\left(\matrix{c_1(1+e\cos t)\cr c_2(1+e\cos t)}\right),\left(\matrix{c_1(1+e\cos t)\cr c_2(1+e\cos t)}\right)\right\>   \nn\\
&=&\left\<{1\over1+e\cos t}\left(I_2+S(t)\right)\left(\matrix{c_1(1+e\cos t)\cr c_2(1+e\cos t)}\right),\left(\matrix{c_1(1+e\cos t)\cr c_2(1+e\cos t)}\right)\right\>  \nn\\
&=&\int_0^{2\pi}2(1+e\cos t)|c_1\cos t+c_2\sin t|^2dt  \nn\\
&=&4\pi(|c_1|^2+|c_2|^2),
\end{eqnarray}
which implies $c_1=c_2=0$,
and hence $\ker\tilde{A}_N(-1,e)=\{0\}$.
Thus $\tilde{A}_N(-1,e)$ is positive definite on $\overline{D}(\om,2\pi)$ for any $\om\in\U$.
Therefore, we have
\begin{eqnarray}
&&i_{\om}(\tilde\ga_{-1,e})=\phi_{\om}(\tilde{A}_N(-1,e))=0,  \lb{Nonconvex:index.of.0e}
\\
&&\nu_{\om}(\tilde\ga_{-1,e}) = \nu_{\om}(\tilde{A}_N(-1,e))= 0, \lb{Nonconvex:null.index.of.-1e}
\end{eqnarray}
for any $\om\in\U$.

\subsection{The $\om$-indices on the boundary segment $[-1,+\infty)\times\{0\}$}

In this case $e=0$, the essential part of the motion (\ref{TwoSmallSec:fundamtal.system.nonconvex}) becomes an ODE
system with constant coefficients:
\be B = B_N(t) = \left(\matrix{1 & 0 & 0 & 1\cr
                             0 & 1 & -1 & 0 \cr
                             0 & -1 & -{7+3\sqrt{9-\beta}\over2} & 0 \cr
                             1 & 0 & 0 & 1+\sqrt{9-\beta} \cr}\right)
             = \left(\matrix{1 & 0 & 0 & 1\cr
                             0 & 1 & -1 & 0 \cr
                             0 & -1 & -{7+3\tilde\beta\over2} & 0 \cr
                             1 & 0 & 0 & 1+\tilde\beta \cr}\right)
                             .  \ee
The characteristic polynomial $\det(JB-\lambda I)$ of $JB$ is given by
\be \lambda^4 - {\tilde\beta+1\over2}\lambda^2-{3\tilde\beta(\tilde\beta+3)\over2}  = 0.  \lb{Nonconvex:ch.poly}\ee
Letting $\aa=\lambda^2$,
the two roots of the quadratic polynomial $\aa^2 - {\tilde\beta+1\over2}\aa -{3\tilde\beta(\tilde\beta+3)\over2}$
are given by $\aa_1=\frac{\tilde\bb+1+\sqrt{25\tilde\bb^2+74\tilde\bb+1}}{4}$
and $\aa_2=\frac{\tilde\bb+1-\sqrt{25\tilde\bb^2+74\tilde\bb+1}}{4}$.
Therefore the four characteristic multipliers of the matrix $\tilde\ga_{\tilde\beta,0}(2\pi)$ are given by
\begin{eqnarray}
\rho_{1,\pm}(\tilde\beta) &=&e^{\pm2\pi\sqrt{\aa_1}},
\\
\rho_{2,\pm}(\tilde\beta) &=&e^{\pm2\pi\sqrt{\aa_2}}.
\end{eqnarray}

When $\tilde\beta\in[-1,{-37+8\sqrt{21}\over25})$, we have $25\tilde\bb^2+74\tilde\bb+1<0$,
and then $\alpha_1,\alpha_2\in\C\backslash\R$.
Hence we have $\rho_{i,\pm}\in\C\backslash(\U\cup\R)$ for $i=1,2$.
When $\tilde\beta\in[{-37+8\sqrt{21}\over25},0)$, we have $\alpha_1,\alpha_2>0$,
and hence $\rho_{i,\pm}\in\R\backslash\{\pm1\}$ for $i=1,2$.
When $\tilde\beta=0$, we have $\alpha_1={1\over2}$ and $\alpha_2=0$.
Hence we have $\rho_{1,\pm}\in\R\backslash\{\pm1\}$ and $\rho_{2,\pm}=1$.
When $\tilde\beta>0$, we always have $\alpha_1>0$ and $\alpha_2<0$.
Hence we have $\rho_{1,\pm}\in\R\backslash\{\pm1\}$ and $\rho_{2,\pm}\in\U$.

For more details, we set
\begin{equation}\label{th.of.bb}
\theta(\tilde\beta)=\sqrt{-\alpha_2}=\sqrt{\sqrt{25\tilde\beta^2+74\tilde\beta+1}-(\tilde\beta+1)\over4},\quad\tilde\beta\ge0.
\end{equation}
We denote by $\tilde\bb_\th\ge0$ the $\tilde\bb$ value satisfying $\th(\tilde\bb)=\th$,
and we obtain
$$
\tilde\beta_\th={\th^2-9+\sqrt{25\th^4-6\th^2+81}\over6},\quad\th\ge0.
$$
Moreover, we have
\begin{equation}
{d\beta_\th\over d\th}=\frac{2\th+{2\th(25\th^2-3)\over\sqrt{25\th^4-6\th^2+81}}}{6}
={\th(25\th^2-3+\sqrt{25\th^4-6\th^2+81})\over3\sqrt{25\th^4-6\th^2+81}}>0,
\end{equation}
when $\theta>0$.
For later use, we write $\tilde\bb_\th$ for $\th=n$ and $\th=n+\frac{1}{2}$, $n\in\N\cup\{0\}$ as
\begin{equation}
\hat\bb_n=\frac{n^2-9+\sqrt{25n^4-6n^2+81}}{6},\quad\quad n=0,\;1,\;2...\label{om_n}
\end{equation}
and
\begin{equation}
\hat\bb_{n+\frac{1}{2}}=\frac{(n+\frac{1}{2})^2-9+\sqrt{25(n+\frac{1}{2})^4-6(n+\frac{1}{2})^2+81}}{6},
    \quad\quad n=0,\;1,\;2...\label{om_n.5}
\end{equation}
where we have used the symbol hat to denote these special values of $\tilde\bb$.
Then we obtain the following results:

(i) When $\tilde\bb\in[-1,\hat\bb_0)$, we have $\sg(\tilde\ga_{\tilde\beta,0}(2\pi))\in\C\backslash\U$.

(ii) When $\tilde\bb=\hat\bb_0=0$, we have $\sg(\tilde\ga_{0,0}(2\pi)) = \{e^{\sqrt{2}\pi}, e^{-\sqrt{2}\pi},1, 1\}$.

(iii) Let $i\in\N\cup\{0\}$.
When $\hat\bb_i<\tilde\bb<\hat\bb_{i+\frac{1}{2}}$, the angle $\th(\tilde\bb)$ in (\ref{th.of.bb}) increases strictly from $i$ to $i+\frac{1}{2}$ as $\tilde\bb$ increases
from $\hat\bb_i$ to $\hat\bb_{i+\frac{1}{2}}$.
Therefore $\rho_{2,+}(\tilde\bb)=e^{2\pi \sqrt{-1}\th(\tilde\bb)}$ runs from $1$ to $-1$ counterclockwise along
the upper semi-unit circle in the complex plane $\C$ as $\tilde\bb$ increases from $\hat\bb_i$ to $\hat\bb_{i+\frac{1}{2}}$.
Correspondingly
$\rho_{2,-}(\tilde\bb)=e^{-2\pi \sqrt{-1}\th(\tilde\bb)}$ runs from $1$ to $-1$ clockwise along the lower semi-unit circle in
$\C$ as $\tilde\bb$ increases from $\hat\bb_i$ to $\hat\bb_{i+\frac{1}{2}}$.
Thus specially we obtain
$\rho_{2,\pm}(\tilde\bb)\subset \U\bs\R$ for all $\tilde\bb\in (\hat\bb_i,\hat\bb_{i+\frac{1}{2}})$.

(iv) When $\tilde\bb=\hat\bb_{i+\frac{1}{2}}$, we have $\th(\hat\bb_{i+\frac{1}{2}})=i+\frac{1}{2}$. Therefore we obtain
$\rho_{2,\pm}(\hat\bb_{i+\frac{1}{2}})=e^{\pm \sqrt{-1} \pi} = -1$.

(v) When $\hat\bb_{i+\frac{1}{2}}<\tilde\bb<\hat\bb_{i+1}$, the angle $\th(\tilde\bb)$ increases strictly from $i+\frac{1}{2}$ to $i+1$ as $\tilde\bb$ increase
from $\hat\bb_{i+\frac{1}{2}}$ to $\hat\bb_{i+1}$. Thus $\rho_{2,+}(\tilde\bb)=e^{2\pi \sqrt{-1}\th(\tilde\bb)}$ runs from $-1$ to $1$
counterclockwise along the lower semi-unit circle in $\C$ as $\tilde\bb$ increases from $\hat\bb_{i+\frac{1}{2}}$ to $\hat\bb_{i+1}$. Correspondingly
$\rho_{2,-}(\tilde\bb)=e^{-2\pi \sqrt{-1}\th(\tilde\bb)}$ runs from $-1$ to $1$ clockwise along the
upper semi-unit circle in $\C$ as $\tilde\bb$ increases from $\hat\bb_{i+\frac{1}{2}}$ to $\hat\bb_{i+1}$.
Thus we obtain
$\rho_{2,\pm}(\tilde\bb) \subset \U\bs\R$ for all $\tilde\bb\in (\hat\bb_{i+\frac{1}{2}},\hat\bb_{i+1})$.

(vi) When $\tilde\bb=\hat\bb_{i+1}$, we obtain $\th(\hat\bb_{i+1})=i+1$, and then we have double eigenvalues
$\rho_{2,\pm}(\hat\bb_{i+1}) = 1$.

Under the similar arguments of $({\bf B})$ and $({\bf C})$ in Section 3.2 of \cite{ZL1},
we have
\begin{eqnarray}
&& i_1(\tilde\ga_{\tilde\bb,0}) = \left\{\matrix{  %\lb{3.37}
                 0, &  {\rm if}\;\;\tilde\bb\in[-1,\hat\bb_0], \cr
                 1, &  {\rm if}\;\;\tilde\bb\in(\hat\bb_0,\hat\bb_1], \cr
                 3, &  {\rm if}\;\;\tilde\bb\in(\hat\bb_1,\hat\bb_2], \cr
                 ...,\cr
                 2n+1, &  {\rm if}\;\;\tilde\bb\in(\hat\bb_n,\hat\bb_{n+1}], \cr
                 ...\cr}\right.\lb{Nonconvex:1-index.of.b0}
\\
&& \nu_1(\tilde\ga_{\tilde\bb,0}) = \left\{\matrix{
                 1, &  {\rm if}\;\;\tilde\bb=\hat\bb_0=0, \cr
                 2, &  {\rm if}\;\;\tilde\bb=\hat\bb_n,\;n\ge1, \cr
                 0, &  {\rm if}\;\;\tilde\bb\ne\hat\bb_0,\hat\bb_1,...\hat\bb_n,... \cr}\right. \lb{Nonconvex:null.1-index.of.b0}
\end{eqnarray}
and
\begin{eqnarray}
&& i_{-1}(\tilde\ga_{\tilde\bb,0}) = \left\{\matrix{
                 0, &  {\rm if}\;\;\tilde\bb\in[-1,\hat\bb_{\frac{1}{2}}], \cr
                 2, &  {\rm if}\;\;\tilde\bb\in(\hat\bb_{\frac{1}{2}},\hat\bb_{\frac{3}{2}}], \cr
                 ...,\cr
                 2n, &  {\rm if}\;\;\tilde\bb\in(\hat\bb_{n-\frac{1}{2}},\hat\bb_{n+\frac{1}{2}}],\;n\ge1, \cr
                 ...\cr}\right. \lb{Nonconvex:-1-index.of.b0}\\
&& \nu_{-1}(\tilde\ga_{\tilde\bb,0}) = \left\{\matrix{
                 2, &  {\rm if}\;\;\tilde\bb=\hat\bb_{n+\frac{1}{2}},\;n\in\N, \cr
                 0, &  {\rm if}\;\;\tilde\bb\ne\hat\bb_{\frac{1}{2}},\hat\bb_{\frac{3}{2}},...\bb_{n+\frac{1}{2}},... \cr}, \right.
\lb{Nonconvex:null.1-index.of.b0}
\end{eqnarray}

\subsection{The degenerate curves}

By Lemma \ref{Nonconvex.Lemma:decreasing.of.index.wrt.beta} and (\ref{Nonconvex:index.of.0e}), we have

\begin{corollary}\label{Nonconvex:C4.2}
For every fixed $e\in[0,1)$ and $\omega\in\U$, the index function $\phi_\om(\tilde{A}_N(\tilde\beta,e))$,
and consequently $i_\om(\tilde\ga_{\tilde\bb,e})$, is non-decreasing as $\tilde\bb$ increase from $-1$ to $+\infty$.
Moreover, they are goes from $0$ to $+\infty$ for every $\om\in\U$.
\end{corollary}

The proof is similar to that of Corollary 4.3 in \cite{ZL1}, thus we omit here.

Because $\tilde{A}_N(\tilde\bb,e)$ is a self-adjoint operator on $\bar{D}(\om,2\pi)$,
and a bounded perturbation of the operator $-\frac{d^2}{dt^2}I_2$, then $\tilde{A}_N(\tilde\bb,e)$
has discrete spectrum on $\bar{D}(\om,2\pi)$.
Thus we can define the $n$-th degenerate point for any $\om$ and $e$:
\be\lb{Nonconvex:bb_s}
\tilde\bb_n(\om,e)=\min\left\{\tilde\bb>-1\;\bigg|\;
  \begin{array}{l}
   [i_\om(\tilde\ga_{\tilde\bb,e})+v_\om(\tilde\ga_{\tilde\bb,e})]\ge n
    \end{array}\right\}.
\ee
By Lemma \ref{Nonconvex.Lemma:decreasing.of.index.wrt.beta} $(iii)$,
$i_\om(\tilde\ga_{\tilde\bb,e})+v_\om(\tilde\ga_{\tilde\bb,e})$ is a right continuous step function with respect to $\tilde\bb$.
Additionally, by Corollary \ref{Nonconvex:C4.2}, $i_\om(\tilde\ga_{\tilde\bb,e})+v_\om(\tilde\ga_{\tilde\bb,e})$ tends to $+\infty$ as
$\tilde\bb\rightarrow+\infty$, the minimum of the right hand side in (\ref{Nonconvex:bb_s}) can be obtained.
Indeed, $\tilde\ga_{\tilde\bb,e}$ is $\om$-degenerate at point $(\tilde\bb_n(\om,e),e)$, i.e.,
\be\label{degenerate.of.bn}
\nu_\om(\tilde\ga_{\tilde\bb_n(\om,e),e})\ge 1.
\ee
Otherwise, if there existed some small enough $\ep>0$ such that $\tilde\bb=\tilde\bb_n(\om,e)-\ep$ would
satisfy $[i_\om(\tilde\ga_{\tilde\bb,e})+v_\om(\tilde\ga_{\tilde\bb,e})]\ge n+1$ in (\ref{Nonconvex:bb_s}),
it would yield a contradiction.

For fixed $\om$ and $n$, $\tilde\bb_n(\om,e)$ actually forms a curve with respect to the eccentricity $e\in [0,1)$
as we shall give below in this section, which we called the $n$-th $\om$-degenerate curve.
By a similar proof of Lemma 4.5 in \cite{ZL1},
we have

\begin{lemma}\label{Nonconvex:continuous}
	For any fixed $n\in\N$ and $\om\in\U$, the $\om$-degenerate curve
	$\tilde\bb_n(\om,e)$ is continuous with respect to $e\in[0,)$.
\end{lemma}

For the first $1$-degenerate curve, we have that

\begin{theorem}\label{first.1.degenerate.curve}
Let $\om=1$.
When $n=1$, we have $\tilde{\bb}_1(1,e)\equiv0$;
when $n>1$, we have $\tilde{\bb}_n(1,e)>0$ for any $e\in[0,1)$.
Therefore, $\tilde{\bb}_1(1,e)=0$ is the first $1$-degenerate curve with multiplicity $1$.
\end{theorem}

{\bf Proof.} When $\tilde\bb=0$, we have
\bea\label{op.estimate}
\tilde{A}_N(0,e)&=&-{d^2\over dt^2}I_2-I_2+{9\over4(1+e\cos t)}\left(I_2+S(t)\right) \nn\\
&\ge&-{d^2\over dt^2}I_2-I_2+{3\over2(1+e\cos t)}\left(I_2+S(t)\right).
\eea
Note that the last operator of (\ref{op.estimate}) was studied in \cite{HLS}
by X.~Hu, S.~Sun and Y.~Long.
By (3.6) and (3.8) of \cite{HLS}, $-{d^2\over dt^2}I_2-I_2+{3\over2(1+e\cos t)}(I_2+S(t))$ is semi-positive on the domain $\overline{D}(1,2\pi)$.
Then $\tilde{A}_N(0,e)$ is non-negative on $\overline{D}(1,2\pi)$.
Thus we have
\be
\phi_1(\tilde{A}_N(0,e))=i_1(\tilde{\ga}(0,e))=0.
\ee
By Corollary \ref{Nonconvex:C4.2}, we have
\be\label{i_1.estimation}
\phi_1(\tilde{A}_N(\tilde\bb,e))=i_1(\tilde{\ga}(\tilde\bb,e))=0,
\qquad\forall\tilde\bb\in[-1,0],
\ee
and
\be\label{nu_1.estimation}
\nu_1(\tilde{A}_N(\tilde\bb,e))=\nu_1(\tilde{\ga}(\tilde\bb,e))=0,
\qquad\forall\tilde\bb\in[-1,0).
\ee

On the other hand, 
for any $x\in\ker(\tilde{A}_N(\tilde\bb,e))$ on $\overline{D}(1,2\pi)$,
(\ref{op.estimate}) implies
\bea
0&=&\<\tilde{A}_N(0,e)x,x\>   \nn\\
&=&\<\Big(-{d^2\over dt^2}I_2-I_2+{9\over4(1+e\cos t)}\left(I_2+S(t)\right)\Big)x,x\> \nn\\
&\ge&\<\Big(-{d^2\over dt^2}I_2-I_2+{3\over2(1+e\cos t)}\left(I_2+S(t)\right)\Big)x,x\> \nn\\
&\ge&0,
\eea
where in the last equality, we used the semi-positiveness of $-{d^2\over dt^2}I_2-I_2+{3\over2(1+e\cos t)}(I_2+S(t))$ on the domain $\overline{D}(1,2\pi)$.
Thus we must have $x\in\ker(-{d^2\over dt^2}I_2-I_2+{3\over2(1+e\cos t)}(I_2+S(t)))$.
Then we have
\bea
0&=&\tilde{A}_N(0,e)x-\Big(-{d^2\over dt^2}I_2-I_2+{3\over2(1+e\cos t)}\left(I_2+S(t)\right)\Big)x   \nn\\
&=&{3\over4(1+e\cos t)}\left(I_2+S(t)\right)x,
\eea
which implies $x=c(\sin t,\cos t)^T$ for some constant $c\ne0$.
Noting that
\begin{eqnarray}
\tilde{A}(0,e)\left(\matrix{-\sin t\cr \cos t}\right)&=&-{d^2\over dt^2}\left(\matrix{-\sin t\cr \cos t}\right)
-\left(\matrix{-\sin t\cr \cos t}\right)
-{9\over4(1+e\cos t)}(I_2+S(t))\left(\matrix{-\sin t\cr \cos t}\right)
\nn\\
&=&\left(\matrix{-\sin t\cr \cos t}\right)-\left(\matrix{-\sin t\cr \cos t}\right)
\nn\\
&=&0,
\end{eqnarray}
thus we have
\be
\ker(\tilde{A}_N(0,e))=\left\{c(-\sin t,\cos t)^T\;|\;c\in\R\right\},
\ee
and hence
\be\label{nu_1.estimation.at.0}
\nu_1(\tilde{A}_N(0,e))=\nu_1(\tilde{\ga}(0,e))=1.
\ee
Then by (\ref{i_1.estimation}), (\ref{nu_1.estimation}), (\ref{nu_1.estimation.at.0}) and the definition of the degenerate curves (\ref{Nonconvex:bb_s}),
we have $\tilde\bb_1(1,e)\equiv0$ and $\tilde\bb_n(1,e)>0$ for any $n>1$
and $e\in[0,1)$.
\hb

By (\ref{Nonconvex:1-index.of.b0}) and (\ref{Nonconvex:null.1-index.of.b0}),
every $1$-degenerate curve starts from the point $(\hat\bb_n,0)$.
Indeed, we have

\begin{lemma}\label{Nonconvex:starting.points}
\begin{eqnarray}
\tilde\bb_n(1,0) = \hat\bb_{\lfloor{n\over2}\rfloor}, \quad {\rm if}\;\;n \ge1. \lb{Nonconvex:bb_n0}
\end{eqnarray}
Here $\lfloor x\rfloor$ denote the greatest integer less than or equal to $x$ for any $x\in\R$.
\end{lemma}

{\bf Proof.} By (\ref{Nonconvex:1-index.of.b0}) and (\ref{Nonconvex:null.1-index.of.b0}),
we have $i_1(\tilde\ga_{0,0})=1$, $\nu_1(\tilde\ga_{0,0})=1$
$\nu_1(\tilde\ga_{\hat\bb_{m},0})=2,m\ge1$ and
\begin{equation}\label{1-index.plus.nullity}
i_1(\tilde\ga_{\tilde\bb,0})+\nu_1(\tilde\ga_{\tilde\bb,0})=
\left\{
\begin{array}{l}
 0, \quad {\rm if}\;\;-1\le\tilde\bb<\hat\bb_0,
\\
 1, \quad {\rm if}\;\;\hat\bb_0\le\tilde\bb<\hat\bb_{1},
\\
\ldots
\\
2m+1, \quad {\rm if}\;\;\hat\bb_{m}\le\tilde\bb<\hat\bb_{m+1},\;\;m\ge1,
\\
\ldots.
\end{array}
\right.
\end{equation}
Then (\ref{Nonconvex:bb_n0}) is obvious for $n=1$.
When $n>1$, we suppose $n=2m$ or $2m+1$ for some $m\in\N$.
By (\ref{1-index.plus.nullity}),
$[i_1(\tilde\ga_{\tilde\bb,0})+\nu_1(\tilde\ga_{\tilde\bb,0})]\ge n+1$ is equivalent to $\bb\ge\hat\bb_{m}$.
Then the minimal value of $\tilde\bb$ in $\{\tilde\bb\in(-1,+\infty)\;|\;\tilde\bb\ge\hat{\bb}_{m}\}$
such that $\tilde{A}_N(\tilde\bb,e)$ is degenerate on
$\overline{D}(1,2\pi)$ is $\hat{\bb}_{m}$. Thus by (\ref{Nonconvex:bb_s}), we obtain (\ref{Nonconvex:bb_n0}). \hb

Moreover, we have the following theorem:
\begin{theorem}\label{Nonconvex:multiplicity}
Every $1$-degenerate curves has even geometric multiplicity except for the first one.
\end{theorem}

{\bf Proof.} The statement has already been proved for $e=0$. We will prove that, if $\tilde{A}(\tilde\bb,e)z=0$
has a solution $z\in\ol{D}(1,2\pi)$ for a fixed value $e\in (0,1)$, there exists a second periodic
solution which is linear independent of $z$. Then the space of solutions of $\tilde{A}(\tilde\bb,e)z=0$ is the direct
sum of two isomorphic subspaces, hence it has even dimension. This method is due to R. Mat\'{i}nez,
A. Sam\`{a} and C. Sim\`{o} in \cite{MSS1}.

Let $z(t)=R(t)(x(t),y(t))^T$ be a nontrivial solution of $\tilde{A}(\tilde\bb,e)z(t)=0$, then it yields
\begin{equation}
\left\{
\begin{array}{l}
(1+e\cos t)x''(t)={9+3\tilde\bb\over2}x(t)+2y'(t)(1+e\cos t),
\\
(1+e\cos t)y''(t)=-\tilde\bb y(t)-2x'(t)(1+e\cos t).
\end{array}
\right.
\end{equation}
By Fourier expansion, $x(t)$ and $y(t)$ can be written as
\bea
x(t) =a_0+\sum_{n\ge1}a_n\cos nt+\sum_{n\ge1}b_n\sin nt,
\\
y(t) =c_0+\sum_{n\ge1}c_n\cos nt+\sum_{n\ge1}d_n\sin nt.
\eea
Then the coefficients must satisfy the following uncoupled sets of recurrences:
\begin{equation}
\left\{
\begin{array}{l}
{9+3\tilde\bb\over2}a_0=-e(d_1+\frac{a_1}{2}),
\\
eA_2\pmatrix{a_2\cr d_2}=B_1\pmatrix{a_1\cr d_1},
\\
eA_{n+1}\pmatrix{a_{n+1}\cr d_{n+1}}
=B_n\pmatrix{a_n\cr d_n}-eA_{n-1}\pmatrix{a_{n-1}\cr d_{n-1}},
\quad n\ge2,
\end{array}
\right.
\label{ad.equations}
\end{equation}
and
\begin{equation}
\label{bc.equations}
\left\{
\begin{array}{l}
-\tilde\bb c_0=e(b_1-\frac{c_1}{2}),
\\
eA_2\pmatrix{b_2\cr -c_2}=B_1\pmatrix{b_1\cr -c_1},
\\
eA_{n+1}\pmatrix{b_{n+1}\cr -c_{n+1}}
=B_n\pmatrix{b_n\cr -c_n}-eA_{n-1}\pmatrix{b_{n-1}\cr -c_{n-1}},
\quad n\ge2,
\end{array}
\right.
\end{equation}
where
\begin{equation}
A_n=-\frac{n}{2}\pmatrix{n&2\cr 2&n},\quad
B_n=\pmatrix{n^2+{9+3\tilde\bb\over2}&2n\cr 2n&n^2-\tilde\bb}.
\end{equation}

%Thus $\det(B_1)=-{(\tilde\bb+3)(3\tilde\bb-1)\over2}\ne0$ for $\tilde\bb>0,\tilde\bb\ne{1\over3}$ and $\det(A_n)\ne0$ when $n\ge3$.
%Thus given $(a_2,d_2)^T$, we can obtain $(a_1,d_1)^T$ uniquely from the second equality of (\ref{ad.equations}),
%and then obtain $(a_n,d_n)^T$ for $n\ge3$ by the last equality of (\ref{ad.equations}).
%
%However, when $\tilde\bb={1\over3}$, we have $\det(B_1)=0$ and

By the non-triviality of $z=z(t)$, both (\ref{ad.equations}) and (\ref{bc.equations}) have solutions
$\{(a_n,d_n)\}_{n=1}^\infty$ and $\{(b_n,c_n)\}_{n=1}^\infty$ respectively.
We assume (\ref{ad.equations}) admits a nontrivial solutions.
Then $\sum_{n\ge1}a_n\cos nt$ and $\sum_{n\ge1}d_n\sin nt$ are convergent.
Thus, $\sum_{n\ge1}a_n\sin nt$ and $-\sum_{n\ge1}d_n\cos nt$ are convergent too.
Moreover, if $\tilde\bb>0$,
by the similar structure between equations (\ref{ad.equations}) and (\ref{bc.equations}),
we can construct a new solution of (\ref{bc.equations}) given below
\bea
\tilde{c}_0&=&-\frac{e}{\tilde\bb}(a_1+\frac{d_1}{2}),
\\
\left(\matrix{\tilde{b}_n\cr\tilde{c}_n}\right)&=&\left(\matrix{a_n\cr -d_n}\right),\quad n\ge1.
\eea
Here $\tilde\bb>0$ is guaranteed by Theorem \ref{first.1.degenerate.curve}
when we consider the $1$-degenerate curves from the second one.
Therefore we can build two independent solutions of $\tilde{A}(\tilde\bb,e)w=0$ as
\bea
w_1(t)&=&R(t)\pmatrix{a_0+\sum_{n\ge1}a_n\cos nt\cr\sum_{n\ge1}d_n\sin nt},
\\
w_2(t)&=&R(t)\pmatrix{\sum_{n\ge1}\tilde{b}_n\sin nt\cr \tilde{b}_0+\sum_{n\ge1}\tilde{c}_n\cos nt}
        =R(t)\pmatrix{\sum_{n\ge1}a_n\sin nt\cr -\frac{e}{\tilde\bb}(a_1+\frac{d_1}{2})-\sum_{n\ge1}d_n\cos nt}.
\eea
\hb

\subsection{The order of the $1$-degenerate curves and $-1$-degenerate curves and the normal forms of $\tilde\ga_{\tilde\bb,e}(2\pi)$}

In the proofs of following theorems, 
we need the results of $\om$-indices and splitting numbers.
Now we give their definition for $2n\times2n$ symplectic matrices:

\begin{definition} (\cite{Lon2}, \cite{Lon4})\label{D2.3}
	For any $M\in\Sp(2n)$ and $\omega\in\U$, choosing $\tau>0$ and $\gamma\in\P_\tau(2n)$ with $\gamma(\tau)=M$,
	we define
	\begin{equation}
	S_M^{\pm}(\omega)=\lim_{\epsilon\rightarrow0^+}\;i_{\exp(\pm\epsilon\sqrt{-1}\omega)}(\gamma)-i_\omega(\gamma).
	\end{equation}
	They are called the splitting numbers of $M$ at $\omega$.
\end{definition}

$\om$-indices and splitting numbers have the following properties:

\begin{lemma} (\cite{Lon4}, pp.147-148)
	For $\omega\in\U$ and $\tau>0$, the $\omega$-index part of the index function defined on paths
	in $\P_\tau(2n)$ is uniquely determined by the following five axioms:
	
	$1^\circ$ ({\bf Homotopy invariant}) For $\gamma_0$ and $\gamma_1\in\P_\tau(2n)$,
	if $\gamma_0\sim_\omega\gamma_1$ on $[0,\tau]$, then
	$$ i_\omega(\gamma_0)=i_\omega(\gamma_1). $$
	
	$2^\circ$ ({\bf Symplectic additivity}) For any $\gamma_j\in\P_\tau(2n_j)$ with $j=0,1$,
	$$ i_\omega(\gamma_0\diamond\gamma_1) = i_\omega(\gamma_0)+i_\omega(\gamma_1). $$
	
	$3^\circ$ ({\bf Clockwise continuity}) For any $\ga\in\P_\tau(2)$ satisfying $\ga(\tau)=N_1(\om,b)$
	with $b=\pm 1$ or $0$ when $\omega=\pm 1$, or $\ga(\tau)=R(\vf)$ with $\om=e^{\sqrt{-1}\vf}\in \U\bs\R$,
	there exists a $\th_0>0$ such that
	$$ i_{\om}([\ga(\tau)\phi_{\tau,-\th}]*\ga) = i_{\om}(\ga),\quad\forall 0<\theta\le\th_0, $$
	where $\phi_{\tau,\th}$ is defined by
	$$ \phi_{\tau,\th}=R\left(\frac{t\th}{\tau}\right),\quad\forall t\in [0,\tau],\;\th\in\R. $$
	
	$4^\circ$ ({\bf Counterclockwise jumping}) For any $\ga\in\P_\tau(2)$ satisfying $\ga(\tau)=N_1(\om,b)$
	with $b=\pm 1$ or $0$ when $\om=\pm 1$, or $\ga(\tau)=R(\vf)$ with $\om=e^{\sqrt{-1}\vf}\in \U\bs\R$,
	there exists a $\th_0>0$ such that
	$$ i_{\om}([\ga(\tau)\phi_{\tau,\th}]*\ga) = i_{\om}(\ga)+1,\quad\forall 0<\th\le\th_0. $$
	
	$5^\circ$ ({\bf Normality}) For $\hat{\aa}_0(t)=D\left(1+\frac{t}{\tau}\right)$ with $0\le t\le \tau$,
	$$ i_{\om}(\hat{\aa}_0)=0. $$
\end{lemma}

For paths in $\Sp(2)$, we have

\begin{lemma} \label{oddevity.of.indices}
	(\cite{Lon4}, pp 179-183)
	Let $\ga\in\mathcal{P}_\tau(2)$. Then one and only one of the following cases must happen.
	
	$1^\circ$ If $\sg(\ga(\tau))=\{1,1\}$, then $\ga(\tau)\approx\left(\matrix{1 & a\cr 0 & 1}\right)$
	and we have
	$$i_1(\ga)\in\left\{
	\begin{array}{l}
	2\Z+1,{\rm\;if\;} a\ge0,\\
	2\Z,\quad\;\;{\rm\;if\;} a<0.
	\end{array}\right.
	$$
	
	$2^\circ$ If $\sg(\ga(\tau))=\{-1,-1\}$, we have
	$$ i_1(\ga)\in2\Z+1. $$
	
	$3^\circ$ If $\sg(\ga(\tau))\cap\U=\emptyset$, we have
	$$ i_1(\ga)\in 2\Z+\aa(\ga(\tau)), $$
	where $\aa(\ga(\tau))$ is the hyperbolic index given by Definition 1.8.1 of \cite{Lon4},
	i.e., $\aa(\ga(\tau))=0$ if $\ga(\tau)\approx D(\lm)$, $\aa(\ga(\tau))=1$ if $\ga(\tau)\approx D(-\lm)$
	for $\lm\in(0,1)\cup (1,+\infty)$.
	
	$4^\circ$ If $\sg(\ga(\tau))\in\U\bs\{1,-1\}$, we have
	$$ i_1(\ga)\in 2\Z+1. $$
\end{lemma}

\begin{lemma}
	(Y.~Long, \cite{Lon4}, pp. 191)
	Splitting numbers $S^{\pm}_M(\omega)$ are well defined,
	i.e., they are independent of the choice of the path $\gamma\in\mathcal{P}_T(2n)$ satisfying $\gamma(T)=M$.
	For $\omega\in\U$ and $M\in\Sp(2n)$,
	splitting numbers $S^{\pm}_N(\omega)$ are constant for all $N=P^{-1}MP$, with $P\in\Sp(2n)$.
\end{lemma}

\begin{lemma}\label{Lemma:property.of.splitting.numbers}
	(Y.~Long, \cite{Lon4}, pp. 198--199)
	For $M\in\Sp(2n)$ and $\omega\in\U$, $\theta\in(0,\pi)$, there hold
	\begin{eqnarray}
	S^{\pm}_M(\omega)&=&0,\quad if\; \omega\not\in\sigma(M),\label{split.num.of.D}\\
	S^{\pm}_M(\omega)&=&S^{\mp}_M(\bar\omega),\\
	0\le S^{\pm}_M(\omega)&\le&{\rm dim}\;\ker(M-\omega I),\\
	S^{+}_M(\omega)+S^{-}_M(\omega)&\le&{\rm dim}\;\ker(M-\omega I)^{2},\;\omega\in\sigma(M),\\
	(S^{+}_{N_1(1,b)}(1),S^{-}_{N_1(1,b)}(1))&=&
	\left\{\matrix{(1,1),\;{\rm if}\;b=0,1,\cr (0,0),\;{\rm if}\;b=-1,}\right. \\
	(S^{+}_{N_1(-1,a)}(-1),S^{-}_{N_1(-1,a)}(-1))&=&
	\left\{\matrix{(1,1),\;{\rm if}\;a=-1,0,\cr (0,0),\;{\rm if}\;a=1,\quad\;\;}\right. \\
	(S^{+}_{R(\theta)}(e^{\sqrt{-1}\theta}),S^{-}_{R(\theta)}(e^{\sqrt{-1}\theta}))&=&(0,1),\label{split.num.of.R}\\
	(S^{+}_{R(2\pi-\theta)}(e^{\sqrt{-1}\theta}),S^{-}_{R(2\pi-\theta)}(e^{\sqrt{-1}\theta}))&=&(1,0).
	\\
	(S^{+}_{N_2(\omega,b)}(\omega),S^{-}_{N_2(\omega,b)}(\omega))&=&
	(1,1)\;{\rm for}\;N_2(\omega,b)\;{\rm being\; non-trivial}, \nonumber\\
	\\
	(S^{+}_{N_2(\omega,b)}(\omega),S^{-}_{N_2(\omega,b)}(\omega))&=&
	(0,0)\;{\rm for}\;N_2(\omega,b)\;{\rm being\; trivial},
	\\
	(S_M^+(\om),S_M^-(\om))&=&(0,0)\; {\rm for}\;\om\in\U\;{\rm and}\; M\in\Sp(2n)\;
	{\rm satisfying}\; \sigma(M)\cap\U=\emptyset.
	\nn\\
	{}
	\end{eqnarray}
\end{lemma}
From the definition and property of splitting numbers, for any $\gamma\in\mathcal{P}_T(2n)$ with
$\gamma(T)=M$, we have
\begin{equation} \label{index.formula}
i_{\omega_0}(\gamma)=i_1(\gamma)+S^+_M(1)+\sum_{\omega}(S^+_M(\omega)-S^-_M(\omega))-S^-_M(\omega_0),
\end{equation}
where the sum runs over all the eigenvalues $\omega$ of $M$ belonging to the part of $\U^+=\{{\rm Re} z\ge0|z\in\U\}$
or $\U^-=\{{\rm Re} z\le0|z\in\U\}$
strictly located between $1$ and $\omega_0$.

\medskip

Now we study the order of the $1$-degenerate curves and $-1$-degenerate curves.
By the similar arguments of Theorem 4.10, 
Theorem 4.11 and Theorem 4.14 of \cite{ZL1} respectively, 
we have
\begin{theorem}
	For any $\tilde\bb>0$ and $0\le e<1$, $i_1(\tilde\ga_{\tilde\bb,e})$ is an odd number.
\end{theorem}

and

\begin{lemma}\label{Nonconvex:not.intersect}
	Any $1$-degenerate curves except the first one and any $-1$-degenerate curves cannot intersect each other.
	That is, for any $0\le e<1$, there do not exist $n_1\ge1$ and $n_2\in\N$
	such that $\tilde\bb_{n_1}(1,e)=\tilde\bb_{n_2}(-1,e)$.
	
	Similarly, for $\om\ne\pm1$, any $\om$-degenerate curves and any $-1$-degenerate curves except the first two cannot intersect each other.
	That is, for any $0\le e<1$, there do not exist $n_1,n_2\in\N$ and $n_2\ge3$
	such that $\tilde\bb_{n_1}(\om,e)=\tilde\bb_{n_2}(-1,e)$.
\end{lemma}

\begin{remark}
	We must exclude the first $1$-degenerate curve because its multiplicity is $1$, and hence the method in the proof of Theorem 4.11 of \cite{ZL1} cannot go through.
	In fact, 
	the first $-1$-degenerate curve intersects the first $1$-degenerate curve $\tilde\bb_1(1,e)\equiv0$ at some point.
\end{remark}

Indeed, we have

\begin{lemma}\label{L:intersection.of.pm1.degenerate.curves}
	There exists $x_0\in\overline{D}(-1,2\pi)$ such that 
	\be\label{A_N.x_0.x_0.ineq}
	\<\tilde{A}_N(0,e)x_0,x_0\><0,\quad\forall e\in[e_0,1)
	\ee 
	for some $e_0\in[0,1)$. Hence, we have
	\be\label{-1.index.ineq}
	i_{-1}(\tilde{\ga}_{0,e})=\phi_{-1}(\tilde{A}_N(0,e))\ge1,
	\quad\forall e\in[e_0,1).
	\ee
	
	Therefore, the first $\pm1$-degenerate curves must intersect each other.
\end{lemma}

{\bf Proof.} We just need to prove the first claim.
Let $x_1=(\cos{t\over2},\sin{t\over2})^T,x_2=(-\sin t,\cos t-1)^T\in\overline{D}(-1,2\pi)$
and $x_0=x_1+{\pi\over4}x_2$.
Then we have
\bea
\<\tilde{A}_N(0,e)x_1,x_1\>&=&-{3\over2}\pi
     +{9\over2}\int_0^{2\pi}{\cos^2{t\over2}\over1+e\cos t}dt,
\nn\\
\<\tilde{A}_N(0,e)x_1,x_2\>&=&-{9\over2}\int_0^{2\pi}{\sin t\cos{t\over2}\over1+e\cos t}dt,
\nn\\
\<\tilde{A}_N(0,e)x_2,x_2\>&=&{9\over2}\int_0^{2\pi}{\sin^2t\over1+e\cos t}dt,
\eea
and hence
\bea\label{A_N.x_0.x_0}
\<\tilde{A}_N(0,e)x_0,x_0\>&=&\<\tilde{A}_N(0,e)x_1,x_1\>+{\pi\over2}\<\tilde{A}_N(0,e)x_1,x_2\>+({\pi\over4})^2\<\tilde{A}_N(0,e)x_2,x_2\>
\nn\\
&=&-{3\over2}\pi+{9\over2}\int_0^{2\pi}{(\cos{t\over2}-{\pi\over4}\sin t)^2\over1+e\cos t}.
\eea

When $e\rightarrow1$, from (\ref{A_N.x_0.x_0}), we have
\bea\label{A_N.x_0.x_0}
\lim_{e\rightarrow1}\<\tilde{A}_N(0,e)x_0,x_0\>
&=&-{3\over2}\pi+{9\over2}\lim_{e\rightarrow1}\int_0^{2\pi}{(\cos{t\over2}-{\pi\over4}\sin t)^2\over1+e\cos t}
\nn\\
&=&-{3\over2}\pi+{9\over2}\int_0^{2\pi}{(\cos{t\over2}-{\pi\over4}\sin t)^2\over1+\cos t}
\nn\\
&=&-{3\over2}\pi+{9\over2}\left({\pi^3\over8}-\pi\right)
\nn\\
&=&{3\over2}\pi\left({3\pi^2\over8}-4\right)
\nn\\
&<&0.
\eea
Therefore, there exists $e_0\in[0,1)$, such that (\ref{A_N.x_0.x_0.ineq}) holds.

By the definition of the degenerate curves, (\ref{-1.index.ineq}) implies
$\tilde\bb_1(-1,e)<0$ as $e>e_0$.
On th other hand, the first $-1$-degenerate $\tilde\bb_1(-1,e)$ starts from
$\hat\bb_{1\over2}={-35+\sqrt{1297}\over24}>0$ as $e=0$.
Then by the continuous of the degenerate curves, 
it must intersects with the $e$-axis.
Noting that $\tilde\bb_1(1,e)\equiv0$, then the first $1$-degenerate curve and
the first $-1$-degenerate curve must intersect at some point.
\hb

Since the intersection of the first $\pm1$-deenerate curves,
therefore, in the following discussion, 
we frequently consider the $1$-degenerate curves from the second one.

Because of the starting points from $\tilde\bb$-axis of 
the $1$-degenerate curves except the first one and $-1$-degenerate curves are
alternatively distributed,
and these curves are continuous by Lemma \ref{Nonconvex:continuous},
then any two different $1$-degenerate curves except the first one (or two $-1$-degenerate curves) starting from different points cannot intersect each
other, otherwise,
one of them must intersect with some $-1$-degenerate curve 
($1$-degenerate curve) and contradicts Lemma \ref{Nonconvex:not.intersect}.
Thus we have the following theorem:

\begin{theorem}\label{Th:orders.of.degenerate.curves}
	The $1$-degenerate curves except the first one $\Ga_n,n\ge2$
	and  $-1$-degenerate curves $\Xi_n,n\ge1$ 
	can be ordered from left to right by
	\be
	\Xi_1,\;\Xi_2,\;\Gamma_2(=\Gamma_3),\;\Xi_3,\;\Xi_4,\;\Gamma_4(=\Gamma_5),\;
	\ldots,\;\Xi_{2m+1},\;\Xi_{2m+2},\;\Gamma_{2m+2}(=\Gamma_{2m+3}),\;\ldots
	\ee
	More precisely, for each $e\in[0,1)$, we have
	\bea
	-1&<&\tilde\bb_{1}(-1,e)\le\tilde\bb_{2}(-1,e)<\tilde\bb_{2}(1,e)=\tilde\bb_{3}(1,e)
	\nn\\
	&<&\tilde\bb_{3}(-1,e)\le\tilde\bb_{4}(-1,e)<\tilde\bb_{4}(1,e)=\tilde\bb_{5}(1,e)
	\nn\\
	&<&\ldots \nn\\
	&<&\tilde\bb_{2m+1}(-1,e)\le\tilde\bb_{2m+2}(-1,e)<\tilde\bb_{2m+2}(1,e)=\tilde\bb_{2m+3}(1,e)<\ldots
	\eea
\end{theorem}

Using numerical computations, we can draw these $\pm1$-degenerate curves, 
see Figure \ref{Nonconvex:bifurcation_curves}.
For the second $1$-degenerate curve, we have

\begin{lemma}\label{Nonconvex:second.degenerate.curve.left.to.1}
	When $\tilde\bb=1$, we have
	\be\label{1.index.>2}
	i_1(\tilde\ga_{1,e})=\phi_1(\tilde{A}_N(1,e))\ge2.
	\ee
	Then by the definition of the degenerate curves, we have
	\be
	\tilde\bb_2(1,e)\le1.
	\ee
\end{lemma}

{\bf Proof.} Let $x_1=\left(\matrix{-\sin t\cr \cos t}\right)$, 
$x_2=\left(\matrix{-\sin t\cr \cos t}\right)\cos t\in\overline{D}(1,2\pi)$,
we have
\bea
\<\tilde{A}_N(1,e)x_1,x_1\>&=&-\int_0^{2\pi}{1\over1+e\cos t}dt<0,
\\
\<\tilde{A}_N(1,e)x_1,x_2\>&=&-\int_0^{2\pi}{\sin t\over1+e\cos t}dt
\nn\\
&=&-\int_0^{\pi}{\sin t\over1+e\cos t}dt-\int_{\pi}^{2\pi}{\sin t\over1+e\cos t}dt
\nn\\
&=&-\int_0^{\pi}{\sin t\over1+e\cos t}dt
   -\int_{\pi}^{0}{-\sin s\over1+e\cos s}(-ds),
\nn\\
&=&0,  \label{A_N.x1.x2}
\\
\<\tilde{A}_N(1,e)x_2,x_2\>&=&\int_0^{2\pi}\left(\cos^2t-{\cos^2t\over1+e\cos t}\right)dt
\nn\\
&=&e\int_0^{2\pi}{\cos^3t\over1+e\cos t}dt
\nn\\
&=&2e\int_0^{\pi}{\cos^3t\over1+e\cos t}dt
\nn\\
&=&2e\int_0^{\pi\over2}{\cos^3t\over1+e\cos t}dt
  +2e\int_{\pi\over2}^{\pi}{\cos^3t\over1+e\cos t}dt
\nn\\
&=&2e\int_0^{\pi\over2}{\cos^3t\over1+e\cos t}dt
  -2e\int_0^{\pi\over2}{\cos^3u\over1-e\cos u}du
\nn\\
&=&-4e^2\int_0^{\pi\over2}{\cos^4t\over1-e^2\cos^2t}dt
\nn\\
&<&0,
\label{A_N.x2.x2}
\eea
where in the second last equality of (\ref{A_N.x1.x2}),
we used $s=2\pi-t$,
and in the third last equality of (\ref{A_N.x2.x2}),
we used $u=\pi-t$.

We define a space
\be
E=\span\{x_1,x_2\}
=\span\left\{\left(\matrix{-\sin t\cr \cos t}\right),\; 
             \left(\matrix{-\sin t\cr \cos t}\right)\cos t\right\}.
\ee
Then for any $x\in E$, there exists $c_1,c_2\in\C$ such that $x=c_1x_1+c_2x_2$,
and hence
\bea
\<\tilde{A}_N(1,e)x,x\>&=&|c_1|^2\<\tilde{A}_N(1,e)x_1,x_1\>
    +(c_1\overline{c}_2+\overline{c}_1c_2)\<\tilde{A}_N(1,e)x_1,x_2\>
    +|c_2|^2\<\tilde{A}_N(1,e)x_2,x_2\>
\nn\\
&=&-|c_1|^2\int_0^{2\pi}{1\over1+e\cos t}dt
   -4e^2|c_2|^2\int_0^{\pi\over2}{\cos^4t\over1-e^2\cos^2t}dt
\nn\\
&<&0,
\eea
when $(c_1,c_2)\ne(0,0)$.
Therefore, we have $\<\tilde{A}_N(1,e)\cdot,\cdot\>$ is negative definite
on the subspace $E$ of $\overline{D}(1,2\pi)$. 
Hence (\ref{1.index.>2}) holds.

The second claim follows by (\ref{1.index.>2}) 
and the definition of the degenerate curves.
\hb

Now we can give the

{\bf Proofs of Theorem \ref{main.theorem.separation.curves}(i)
	and (v)-(vii).}
(i) is follows from Theorem \ref{first.1.degenerate.curve};
(v) is follows from Lemma \ref{L:intersection.of.pm1.degenerate.curves};
(vi) is follows from Theorem \ref{first.1.degenerate.curve}
and Lemma \ref{Nonconvex:second.degenerate.curve.left.to.1};
(vii) is follows from Theorem \ref{Th:orders.of.degenerate.curves}.
\hb

\medskip

{\bf Proof of Theorem \ref{Th:normal.forms.decomposition}.}
(i) Since $\nu_1(\tilde\ga_{\tilde\bb,e})=2$,
we can suppose $\tilde\ga_{\tilde\bb,e}(2\pi)\approx I_2\diamond M$ for some
basic normal form $M$ in $\Sp(2)$.
By Lemma 3.1 of \cite{ZL1}, there exists two paths $\ga_1$ and $\ga_2$
in $\mathcal{P}_{2\pi}(2)$ such that $\ga_1(2\pi)=I_2$,
$\ga_2(2\pi)=M$, $\tilde\ga_{\tilde\bb,e}\sim_1\ga_1\diamond\ga_2$,
and $i_1(\tilde\ga_{\tilde\bb,e})=i_1(\ga_1)+i_2(\ga_2)$.
Thus one of $i_1(\ga_1)$ and $i_1(\ga_2)$ must be odd,
and the other is even.

By Lemma \ref{oddevity.of.indices}, $i_1(\ga_1)$ is odd,
then $i_1(\ga_2)$ must be even.
Since $1\ne\sg(M)$,
by Lemma \ref{oddevity.of.indices}, 
we must have $M\approx D(2)$.

(ii) Firstly, if $\tilde\ga_{\tilde\bb,e}\approx N_2(e^{\sqrt{-1}\th},b)$
for some $\th\in(0,\pi)\cup(\pi,2\pi)$,
by $\<6\>,\<7\>$ of Lemma \ref{Lemma:property.of.splitting.numbers},
we have
\be
i_{-1}(\tilde\ga_{\tilde\bb,e})
=i_1(\tilde\ga_{\tilde\bb,e})
 -S_{N_2(e^{\sqrt{-1}\theta},b)}^-(e^{\sqrt{-1}\theta})
 +S_{N_2(e^{\sqrt{-1}\theta},b)}^+(e^{\sqrt{-1}\theta})
=i_1(\tilde\ga_{\tilde\bb,e})
\ee
or
\be
i_{-1}(\tilde\ga_{\tilde\bb,e})
=i_1(\tilde\ga_{\tilde\bb,e})
 -S_{N_2(e^{\sqrt{-1}\theta},b)}^-(e^{\sqrt{-1}(2\pi-\theta)})
 +S_{N_2(e^{\sqrt{-1}\theta},b)}^+(e^{\sqrt{-1}(2\pi-\theta)})
=i_1(\tilde\ga_{\tilde\bb,e}),
\ee
which contradicts $i_1(\tilde\ga_{\tilde\bb,e})=2n+1$ 
and $i_{-1}(\tilde\ga_{\tilde\bb,e})=2n$.

Then we can suppose $\tilde\ga_{\tilde\bb,e}(2\pi)\approx M_1\diamond M_2$
where $M_1$ and $M_2$ are two basic normal forms $\Sp(2)$.
Moreover, $1,-1\ne\sg(M_1)\cup\sg(M_2)$ 
since $\nu_{\pm1}(\tilde\ga_{\tilde\bb,e})=0$.
By Lemma 3.1 of \cite{ZL1}, there exists two paths $\ga_1$ and $\ga_2$
in $\mathcal{P}_{2\pi}(2)$ such that $\ga_1(2\pi)=M_1$,
$\ga_2(2\pi)=M_2$, $\tilde\ga_{\tilde\bb,e}\sim_1\ga_1\diamond\ga_2$,
and $i_1(\tilde\ga_{\tilde\bb,e})=i_1(\ga_1)+i_2(\ga_2)$.
Thus one of $i_1(\ga_1)$ and $i_1(\ga_2)$ must be odd,
and the other is even.

Without loss of generality, we
Since $1\not\in \sg(M_2)$,
by Lemma \ref{oddevity.of.indices}, we must have $\sg(M_2)\cap\U=\emptyset$ and $\alpha(M_2)=0$.
Thus, $M_2=D(2)$.

Similarly, 
by Lemma \ref{oddevity.of.indices}, together with $\pm 1\not\in\sg(M_1)$,
we have $M_1=D(-2)$ or $M_1=R(\theta)$
for some $\theta\in(0,\pi)\cup(\pi,2\pi)$.
If $M_1=D(-2)$, by the properties of splitting numbers in Chapter 9 of \cite{Lon4}, specially (9.3.3) on p.204, we
obtain $i_{-1}(\ga_{\bb,e})=i_1(\ga_{\bb,e})$, 
which contradicts $i_1(\tilde\ga_{\tilde\bb,e})=2n+1$ 
and $i_{-1}(\tilde\ga_{\tilde\bb,e})=2n$. 
Therefore, we must have $M_1=R(\theta)$.

If $\theta\in(0,\pi)$, we have
$i_{-1}(\tilde\ga_{\tilde\bb,e})=i_1(\tilde\ga_{\tilde\bb,e})-S_{R(\theta)}^-(e^{\sqrt{-1}\theta})+S_{R(\theta)}^+(e^{\sqrt{-1}\theta})=2n$.
When $\theta\in(\pi,2\pi)$, we obtain
$i_{-1}(\tilde\ga_{\tilde\bb,e})=i_1(\tilde\ga_{\tilde\bb,e})-S_{R(\theta)}^-(e^{\sqrt{-1}(2\pi-\theta)})+S_{R(\theta)}^+(e^{\sqrt{-1}(2\pi-\theta)})=2n+2$
contradicting $i_{-1}(\tilde\ga_{\tilde\bb,e})=2n$. 
Therefore, we have $\theta\in(0,\pi)$, and then $\tilde\ga_{\tilde\bb,e}(2\pi)\approx R(\theta)\diamond D(2)$. 

(v) If $\tilde\ga_{\tilde\bb,e}(2\pi)\approx N_2(e^{\sqrt{-1}\theta},b)$ 
for some
$\theta\in(0,\pi)\cup(\pi,2\pi)$, we now cannot use the method in (ii) directly to obtain the contradiction
because of $i_1(\tilde\ga_{\tilde\bb,e})=i_{-1}(\tilde\ga_{\tilde\bb,e})$.

On the one hand, $\tilde\ga_{\tilde\bb,e}(2\pi)\approx N_2(e^{\sqrt{-1}\theta},b)$ implies that
$(\tilde\bb,e)$ is on some $\om$-degenerate curve $\Th_{\om}$ where $\om\ne\pm1$.
On the other hand, 
$\tilde\bb_{2n+1}(-1,e)<\tilde\bb<\tilde\bb_{2n+2}(-1,e)$ implies that
$(\tilde\bb,e)$ is between the two $-1$-degenerate curves $\Xi_1^{\pm}$ which
start from the same point $(\hat\bb_{n+{1\over 2}},0)$. But $\Th_{\om}$ is a continuous curve defined on
the interval $[0,1)$ by Lemma \ref{Nonconvex:continuous}. Thus $\Th_{\om}$ must come down from the point
$(\tilde\bb,e)$ to the horizontal axis of $e=0$, and then it must intersect with at least one of $\Xi_{2n+1}$ and $\Xi_{2n+2}$,
which contradicts Theorem \ref{Nonconvex:not.intersect}.

Then we can suppose $\tilde\ga_{\tilde\bb,e}(2\pi)\approx M_1\diamond M_2$, and following similar steps in (ii),
we can obtain $\tilde\ga_{\tilde\bb,e}(2\pi)\approx D(-2)\diamond D(2)$.

(iii)-(iv) and (vi)-(vii) can be proved similarly.
For such items,
we first prove that $\tilde\ga_{\tilde\bb,e}(2\pi)\approx N_2(\om,b)$ 
is impossible by a method similar to that in the proof of (ii) or (v). 
Then $\ga_{\bb,e}(2\pi)\approx M_1\dm M_2$ must hold.
Then we use the information of $\pm1$-indices, null $\pm1$-indices, Lemma \ref{oddevity.of.indices} and Formula (\ref{index.formula})
to determine the basic normal forms of $M_1$ and $M_2$. 
Here the details are omitted. 
\hb

%\begin{remark}
%	By numerical computations, we can see that
%	the first $-1$-degenerate curve just intersect the $\tilde\bb$-axis (i.e., the first $1$-degenerate curve) at only one point $(0,e_0)$ 
%	for some $e_0\in(0,1)$,
%	and the second $-1$-degenerate curve is right to the $\tilde\bb$-axis.
%	Under these notation, we have:
%	
%	(i) If $-1<\tilde\bb<\tilde\bb_{1}(-1,e)<0$,
%	we have $i_1(\tilde\ga_{\tilde\bb,e})=0,\nu_1(\tilde\ga_{\tilde\bb,e})=0$,
%	$i_{-1}(\tilde\ga_{\tilde\bb,e})=0,\nu_{-1}(\tilde\ga_{\tilde\bb,e})=0$
%	and $\tilde\ga_{\tilde\bb,e}(2\pi)\approx ?\diamond ?$;
%	
%	(ii)
%\end{remark}

\subsection{The other properties of the degenerate curves}

Recall $\tilde{A}_N(-1,e)$ is positive definite on $\overline{D}(\om,2\pi)$
for any $\om\in\U$.
Now we set
\be
B(e,\om)=\tilde{A}_N(-1,e)^{-\frac{1}{2}}
               \left({I_2+5S(t)\over4(1+e\cos t)}\right)
               \tilde{A}_N(-1,e)^{-\frac{1}{2}}.
\ee
Because $\tilde{A}_N(-1,e)$ and $\frac{I_2+5S(t)}{4(1+e\cos t)}$ are self-adjoint, $B(e,\om)$ is also self-adjoint.
Moreover, $\tilde{A}_N(-1,e)$ and $\tilde{A}_NA(-1,e)^{-\frac{1}{2}}$ are compact operators, and hence by Theorem 4.8
in p.158 of \cite{Ka}, $B(e,\om)$ is a compact operator.
Then we have

\begin{lemma}\label{Nonconvex:B.operator}
	For $0\le e<1$, $\tilde{A}_N(\tilde\bb,e)$ is $1$-degenerate if and only if $-\frac{1}{\tilde\bb+1}$ is an eigenvalue of $B(e,\om)$.
\end{lemma}
{\bf Proof.} Suppose $\tilde{A}_N(\tilde\bb,e)x=0$ holds for some $x\in\overline{D}(1,2\pi)$.
Let $y=\tilde{A}_N(-1,e)^{\frac{1}{2}}x$. Then by (\ref{tilde.A_N}) we obtain
\bea
&&\tilde{A}_N(-1,e)^{\frac{1}{2}}\left(\frac{1}{\tilde\bb+1}+B(e,\om)\right)y(t)
\nonumber\\
&&=\left(\frac{\tilde{A}_N(-1,e)}{\tilde\bb+1}+\frac{I_2+5S(t)}{4(1+e\cos t)}\right)x(t)
\nonumber\\
&&=\frac{1}{\tilde\bb+1}A(\tilde\bb,e)x
\nonumber\\
&&=0.    \label{A.B}
\eea

Conversely, if $(\frac{1}{\tilde\bb+1}+B(e,\om))y=0$,
then $x=\tilde{A}_N(-1,e)^{-\frac{1}{2}}y$ is an eigenfunction of $\tilde{A}_N(\tilde\bb,e)$ belonging to the eigenvalue $0$
by our computation (\ref{A.B}).
\hb

Although $e<0$ does not have physical meaning, we can extend the fundamental solution to the case $e\in(-1,1)$
mathematically and all the above results which holds for $e>0$ also holds for $e<0$.
Then by the similar arguments of Theorem 4.16 and Theorem 4.17 in \cite{ZL1},
we have

\begin{theorem}\label{Th:analytic.om=1}
	Every $1$-degenerate curves $(\tilde\bb_N(1,e),e)$ in $e\in(-1,1)$
	is a real analytic function.
	Moreover, every $1$-degenerate curve starting from a point $(\hat\bb_n,0)$
	with $n\ge0$ must be orthogonal to the $\tilde\bb$-axis.
\end{theorem}

For $\om\ne1$, we have

\begin{theorem}\label{Th:analytic.om.orign}
	For $\om\ne1$, there exist two analytic $\om$-degenerate curves $(h_i(e),e)$ in $e\in(-1,1)$ with $i=1,2$
	such that $-1<h_i(e)<\tilde\bb_2(1,e)$.
	Specially, each $h_i(e)$ is a real analytic function in $e\in(-1,1)$ and $-1<h_i(e)<\tilde\bb_{2}(1,e)$.
	In fact, $\tilde\ga_{h_i(e),e}(2\pi)$ is $\om$-degenerate for $\om\in\U\backslash\{1\}$ and $i=1,2$.
	
	Moreover, for $\om\ne1$, there exist two analytic $\om$-degenerate curves $(h_i(e),e)$ in $e\in(-1,1)$ with $i=2n+1,2n+2$
	such that $\tilde\bb_{2n+1}(1,e)<h_i(e)<\tilde\bb_{2n+2}(1,e),n\ge1$.
	Specially, each $h_i(e)$ is a real analytic function in $e\in(-1,1)$ and $\tilde\bb_{2n+1}(1,e)<h_i(e)<\tilde\bb_{2n+2}(1,e)$.
	In fact, $\tilde\ga_{h_i(e),e}(2\pi)$ is $\om$-degenerate for $\om\in\U\backslash\{1\}$ and $i=2n+1,2n+2$.
\end{theorem}

{\bf Proof.} For $\tilde\bb\in(-1,\tilde\bb_2(1,e))$,
we have
\bea
i_\om(\tilde\ga_{\tilde\bb_{2}(1,e),e})&=&i_1(\tilde\ga_{\tilde\bb_{2}(1,e),e})+S_{\tilde\ga_{\tilde\bb_{2}(1,e),e}(2\pi)}^+(1)
\nn\\
&=&1+S_{I_2}^+(1)
\nn\\
&=&2.
\eea
Moreover, by Theorem \ref{Th:normal.forms.decomposition},
when $\tilde\bb=\tilde\bb_2(1,e)$, we have
\be\label{boundary1}
\tilde\ga_{\tilde\bb,e}\approx I_2\diamond D(2).
\ee
Therefore, by Lemma \ref{Nonconvex.Lemma:decreasing.of.index.wrt.beta}, 
it shows that, for fixed $e\in(-1,1)$,
there are exactly two values $\tilde\bb=h_{1}(e)$ and $h_{2}(e)$ 
in the interval $(-1,\tilde\bb_{2}(1,e)]$
at which (\ref{A.B}) is satisfied, and then $\tilde{A}_N(\tilde\bb,e)$ 
at these two values is $\om$-degenerate.
Note that these two $\tilde\bb$ values are possibly equal to 
each other at some $e$.
Moreover, (\ref{boundary1}) implies that $h_i(e)\ne \tilde\bb_{2}(1,e)$ 
for $i=1, 2$.

By Lemma \ref{Nonconvex:B.operator}, 
$-\frac{1}{h_i(e)+1}$ is an eigenvalue of ${B}(e,\om)$.
Note that ${B}(e,\om)$ is a compact operator and self adjoint when $e$ is real.
Moreover, it depends analytically on $e$. 
By \cite{Ka} (Theorem 3.9 in p.392), we know that
$-\frac{1}{h_i(e)+1}$ is analytic in $e$ for each $i$. 
This in turn implies that
both $h_{1}(e)$ and $h_{2}(e)$ are real analytic functions in $e$. 

On the second claim, when $\tilde\bb\in(\tilde\bb_{2n+1}(1,e),\tilde\bb_{2n+2}(1,e)),\;n\ge1$, 
from Theorem \ref{Th:normal.forms.decomposition},
we have
\be
i_1(\tilde\ga_{\tilde\bb,e})=2n+1,\quad \nu_1(\tilde\ga_{\tilde\bb,e})=0.
\ee
Moreover, we have
\be\label{boundary}
\tilde\ga_{\tilde\bb,e}\approx I_2\diamond D(2),\quad \tilde\bb=\tilde\bb_{2n+1}(1,e)\;\;{\rm or}\;\;\tilde\bb_{2n+2}(1,e).
\ee
Then for $\om\in\U\backslash\{1\}$, we have
\bea
i_\om(\tilde\ga_{\tilde\bb_{2n+1}(1,e),e})&=&i_1(\tilde\ga_{\tilde\bb_{2n+1}(1,e),e})+S_{\tilde\ga_{\tilde\bb_{2n+1}(1,e),e}(2\pi)}^+(1)
\nn\\
&=&2n-1+S_{I_2}^+(1)
\nn\\
&=&2n. \label{left.om.index}
\eea
Similarly, we have
\be
i_\om(\tilde\ga_{\tilde\bb_{2n+2}(1,e),e})=2n+2.\label{right.om.index}
\ee
Therefore, by Lemma \ref{Nonconvex.Lemma:decreasing.of.index.wrt.beta}, 
it shows that, for fixed $e\in(-1,1)$,
there are exactly two values $\tilde\bb=h_{2n+1}(e)$ and $h_{2n+2}(e)$ 
in the interval $[\tilde\bb_{2n+1}(1,e),\tilde\bb_{2n+2}(1,e)]$
at which (\ref{A.B}) is satisfied, and then $\tilde{A}_N(\tilde\bb,e)$ 
at these two values is $\om$-degenerate.
Note that these two $\tilde\bb$ values are possibly equal to 
each other at some $e$.
Moreover, (\ref{boundary}) implies that $h_i(e)\ne \tilde\bb_{2n+1}(1,e)$ 
and $\tilde\bb_{2n+2}(1,e)$ for $i=1, 2$.

By Lemma \ref{Nonconvex:B.operator}, 
$-\frac{1}{h_i(e)+1}$ is an eigenvalue of ${B}(e,\om)$.
Note that ${B}(e,\om)$ is a compact operator and self adjoint when $e$ is real.
Moreover, it depends analytically on $e$. 
By \cite{Ka} (Theorem 3.9 in p.392), we know that
$-\frac{1}{h_i(e)+1}$ is analytic in $e$ for each $i$. 
This in turn implies that
both $h_{2n+1}(e)$ and $h_{2n+2}(e)$ are real analytic functions in $e$. 
\hb

By the definition of $\tilde\bb_n(\om,e)$ in (\ref{Nonconvex:bb_s}), 
together with (\ref{left.om.index}) and (\ref{right.om.index}),
we have
\bea
\tilde\bb_{2n+1}(\om,e)=\min\{h_{2n+1}(e),h_{2n+2}(e)\}, \\
\tilde\bb_{2n+2}(\om,e)=\max\{h_{2n+1}(e),h_{2n+2}(e)\}.
\eea
Thus we have the following theorem:

\begin{theorem}\label{Th:analytic.om}
	For $\om\ne 1$, every $\om$-degenerate curve $(\tilde\bb_n(\om,e),e)$ in $e\in (-1,1)$ is a piecewise analytic function.
	The set of $e\in(-1,1)$ such that $\tilde\bb_{2n+1}(\om,e)=\tilde\bb_{2n+2}(\om,e)$ is discrete or equal to the whole interval
	$(-1,1)$. In the first case the functions $e\mapsto\tilde\bb_i(\om,e)$ with $i=2n+1$ and $2n+2$ are analytic
	for those $e$ when $\tilde\bb_{2n+1}(\om,e)<\bb_{2n+2}(\om,e)$. In the second case,
	$e\mapsto\tilde\bb_{2n+1}(\om,e)=\tilde\bb_{2n+2}(\om,e)$ is analytic everywhere.
\end{theorem}

In particular, we consider the $-1$-degenerate curves.
For $\hat{\bb}_{n+\frac{1}{2}}$ defined by (\ref{om_n.5}),
$\tilde{A}_N(\hat\bb_{n+\frac{1}{2}},0)$ is degenerate 
and by (\ref{Nonconvex:null.1-index.of.b0}),
$\dim\ker\tilde{A}_N(\hat\bb_{n+\frac{1}{2}},0)=v_{-1}(\tilde\ga_{\hat\bb_{n+\frac{1}{2}},0})=2$.
We set
$R(t)\left(\matrix{{a}_n\sin (n+\frac{1}{2})t\cr \cos (n+\frac{1}{2})t}\right)\in\ol{D}(-1,2\pi)$
for some constant ${a}_n$.

Moreover,
$\tilde{A}_N(\tilde\bb,0)R(t)\left(\matrix{{a}_n\sin (n+\frac{1}{2})t\cr \cos (n+\frac{1}{2})t}\right)=0$ reads
\be
\left\{
\begin{array}{cr}
	(n+\frac{1}{2})^2{a}_n-2(n+\frac{1}{2})+{3(\tilde\bb+3)\over2}{a}_n&=0,
	\\
	(n+\frac{1}{2})^2-2(n+\frac{1}{2}){a}_n-\tilde\bb&=0.
\end{array}
\right.
\ee
Then $3\tilde\bb^2-((n+\frac{1}{2})^2-9)\tilde\bb-(n+\frac{1}{2})^2[2(n+\frac{1}{2})^2+1]=0$ which yields $\tilde\bb=\hat\bb_{n+\frac{1}{2}}$ again and
\be\lb{a_n}
{a}_n=\frac{(n+\frac{1}{2})^2-\hat\bb_{n+\frac{1}{2}}}{2n+1}.
\ee
Then we have
$R(t)\left(\matrix{{a}_n\sin (n+\frac{1}{2})t\cr \cos (n+\frac{1}{2})t}\right)\in\ker A(\hat\bb_{n+\frac{1}{2}},0)$.
Similarly
$R(t)\left(\matrix{{a}_n\cos (n+\frac{1}{2})t\cr -\sin (n+\frac{1}{2})t}\right)\in\ker A(\hat\bb_{n+\frac{1}{2}},0)$,
therefore we have
\begin{equation}
\ker A(\hat\bb_{n+\frac{1}{2}},0)=\span\left\{\;
R(t)\left(\matrix{{a}_n\sin (n+\frac{1}{2})t\cr \cos (n+\frac{1}{2})t}\right),\quad
R(t)\left(\matrix{{a}_n\cos (n+\frac{1}{2})t\cr -\sin (n+\frac{1}{2})t}\right)
\;\right\}.
\lb{Nonconvex:ker.A.of.-1}
\end{equation}

Denote by $g$ the following operator
\begin{equation}
g(z)(t)=Nz(2\pi-t),
\end{equation}
where $N=\left(\matrix{1&0\cr 0&-1}\right)$.
Obviously, $g^2=Id$ and $g$ is unitary on $L^2([0,2\pi],\R^2)$.
One can check directly that
\begin{equation}
\tilde{A}_N(\tilde\beta,e)g=g\tilde{A}_N(\tilde\beta,e).
\end{equation}
Recall $E=\overline{D}(-1,2\pi)$ is given by (\ref{closure.of.D.om}),
and let $E_+=\ker(g+I)$,$E_-=\ker(g-I)$.
Following the studies in Section 2.2 and especially the proof in Theorem 1.1 in \cite{HS1},
the subspaces $E_+$ and $E_-$ are $\tilde{A}_N(\beta,e)$-orthogonal, 
and $E=E_+\bigoplus E_-$.
In fact, the subspaces $E=E_-$ and $E=E_+$ are isomorphic
to the following subspaces $E_1$ and $E_2$ respectively:
\begin{eqnarray}
E_1&=&\{z=(x,y)^T\in W^{2,2}([0,\pi],\R^2)\;|\;x(0)=0,y(\pi)=0\},\\
E_2&=&\{z=(x,y)^T\in W^{2,2}([0,\pi],\R^2)\;|\;x(\pi)=0,y(0)=0\}.\label{Nonconvex:E_2}
\end{eqnarray}
For $(\tilde\beta,e)\in(-1,+\infty)\times[0,1)$, restricting $\tilde{A}_N(\beta,e)$ to $E_1$ and $E_2$ respectively, we then obtain
\begin{eqnarray}
\phi_{-1}(\tilde{A}_N(\tilde\beta,e))&=&
  \phi_{-1}(\tilde{A}_N(\tilde\beta,e)|_{E_1})
  +\phi_{-1}(\tilde{A}_N(\tilde\beta,e)|_{E_2}),\\
\nu_{-1}(\tilde{A}_N(\tilde\beta,e))&=&
  \nu_{-1}(\tilde{A}_N(\tilde\beta,e)|_{E_1})+\nu_{-1}(\tilde{A}_N(\tilde\beta,e)|_{E_2}).
\end{eqnarray}
Similar to Proposition 7.1 in \cite{HLS}, we have

\begin{proposition}
	For any $n\ge0$,
	the $\om=-1$ degeneracy curve $(\tilde\beta_{2n+i}(e,-1),e)$ is precisely the
	degeneracy curve of $\tilde{A}_N(\tilde\beta,e)|_{E_i}$ for $i = 1$ or $2$
	when we restricted $\tilde\bb$ on the open interval $(\tilde\bb_{2n+1}(1,e),\tilde\bb_{2n+2}(1,e))$ 
	(when $n=0$, the interval is  $(-1,\tilde\bb_2(1,e))$).
\end{proposition}

Then we have the following theorem:

\begin{theorem}\label{Nonconves:orth.om}
	
	Every $-1$-degenerate curve must start from a point $(\hat\bb_{n+\frac{1}{2}},0)$ with some $n\ge0$.
	
	The tangent directions of the first two $-1$-degenerate curves are given by
	\bea
	{\partial\tilde\bb_1(-1,e)\over\partial e}\bigg|_{e=0}&=&
	-{41+5\sqrt{1297}\over48\sqrt{1297}}, \nn\\
	%-{2\sqrt{1297}(5\sqrt{1297}-41)\over1281},  \nn\\
	{\partial\tilde\bb_2(-1,e)\over\partial e}\bigg|_{e=0}&=&
	{41+5\sqrt{1297}\over48\sqrt{1297}}.
	%{2\sqrt{1297}(5\sqrt{1297}-41)\over1281}.
	\eea
	Any other $-1$-degenerate curve must be orthogonal to the $\tilde\bb$-axis.
\end{theorem}

{\bf Proof.} Similarly to Lemma \ref{Nonconvex:starting.points}, we have
\begin{equation}
\tilde\bb_n(-1,0)=
\left\{
\begin{array}{l}
\hat\bb_{m-\frac{1}{2}}, \quad {\rm if}\;\;n = 2m-1,
\\
\hat\bb_{m-\frac{1}{2}}, \quad {\rm if}\;\;n = 2m.
\end{array}
\right.
\lb{bb_n0.of.-1}
\end{equation}
Thus every $-1$-degenerate curve $(\tilde\bb(-1,e),e)$ must start from some point $(\hat\bb_{n-\frac{1}{2}},0)$.

Now we only consider the first two $-1$-degenerate curves. The others can be treated by the similar arguments.
Let $(\tilde\bb(e),e)$ be one of such curves (i.e., one of $(\tilde\bb_i(-1,e),e)$ with $i=1$ or $2$.)
which starts from $\tilde\bb(0)=\hat\bb_{n+\frac{1}{2}}$ with $e\in(-\epsilon,\epsilon)$ for some small $\epsilon>0$,
and $x_e\in\ol{D}(-1,2\pi)$ be the corresponding eigenvector, that is,
\be
\tilde{A}_N(\tilde\bb(e),e)x_e=0.
\ee
%Without loose of generality, by (\ref{ker.A.of.-1}), we suppose
By (\ref{Nonconvex:ker.A.of.-1}) and (\ref{Nonconvex:E_2}),
we have
\be
\ker(\tilde{A}_N(\tilde\bb,e)|_{E_1})
=\ker(\tilde{A}_N(\tilde\bb,e))\cap E_1
=\span\left\{\;R(t)\left(\matrix{{a}_0\sin\frac{1}{2}t\cr \cos\frac{1}{2}t}\right)\;\right\}.
\ee
Without loose of generality, we suppose
\be
z=({a}_0\sin\frac{1}{2}t,\cos\frac{1}{2}t)^T
\ee
and
\be\label{x0}
x_0=R(t)z=R(t)({a}_0\sin\frac{1}{2}t,\cos\frac{1}{2}t)^T.
\ee
There holds
\be
\<\tilde{A}_N(\tilde\bb(e),e)x_e,x_e\>=0.\label{Axx-1}
\ee

Differentiating both side of (\ref{Axx-1}) with respect to $e$ yields
$$ \tilde\bb'(e)\<\frac{\pt}{\pt \tilde\bb}\tilde{A}_N(\tilde\bb(e),e)x_e,x_e\> + (\<\frac{\pt}{\pt e}\tilde{A}_N(\tilde\bb(e),e)x_e,x_e\>
+ 2\<\tilde{A}_N(\tilde\bb(e),e)x_e,x'_e\> = 0,  $$
where $\tilde\bb'(e)$ and $x'_e$ denote the derivatives with respect to $e$. 
Then evaluating both sides at $e=0$ yields
\be  \tilde\bb'(0)\<\frac{\pt}{\pt \tilde\bb}\tilde{A}_N(\hat\bb_{\frac{1}{2}},0)x_0,x_0\>
+ \<\frac{\pt}{\pt e}\tilde{A}_N(\hat\bb_{\frac{1}{2}},0)x_0,x_0\> = 0.
\lb{4.73}
\ee
Then by the definition (\ref{tilde.A_N}) of $\tilde{A}_N(\tilde\bb,e)$ we have
\bea
\left.\frac{\pt}{\pt\tilde\bb}\tilde{A}_N(\tilde\bb,e)\right|_{(\tilde\bb,e)=(\hat\bb_{\frac{1}{2}},0)}
&=& \left.R(t)\frac{\pt}{\pt\tilde\bb}K_{N;\tilde\bb,e}(t)\right|_{(\tilde\bb,e)=(\hat\bb_{\frac{1}{2}},0)}R(t)^T,  \lb{4.74}\\
\left.\frac{\pt}{\pt e}\tilde{A}_N(\tilde\bb,e)\right|_{(\tilde\bb,e)=(\hat\bb_{\frac{1}{2}},0)}
&=& \left.R(t)\frac{\pt}{\pt e}K_{N;\tilde\bb,e}(t)\right|_{(\tilde\bb,e)=(\hat\bb_{\frac{1}{2}},0)}R(t)^T,  \lb{4.75}\eea
where $R(t)$ is given in \S 2.1. By direct computations from the definition of $K_{N;\tilde\bb,e}(t)$ in
(\ref{Nonconvex:K_tbe}), we obtain
\bea
&& \frac{\pt}{\pt\tilde\bb}K_{N;\tilde\bb,e}(t)\left|_{(\tilde\bb,e)=(\hat\bb_{\frac{1}{2}},0)}
= \left(\matrix{{3\over2} & 0\cr
	0 &  -1\cr}\right),\right.   \lb{4.76}\\
&& \frac{\pt}{\pt e}K_{N;\tilde\bb,e}(t)\left|_{(\tilde\bb,e)=(\hat\bb_{\frac{1}{2}},0)}
= {-\cos t}\left(\matrix{(9+3\hat\bb_{\frac{1}{2}})\slash2 & 0 \cr
	0 & -\hat\bb_{\frac{1}{2}}\cr}\right).\right.   \lb{4.77}\eea
Therefore from (\ref{x0}) and (\ref{4.74})-(\ref{4.77}) we have
\bea
\<\frac{\pt}{\pt\tilde\bb}\tilde{A}_N(\hat\bb_{\frac{1}{2}},0)x_0,x_0\>
&=& \<\frac{\pt}{\pt\tilde\bb}\tilde{A}_N(\hat\bb_{\frac{1}{2}},0)R(t)z,R(t)z\>
    \nn\\
&=& \<\frac{\pt}{\pt\tilde\bb}K_{N;\hat\bb_{\frac{1}{2}},0}z,z\>    \nn\\
&=& \int_0^{2\pi}[{3\over2}{a}_0^2\sin^2\frac{t}{2}
-\cos^2\frac{t}{2}]dt  \nn\\
&=& \pi({3\over2}{a}_0^2-1) \nn\\
&=&-{(41-\sqrt{1297})\sqrt{1297}\over192}\pi.  \lb{4.78}
\eea
Similarly, for $n\ge1$, we have
\bea  \<\frac{\pt}{\pt e}\tilde{A}_N(\hat\bb_{\frac{1}{2}},0)x_0,x_0\>
&=& {-1201+41\sqrt{1297}\over2304}.  \lb{4.79}\eea
Therefore by (\ref{4.73}) and (\ref{4.78})-(\ref{4.79}),
we obtain
\be  \beta'(0) = {41+5\sqrt{1297}\over48\sqrt{1297}}.\ee
Thus the theorem is proved.
\hb

Now we can give the

{\bf Proofs of Theorem \ref{main.theorem.separation.curves}(ii)-(iv).}
(ii) follows from Lemma \ref{Nonconvex:starting.points}, Theorem \ref{Nonconvex:multiplicity} and Theorem \ref{Th:analytic.om=1}.
(iii) and (iv) follow from
from Theorem \ref{Th:analytic.om} and Theorem \ref{Nonconves:orth.om}.
\hb

\setcounter{equation}{0}
\section{The convex ERE}

\subsection{The corresponding second order differential operator and the first estimation of the hyperbolic region}

In the convex case, by the argument below (\ref{direction.of.34}), we have
\be
\beta_{2,0}=\lambda_2={3-\sqrt{9-\beta}\over2},
\ee
and hence by (\ref{TwoSmallSec:bb_120}), we similarly have
\bea
\beta_{22,0}
&=&\frac{3[1-\sqrt{3}(1-2m)\sqrt{-1}]}{4}\left(-1+{3(3-\sqrt{9-\beta})\over2}\cdot\frac{1}{\sqrt{9-\beta}}\right)
\nonumber\\
&=&\frac{3[1-\sqrt{3}(1-2m)\sqrt{-1}]}{4}\cdot{9-5\sqrt{9-\beta}\over2\sqrt{9-\beta}}.
\eea
Moreover, we have
\bea
\lambda_{3}&=&{9-3\sqrt{9-\beta}\over2},
\\
\lambda_{4}&=&\sqrt{9-\beta}.
\eea
Here we don't care the order of the two roots.

Now let $\ga_C=\ga_{C;\beta,e}(t)$ is the fundamental solution of system (\ref{TwoSmallSec:linearized.system.sep_2}), i.e.,
\be  \label{TwoSmallSec:fundamtal.system.convex}
\left\{
\matrix{
\dot\ga(t)&=&JB_C(t)\ga(t),
\cr
\ga(0)&=&I_4,
}
\right.
\ee
with
\be
B_C(t)=\left(\matrix{1&  0&  0& 1\cr
                0&  1& -1& 0\cr
                0& -1& 1-\frac{9-3\sqrt{9-\bb}}{2(1+e\cos(t))}& 0\cr
                1&  0&   0& 1-\frac{\sqrt{9-\bb}}{1+e\cos(t)}
                }\right).
\ee
If there is no confusion, we will omit the subscript ``C", which indicates the ``convex case".

Let
\be  J_2=\left(\matrix{ 0 & -1 \cr 1 & 0 \cr}\right), \qquad
    K_{C;\bb,e}(t)=\left(\matrix{\frac{9-3\sqrt{9-\bb}}{2(1+e\cos(t))} & 0 \cr
                                     0 & \frac{\sqrt{9-\bb}}{1+e\cos(t)} \cr}\right),  \lb{Convex:K_be}\ee
and set
\be L(t,x,\dot{x})=\frac{1}{2}\|\dot{x}\|^2 + J_2x(t)\cdot\dot{x}(t) + \frac{1}{2}K_{C;\bb,e}(t)x(t)\cdot x(t),
       \qquad\quad  \forall\;x\in W^{1,2}(\R/2\pi\Z,\R^2),  \lb{TwoSmallSec:Lagrangian}\ee
where $a\cdot b$ denotes the inner product in $\R^2$. Obviously the origin in the configuration space is a
solution of the corresponding Euler-Lagrange system. By Legendrian transformation, the corresponding
Hamiltonian function is
$$   H(t,z)=\frac{1}{2}B_C(t)z\cdot z,\qquad \forall\; z\in\R^4.  $$

With a similar discussion in Section 4,
for $(\bb,e)\in [0,{27\over4}]\times [0,1)$, the second order differential operator corresponding to (\ref{TwoSmallSec:fundamtal.system.convex}) is given by
\bea  A_C(\bb,e)
&=& -\frac{d^2}{dt^2}I_2-I_2+R(t)K_{C;\bb,e}(t)R(t)^T  \nn\\
&=& -\frac{d^2}{dt^2}I_2-I_2+\frac{1}{1+e\cos t}\left({9-\sqrt{9-\beta}\over4}I_2+\left|{9-5\sqrt{9-\beta}\over4}\right|S(t)\right),  \lb{A_C}\eea
where $S(t)=\left(\matrix{ \cos 2t & \sin 2t \cr
                           \sin 2t & -\cos 2t \cr}\right)$, defined on the domain $\ol{D}(\omega,2\pi)$
in (\ref{closure.of.D.om}) below with $n=2$. Then it is self-adjoint and depends on the parameters $\bb$ and $e$.
As a similar to the argument in the non-convex case, we have for any $\bb$ and $e$, the Morse index $\phi_{\om}(A_C(\bb,e))$ and nullity $\nu_{\om}(A_C(\bb,e))$
of the operator $A_C(\bb,e)$ on the domain $\ol{D}(\omega,2\pi)$ satisfy
\be  \phi_{\om}(A_C(\bb,e)) = i_{\om}(\ga_{\bb,e}), \quad
      \nu_{\om}(A_C(\bb,e)) = \nu_{\om}(\ga_{\bb,e}), \qquad
           \forall \,\om\in\U. \ee
Then we have
\begin{lemma}\label{Convex:Lemma:hyperbolic}
If $\beta>{32\over9}$, for any $\om$ boundary conditions,
$A_C(\tilde\tau,\beta,e)$ is a positive operator.
Therefore,
\begin{equation}
i_\omega(\ga_{\beta,e})=0,\quad \beta\in\left({32\over9},{27\over4}\right],
\end{equation}
for any $e\in[0,1),\om\in\U$.
\end{lemma}

{\bf Proof.}
If $\beta\ge{144\over25}$, we have ${9-5\sqrt{9-\beta}\over4}\ge0$.
By (\ref{A_C}), we obtain
\bea\lb{A_C.decomposition}
A_C(\bb,e)
&=&-\frac{d^2}{dt^2}I_2-I_2+\frac{1}{1+e\cos t}\left({9-\sqrt{9-\beta}\over4}I_2+{9-5\sqrt{9-\beta}\over4}S(t)\right)
\nonumber\\
&=&-\frac{d^2}{dt^2}I_2-I_2+\frac{\sqrt{9-\beta}}{1+e\cos t}I_2
   +\frac{9-5\sqrt{9-\beta}}{4(1+e\cos t)}(I_2+S(t)).
\eea

Note that the range of $\beta$ is $[0,{27\over4}]$ and hence $\sqrt{9-\beta}>1$.
Then by Lemma 4.1 in \cite{ZL1},
$-\frac{d^2}{dt^2}I_2-I_2+\frac{\sqrt{9-\beta}}{1+e\cos t}I_2$ is positive definite
for any $\om$ boundary conditions.
Therefore, when $\beta\in\left[{144\over25},{27\over4}\right]$
by $\frac{9-5\sqrt{9-\beta}}{4(1+e\cos t)}\ge0$,
the non-negative definite of $I_2{\pm}S(t)$ and (\ref{A_C.decomposition}),
$A_C$ is positive definite for any $\om$ boundary conditions.

If $\beta<{144\over25}$, we have ${9-5\sqrt{9-\beta}\over4}<0$.
By (\ref{A_C}), we obtain
\bea\lb{A_C.decomposition.2}
A_C(\bb,e)
&=&-\frac{d^2}{dt^2}I_2-I_2+\frac{1}{1+e\cos t}\left({9-\sqrt{9-\beta}\over4}I_2+{5\sqrt{9-\beta}-9\over4}S(t)\right)
\nonumber\\
&=&-\frac{d^2}{dt^2}I_2-I_2+\frac{9-3\sqrt{9-\beta}}{2(1+e\cos t)}I_2
   +\frac{5\sqrt{9-\beta}-9}{4(1+e\cos t)}(I_2+S(t)).
\eea
When $\beta>{32\over9}$, we have ${9-3\sqrt{9-\beta}\over2}>1$.
Then by Lemma 4.1 in \cite{ZL1},
$-\frac{d^2}{dt^2}I_2-I_2+\frac{9-3\sqrt{9-\beta}}{2(1+e\cos t)}I_2$ is positive definite
for any $\om$ boundary conditions.
Therefore, when $\beta\in\left({32\over9},{144\over25}\right]$
by $\frac{5\sqrt{9-\beta}-9}{4(1+e\cos t)}\ge0$,
the non-negative definite of $I_2{\pm}S(t)$ and (\ref{A_C.decomposition.2}),
$A_C$ is positive definite for any $\om$ boundary conditions.
\hb

\begin{remark}
In particular, $5\in\left({32\over9},{144\over25}\right]$,
and hence $A_C(5,e)$ is positive definite for any $\om$ boundary conditions.
For later use, we write its explicit expression:
\begin{equation}
A_C(5,e)=-\frac{d^2}{dt^2}I_2-I_2+\frac{7I_2+S(t)}{4(1+e\cos t)}.
\end{equation}
\end{remark}

\subsection{The monotonicity of the Morse indices $\phi_\om(A_C(\beta,e))$ with respect to $\beta$}

Now by Lemma \ref{Convex:Lemma:hyperbolic},
we just need to study the operator $A_C(\beta,e)$ for $\beta\in[0,{32\over9}]\subset[0,5)$.
In such a case, we can rewrite $A_C(\bb,e)$ as follows
\begin{eqnarray}
A_C(\bb,e) &=& -\frac{d^2}{dt^2}I_2-I_2+\frac{1}{1+e\cos t}\left({9-\sqrt{9-\beta}\over4}I_2+{5\sqrt{9-\beta}-9\over4}S(t)\right)
\nonumber\\
&=&-\frac{d^2}{dt^2}I_2-I_2+\frac{7I_2+S(t)}{4(1+e\cos t)}+{\sqrt{9-\beta}-2\over4(1+e\cos t)}(-I_2+2S(t))
\nn\\
&=&A_C(5,e)+{\sqrt{9-\beta}-2\over4(1+e\cos t)}(-I_2+2S(t))
\nn\\
&=& (\sqrt{9-\beta}-2)\bar{A}_C(\bb,e), \lb{4.11}
\end{eqnarray}
where we define
\be\label{bar.A_C}
\bar{A}_C(\bb,e)=\frac{A_C(5,e)}{\sqrt{9-\beta}-2}
                          +\frac{-I_2+2S(t)}{4(1+e\cos t)}
.
\ee
Therefore we have
\bea
\phi_\omega(A_C(\bb,e)) &=& \phi_\omega(\bar{A}_C(\bb,e)), \lb{4.12}\\
\nu_\omega(A_C(\bb,e)) &=& \nu_\omega(\bar{A}_C(\bb,e)).   \lb{4.13}\eea

Now motivated by Lemma 4.4 in \cite{HLS} or Lemma 4.2 in \cite{ZL1}, and modifying its proof to our case,
we get the following lemma:

\begin{lemma}\label{Lemma:decreasing.of.index}
(i) For each fixed $e\in [0,1)$, the operator $\bar{A}_C(\bb,e)$ is increasing
with respect to $\beta\in [0,5)$ for any fixed $\omega\in\U$. Specially
\be  \frac{\pt}{\pt\beta}\bar{A}_C(\beta,e)|_{\bb=\bb_0}
     = \frac{1}{2\sqrt{9-\beta_0}(\sqrt{9-\beta_0}-2)^2}A_C(5,e),  \lb{4.14}\ee
is a positive definite operator for every $\bb_0\in [0,5)$.

(ii) For every eigenvalue $\lm_{\bb_0}=0$ of $\bar{A}_C(\bb_0,e_0)$ with $\om\in\U$ for some
$(\bb_0,e_0)\in[0,5)\times[0,1)$, there holds
\be \frac{d}{d\bb}\lm_{\bb}|_{\bb=\bb_0} > 0.  \lb{4.15}\ee

(iii)  For every $e_0\in[0,1),\bb_0\in[0,5)$ and $\om\in\U$,
there exist $\epsilon_0=\epsilon_0(\bb_0,e_0)>0$ small enough such that for all $\epsilon\in(0,\epsilon_0)$ there holds
\be
i_\om(\ga_{\bb_0,e_0}) - i_\om(\ga_{\bb_0+\epsilon,e_0})=\nu_\om(\ga_{\bb_0,e_0}).
\ee
\end{lemma}

\subsection{The $\om$-indices on the boundary segments $\{0\}\times[0,1)$ and $[0,5)\times\{0\}$}

Noting that when $\beta=0$, we have
\begin{equation}
A_C(0,e)=-{d^2\over dt^2}I_2-I_2+{3\over2(1+e\cos t)}(I_2+S(t)).
\end{equation}
This is just the same case which has been discussed in Section 3.1 of \cite{HLS}.
We just cite the results here:
\begin{eqnarray}
i_\om(\ga_{0,e})&=&
\left\{
\begin{array}{l}
 0, \quad {\rm if}\;\;\om=1,
\\
 2, \quad {\rm if}\;\;\om\in\U\backslash\{1\},
\end{array}
\right.  \label{Convex:index.of.0e}
\\
\nu_\om(\ga_{0,e})&=&
\left\{
\begin{array}{l}
 3, \quad {\rm if}\;\;\om=1,
\\
 0, \quad {\rm if}\;\;\om\in\U\backslash\{1\}.
\end{array}
\right.   \label{Convex:nullity.of.0e}
\end{eqnarray}

Next, we consider the case $e=0$.
The system (\ref{TwoSmallSec:fundamtal.system.convex}) becomes an ODE
system with constant coefficients:
\be B_C = B_C(t) = \left(\matrix{1 & 0 & 0 & 1\cr
                             0 & 1 & -1 & 0 \cr
                             0 & -1 & -{7-3\sqrt{9-\beta}\over2} & 0 \cr
                             1 & 0 & 0 & 1-\sqrt{9-\beta} \cr}\right)
                             .  \ee
The characteristic polynomial $\det(JB_C-\lambda I)$ of $JB_C$ is given by
\be \lambda^4 + {\sqrt{9-\beta}-1\over2}\lambda^2+{3\sqrt{9-\beta}(3-\sqrt{9-\beta})\over2}  = 0.  \lb{Convex:ch.poly}\ee
Letting $\aa=\lambda^2$,
the two roots of the quadratic polynomial $\aa^2 + {\sqrt{9-\beta}-1\over2}\aa +{3\sqrt{9-\beta}(3-\sqrt{9-\beta})\over2}$
are given by $\aa_1=\frac{-(\sqrt{9-\beta}-1)+\sqrt{25({9-\bb})-74\sqrt{9-\bb}+1}}{4}$
and $\aa_2=\frac{-(\sqrt{9-\beta}-1)-\sqrt{25({9-\bb})-74\sqrt{9-\bb}+1}}{4}$.
Therefore the four characteristic multipliers of the matrix $\ga_{\beta,0}(2\pi)$ are given by
\begin{eqnarray}
\rho_{1,\pm}(\beta) &=&e^{\pm2\pi\sqrt{\aa_1}},
\\
\rho_{2,\pm}(\beta) &=&e^{\pm2\pi\sqrt{\aa_2}}.
\end{eqnarray}

When $\beta\in[0,{16\over625}(182-37\sqrt{21})]$,
we have $\sqrt{9-\beta}\in[{37+8\sqrt{21}\over25},3]$ and $25(9-\bb)-74\sqrt{9-\bb}+1\ge0$,
and then $\alpha_1,\alpha_2\in\R^-$.
Hence we have $\rho_{i,\pm}\in\U$ for $i=1,2$.
When $\tilde\beta\in[{16\over625}(182-37\sqrt{21}),5)$,
we have $25(9-\bb)-74\sqrt{9-\bb}+1<0$ and hence $\alpha_1,\alpha_2\in\C\backslash\R$.
Then $\rho_{i,\pm}\in\C\backslash(\U\cup\R)$ for $i=1,2$.

For more details, if $\beta\in[0,{16\over625}(182-37\sqrt{21})]$, we set
\begin{eqnarray}
\theta_1(\beta)=\sqrt{-\alpha_1}&=&\sqrt{(\sqrt{9-\bb}-1)-\sqrt{25(9-\bb)-74\sqrt{9-\bb}+1}\over4},
\label{th1.of.bb}
\\
\theta_2(\beta)=\sqrt{-\alpha_2}&=&\sqrt{(\sqrt{9-\bb}-1)+\sqrt{25(9-\bb)-74\sqrt{9-\bb}+1}\over4}.  \label{th2.of.bb}
\end{eqnarray}
Moreover, we have $0<\sqrt{25(9-\bb)-74\sqrt{9-\bb}+1}\le2$ and hence
\begin{eqnarray}
{d\th_1(\bb)^2\over d\bb}&=&-{d\alpha_1\over d\bb}
   =-{1\over8\sqrt{9-\beta}}\left(1-{25\sqrt{9-\bb}-37\over\sqrt{25(9-\bb)-74\sqrt{9-\bb}+1}}\right)
   >0,
\\
{d\th_2(\bb)^2\over d\bb}&=&-{d\alpha_2\over d\bb}
   =-{1\over8\sqrt{9-\beta}}\left(1+{25\sqrt{9-\bb}-37\over\sqrt{25(9-\bb)-74\sqrt{9-\bb}+1}}\right)
<0,
\end{eqnarray}
when $\bb\in\left[0,{16\over625}(182-37\sqrt{21}\;)\right)$.
Letting $\bb^*$ be the $\bb$ such that $\th_1(\bb^*)={1\over2}$, then we have
\begin{equation}
\bb^*={1331-35\sqrt{1297}\over288}.
\end{equation}

Then we obtain the following results:

(i) When $\bb=0$, we have $\sg(\ga_{0,0}(2\pi)) = \{1, 1, 1, 1\}$.

(ii) When $0<\bb<{1331-35\sqrt{1297}\over288}$,
the angle $\th_1(\bb)$ in (\ref{th1.of.bb}) increases strictly from $0$ to $\frac{1}{2}$ as $\bb$ increases
from $0$ to ${1331-35\sqrt{1297}\over288}$.
Therefore $\rho_{1,+}(\bb)=e^{2\pi \sqrt{-1}\th_1(\bb)}$ runs from $1$ to $-1$ counterclockwise along
the upper semi-unit circle in the complex plane $\C$ as $\bb$ increases from $0$ to ${1331-35\sqrt{1297}\over288}$.
Correspondingly
$\rho_{1,-}(\bb)=e^{-2\pi \sqrt{-1}\th_1(\bb)}$ runs from $1$ to $-1$ clockwise along the lower semi-unit circle in
$\C$ as $\bb$ increases from $0$ to ${1331-35\sqrt{1297}\over288}$.
At the same time, because $\th_2(\bb)$ decreases strictly from $1$ to ${\sqrt{3\sqrt{1297}-3}\over12}$,
therefore $\rho_{2,+}(\bb)=e^{2\pi \sqrt{-1}\th_2(\bb)}$ runs from $1$ to $e^{\sqrt{-1}2\pi{\sqrt{3\sqrt{1297}-3}\over12}}$
clockwise along the lower semi-unit circle in the complex plane $\C$
as $\bb$ increases from $0$ to ${1331-35\sqrt{1297}\over288}$.
Correspondingly
$\rho_{2,-}(\bb)=e^{-2\pi \sqrt{-1}\th_2(\bb)}$ runs from $1$ to $e^{-\sqrt{-1}2\pi{\sqrt{3\sqrt{1297}-3}\over12}}$
counterclockwise along the upper semi-unit circle in $\C$
as $\bb$ increases from $0$ to ${1331-35\sqrt{1297}\over288}$.
Thus specially we obtain
$\sg(\ga_{\bb,0}(2\pi))\subset \U\bs\R$ for all $\bb\in (0,{1331-35\sqrt{1297}\over288})$.

(iii) When $\bb={1331-35\sqrt{1297}\over288}$, we have $\th_1(\bb)={1\over2}$ and $\th_2(\bb)={\sqrt{3\sqrt{1297}-3}\over12}$.
Therefore we obtain $\rho_{1,\pm}(\bb)=-1$ and $\rho_{2,\pm}(\bb)\in\U\backslash\{\pm1\}$.

(iv) When ${1331-35\sqrt{1297}\over288}<\bb<{16\over625}(182-37\sqrt{21})$,
the angle $\th_1(\bb)$ in (\ref{th1.of.bb}) increases strictly from $\frac{1}{2}$ to ${\sqrt{3+2\sqrt{21}}\over5}$ as $\bb$ increases
from ${1331-35\sqrt{1297}\over288}$ to ${16\over625}(182-37\sqrt{21})$.
Therefore $\rho_{1,+}(\bb)=e^{2\pi \sqrt{-1}\th_1(\bb)}$ runs from $-1$ to $e^{\sqrt{-1}2\pi{\sqrt{3+2\sqrt{21}}\over5}}$
counterclockwise along the lower semi-unit circle in the complex plane $\C$
as $\bb$ increases from ${1331-35\sqrt{1297}\over288}$ to ${16\over625}(182-37\sqrt{21})$.
Correspondingly
$\rho_{1,-}(\bb)=e^{-2\pi \sqrt{-1}\th_1(\bb)}$ runs from $-1$ to $e^{-\sqrt{-1}2\pi{\sqrt{3+2\sqrt{21}}\over5}}$
clockwise along the upper semi-unit circle in $\C$
as $\bb$ increases from ${1331-35\sqrt{1297}\over288}$ to ${16\over625}(182-37\sqrt{21})$.
At the same time, because $\th_2(\bb)$ decreases strictly from ${\sqrt{3\sqrt{1297}-3}\over12}$ to ${\sqrt{3+2\sqrt{21}}\over5}$,
therefore $\rho_{2,+}(\bb)=e^{2\pi \sqrt{-1}\th_2(\bb)}$ runs from $e^{\sqrt{-1}2\pi{\sqrt{3\sqrt{1297}-3}\over12}}$
to $e^{\sqrt{-1}2\pi{\sqrt{3+2\sqrt{21}}\over5}}$
clockwise along the lower semi-unit circle in the complex plane $\C$
as $\bb$ increases from ${1331-35\sqrt{1297}\over288}$ to ${16\over625}(182-37\sqrt{21})$.
Correspondingly
$\rho_{2,-}(\bb)=e^{-2\pi \sqrt{-1}\th_2(\bb)}$ runs from $e^{-\sqrt{-1}2\pi{\sqrt{3\sqrt{1297}-3}\over12}}$
to $e^{-\sqrt{-1}2\pi{\sqrt{3+2\sqrt{21}}\over5}}$
counterclockwise along the upper semi-unit circle in $\C$
as $\bb$ increases from ${1331-35\sqrt{1297}\over288}$ to ${16\over625}(182-37\sqrt{21})$.
Thus specially we obtain
$\sg(\ga_{\bb,0}(2\pi))\subset \U\bs\R$ for all $\bb\in ({1331-35\sqrt{1297}\over288},{16\over625}(182-37\sqrt{21}))$.

(v) When $\bb={16\over625}(182-37\sqrt{21})$, we obtain $\th_1(\bb)=\th_2(\bb)={\sqrt{3+2\sqrt{21}}\over5}$,
and then we have double eigenvalues
$\rho_{1,\pm}(\bb)=\rho_{2,\pm}(\bb)=e^{\pm\sqrt{-1}2\pi{\sqrt{3+2\sqrt{21}}\over5}}\in\U\backslash\{\pm1\}$.

(vi) When ${16\over625}(182-37\sqrt{21})<\tilde\beta<5$,
we have $25(9-\bb)-74\sqrt{9-\bb}+1<0$ and hence $\alpha_1,\alpha_2\in\C\backslash\R$.
Then $\rho_{i,\pm}\in\C\backslash(\U\cup\R)$ for $i=1,2$.

Under the similar arguments of $({\bf B})$ and $({\bf C})$ in Section 3.2 of \cite{ZL1},
we have
\begin{eqnarray}
&& i_1(\ga_{\bb,0}) = 0,\qquad\forall\;\bb\in[0,5],    \lb{Convex:1-index.of.b0}
\\
&& \nu_1(\ga_{\bb,0}) = \left\{\matrix{
                 3, &  {\rm if}\;\;\bb=0,\quad \cr
                 0, &  {\rm if}\;\;\bb\in(0,5]\cr}\right. \lb{Convex:null.1-index.of.b0}
\end{eqnarray}
and
\begin{eqnarray}
&& i_{-1}(\ga_{\bb,0}) = \left\{\matrix{
                 2, &  {\rm if}\;\;\bb\in[0,\bb^*), \cr
                 0, &  {\rm if}\;\;\bb\in(\bb^*,5],\cr}\right. \lb{Convex:-1-index.of.b0}\\
&& \nu_{-1}(\ga_{\bb,0}) = \left\{\matrix{
                 0, &  {\rm if}\;\;\bb\in[0,5]\backslash\{\bb^*\}, \cr
                 2, &  {\rm if}\;\;\bb=\bb^*.\qquad\quad \cr} \right.
\lb{Convex:null.-1-index.of.b0}
\end{eqnarray}

\subsection{The degenerate curves}

By (\ref{Convex:index.of.0e}), (\ref{Convex:nullity.of.0e}) and Lemma \ref{Lemma:decreasing.of.index}, we have

\begin{corollary}\label{Convex:C5.4} For every fixed $e\in [0,1)$ and $\omega\in \U$, the index function
$\phi_{\omega}(A_C(\beta,e))$, and consequently $i_{\omega}(\gamma_{\beta,e})$, is non-increasing
as $\beta$ increases from $0$ to $5$.
When $\omega=1$, these index functions are constantly equal to $0$,
and when $\omega\in\U\setminus\{1\}$, they are decreasing and tends from $2$ to $0$.
\end{corollary}

{\bf Proof.}
For $0<\bb_1<\bb_2\le1$ and fixed $e\in [0,1)$, when $\bb$ increases from $\bb_1$ to
$\bb_2$, it is possible that negative eigenvalues of $\bar{A}_C(\bb_1,e)$ pass through $0$ and become
positive ones of $\bar{A}_C(\bb_2,e)$, but it is impossible that positive eigenvalues of
$\bar{A}_C(\bb_2,e)$ pass through $0$ and become negative by (ii) of Lemma \ref{Lemma:decreasing.of.index}.
\hb

By a similar analysis to the proof of Proposition 6.1 in \cite{HLS},
for every $e\in[0,1)$ and $\omega\in\U\backslash\{1\}$,
the total multiplicity of $\omega$-degeneracy of $\gamma_{\bb,e}(2\pi)$ for $\bb\in[0,5]$ is always precisely 2, i.e.,
\begin{equation}
    \sum_{\bb\in[0,5]}v_\omega(\gamma_{\bb,e}(2\pi))=2,\quad \forall \omega\in\U\backslash\{1\}.
\end{equation}

Consequently,
together with the positive definiteness of $A_C(5,e)$ for the $\omega\in\U\backslash\{1\}$ boundary condition,
we have

\begin{theorem}\label{Convex:Th:om.degenerate.curves}
For any $\omega\in\U\backslash\{1\}$, there exist two analytic $\omega$-degenerate curves $(\bb_i(e,\omega),e)$
in $e\in[0,1)$ with $i=1,2$.
Specially, each $\bb_i(e,\omega)$ is a real analytic function in $e\in[0,1)$,
and $0<\bb_i(e,\omega)<5$ and $\gamma_{\bb_i(e,\omega),e}(2\pi)$ is $\omega$-degenerate for $\omega\in\U\backslash\{1\}$
and $i=1,2$.
\end{theorem}

{\bf Proof.}
By Lemma \ref{Convex:Lemma:hyperbolic},
we have $i_\omega(\gamma_{\bb,e})=0$ when $\bb$ is near $5$.
Then under similar steps to those of Lemma 6.2 and Theorem 6.3 in \cite{HLS},
we can prove the theorem.
\hb

Specially, for $\omega=-1$, $e\in[0,1)$, we define
\begin{equation}
    \mu_l(e)=\min\{\bb_1(e,-1),\bb_2(e,-1)\},\quad
    \mu_m(e)=\max\{\bb_1(e,-1),\bb_2(e,-1)\},
\end{equation}
where $\bb_i(e,-1)$ are the two $-1$-dgenerate curves as in Theorem \ref{Convex:Th:om.degenerate.curves}.

%For the $\om=-1$ boundary condition,
%denote by $g$ the following operator
%\begin{equation}
%g(z)(t)=-Nz(2\pi-t),
%\end{equation}
%where $N=\left(\matrix{1& 0\cr 0& -1}\right)$.
%Obviously, $g^2=Id$ and $g$ is unitary on $L^2([0,2\pi],\R^2)$
%One can check directly that
%\begin{equation}
%A_C(\beta,e)g=gA_C(\beta,e).
%\end{equation}
Recall $E=\overline{D}(-1,2\pi)$ which is given by (\ref{closure.of.D.om}), and 
%let $E_+=\ker(g+I)$,$E_-=\ker(g-I)$.
%Following the studies in Section 2.2 and especially the proof in Theorem 1.1 in \cite{HS1},
%the subspaces $E_+$ and $E_-$ are $A_C(\beta,e)$-orthogonal, and $E=E_+\bigoplus E_-$.
%In fact, the subspaces $E=E_-$ and $E=E_+$ are isomorphic
%to the following subspaces $E_1$ and $E_2$ respectively:
\begin{eqnarray}
E_1&=&\{z=(x,y)^T\in W^{2,2}([0,\pi],\R^2)\;|\;x(0)=0\;y(\pi)=0\},\nn\\
E_2&=&\{z=(x,y)^T\in W^{2,2}([0,\pi],\R^2)\;|\;x(\pi)=0,\;y(0)=0\}.\nn
\end{eqnarray}
For $(\beta,e)\in[0,5)\times[0,1)$, restricting $A_C(\beta,e)$ to $E_1$ and $E_2$ respectively, we then obtain
\begin{eqnarray}
\phi_{-1}(A_C(\beta,e))&=&\phi_{-1}(A_C(\beta,e)|_{E_1})+\phi_{-1}(A_C(\beta,e)|_{E_2}),\\
\nu_{-1}(A_C(\beta,e))&=&\nu_{-1}(A_C(\beta,e)|_{E_1})+\nu_{-1}(A_C(\beta,e)|_{E_2}),
\end{eqnarray}
where the left hand sides are the Morse index and nullity of the operator $A_C(\bb,e)$ on the space $\overline{D}(-1,2\pi)$,
that is, the $-1$ index and nullity of $\ga_{C;\bb,e}$;
on the right hand side, we denote by $\phi(A_C(\bb,e)|_{E_i})$ and $\nu(A_C(\bb,e)|_{E_i})$ the usual Morse index
and nullity of the operator $A_C(\bb,e)|_{E_i}$ on the space $E_i$.

Similar to Proposition 7.1 in \cite{HLS}, we have

\begin{proposition}
The $\om=-1$ degeneracy curve $(\beta_i(e,-1),e)$ is precisely the
degeneracy curve of $A_C(\beta,e)|_{E_i}$ for $i = 1$ or $2$.
\end{proposition}

By (\ref{Convex:null.-1-index.of.b0}), $-1$ is a double eigenvalue of the matrix $\gamma_{\bb^*,e}(2\pi)$,
then the two curves bifurcation out from $(\bb^*,0)$ when $e>0$ is small enough.

Recall that $A_C(\bb^*,0)$ is $-1$-degenerate. Then, by (\ref{Convex:null.-1-index.of.b0}), we find that
$\dim\ker A_C(\bb^*,0)=\nu_{-1}(\gamma_{\bb^*,0})=2$.
By the definition of (\ref{closure.of.D.om}),
we have $R(t)\left(\matrix{\tilde{a}_n\sin (n+\frac{1}{2})t\cr \cos (n+\frac{1}{2})t}\right)\in\overline{D}(-1,2\pi)$
for any constant $n\in\N$ and $\tilde{a}_n\in\C$.

Moreover,
$A_C(\bb,0)R(t)\left(\matrix{\tilde{a}_n\sin (n+\frac{1}{2})t\\ \cos (n+\frac{1}{2})t}\right)=0$ reads
\begin{equation}
    \left\{
    \begin{array}{cr}
    	(n+\frac{1}{2})^2\tilde{a}_n-2(n+\frac{1}{2})+{9-3\sqrt{9-\bb}\over2}\tilde{a}_n&=0,\\
    	(n+\frac{1}{2})^2-2(n+\frac{1}{2})\tilde{a}_n+\sqrt{9-\bb}&=0.
    \end{array}
    \right.
\end{equation}
Then $(n+{1\over2})^4-{\sqrt{9-\beta}-1\over2}(n+{1\over2})^2+{3\sqrt{9-\bb}(3-\sqrt{9-\bb})\over2}=0$ holds only when $n=0$
and $\bb={1331-35\sqrt{1297}\over288}=\bb^*$, and hence
\begin{equation}\label{tilde.a}
    \tilde{a}_0={1\over4}+\sqrt{9-\bb^*}={41+\sqrt{1297}\over24}.
\end{equation}
Then we have
$R(t)\left(\matrix{\tilde{a}_0\sin\frac{t}{2}\\ \cos\frac{t}{2}\ }\right)\in\ker A_C(\bb^*,0)$.
Similarly, we have
$R(t)\left(\matrix{\tilde{a}_0\cos\frac{t}{2}\cr -\sin\frac{t}{2}}\right)\in\ker A_C(\bb^*,0)$,
and hence
\begin{equation}
    \ker A(\bb^*,0)={\rm span}\left\{
	R(t)\left(\matrix{\tilde{a}_0\sin\frac{t}{2}\cr \cos\frac{t}{2}}\right),\quad
	R(t)\left(\matrix{\tilde{a}_0\cos\frac{t}{2}\cr -\sin\frac{t}{2}}\right)
    \right\}.
\label{ker.A.of.-1}
\end{equation}
Then we have the following theorem:

\begin{theorem}\label{Convex:Th:tangent.direction}
The tangent directions of the two curves $\Gamma_l$ and $\Gamma_m$ at the same bifurcation point $(\bb^*,0)$
are given by
\begin{equation}
    \bb_l'(e)|_{e=0}=-\frac{2525+67\sqrt{1297}}{96\sqrt{1297}},\quad
    \bb_m'(e)|_{e=0}=\frac{2525+67\sqrt{1297}}{96\sqrt{1297}}.
\end{equation}
\end{theorem}

{\bf Proof.}
Now let $(\bb(e),e)$ be one of such curves
(say, the $E_1$ degenerate curve)
which starts from $\bb^*$ with $e\in[0,\epsilon)$ for some small $\epsilon>0$ and $x_e\in E_1\subset\bar{D}(-1,2\pi)$
being the corresponding eigenvector, that is
\begin{equation}
    A_C(\bb(e),e)x_e=0.
\end{equation}
By (\ref{E_2}) and (\ref{ker.A.of.-1}), we have
\begin{equation}
\ker(A_C(\bb,e)|_{E_2})=\ker(A_C(\bb,e))\cap E_1
     = \span\left\{R(t)\left(\matrix{\tilde{a}_0\sin\frac{t}{2}\cr \cos\frac{t}{2}}\right)\right\}.
\end{equation}
Without loss of generality, by (\ref{ker.A.of.-1}), we suppose
$$
z = (\tilde{a}_0\sin\frac{t}{2},\cos\frac{t}{2})^T
$$
and
\begin{equation}
    x_0=R(t)z=R(t)(\tilde{a}_0\sin\frac{t}{2},\cos\frac{t}{2})^T.  \label{Convex:x_0}
\end{equation}
There holds
\begin{equation}
    \<A_C(\bb(e),e)x_e,x_e\>=0.\label{Convex:Axx-1}
\end{equation}

Differentiating both side of (\ref{Convex:Axx-1}) with respect to $e$ yields
$$ \bb'(e)\<\frac{\partial}{\partial \bb}A_C(\bb(e),e)x_e,x_e\> + (\<\frac{\partial}{\partial e}A_C(\bb(e),e)x_e,x_e\>
       + 2\<A_C(\bb(e),e)x_e,x'_e\> = 0,  $$
where $\bb'(e)$ and $x'_e$ denote the derivatives with respect to $e$. Then evaluating both
sides at $e=0$ yields
\begin{equation}  \label{Convex:equality.of.bb'}
    \bb'(0)\<\frac{\partial}{\partial \bb}A_C(\bb^*,0)x_0,x_0\>
      + \<\frac{\partial}{\partial e}A_C(\bb^*,0)x_0,x_0\> = 0.
\end{equation}
Then by the definition (\ref{A_C}) of $A_C(\bb,e)$ we have
\begin{eqnarray}
\left.\frac{\partial}{\partial\bb}A_C(\bb,e)\right|_{(\bb,e)=(\bb^*,0)}
    &=& \left.R(t)\frac{\partial}{\partial\bb}K_{\bb,e}(t)\right|_{(\bb,e)=(\bb^*,0)}R(t)^T,  \label{Convex:pA.pb.1}\\
\left.\frac{\partial}{\partial e}A_C(\bb,e)\right|_{(\bb,e)=(\bb^*,0)}
    &=& \left.R(t)\frac{\partial}{\partial e}K_{\bb,e}(t)\right|_{(\bb,e)=(\bb^*,0)}R(t)^T.  \label{Convex:pA.pe.1}
\end{eqnarray}
By direct computations from the definition of $K_{\bb,e}(t)$ in
(\ref{Convex:K_be}), we obtain
\begin{eqnarray}
    \left.\frac{\partial}{\partial\bb}K_{\bb,e}(t)\right|_{(\bb,e)=(\bb^*,0)}
       &=& \left(\matrix{{3\over4\sqrt{9-\bb^*}}& 0\cr 0&  -{1\over2\sqrt{9-\bb^*}}}\right),   \label{Convex:pA.pb.2}\\
    \left.\frac{\partial}{\partial e}K_{\bb,e}(t)\right|_{(\bb,e)=(\bb^*,0)}
       &=& {-\cos t}\left(\matrix{{9-3\sqrt{9-\bb^*}\over2} & 0 \cr 0 & \sqrt{9-\bb^*}}\right).   \label{Convex:pA.pe.2}
 \end{eqnarray}
Therefore from (\ref{Convex:x_0}) and (\ref{Convex:pA.pb.1})-(\ref{Convex:pA.pe.2}) we have
\begin{eqnarray}
    \<\frac{\partial}{\partial\bb}A_C(\bb^*,0)x_0,x_0\>
    &=& \<\frac{\partial}{\partial\bb}K_{\bb^*,0}z,z\>    \nonumber\\
    &=& \int_0^{2\pi}\left[{3\over4\sqrt{9-\bb^*}}\tilde{a}_0^2\sin^2\frac{t}{2}
                -{1\over2\sqrt{9-\bb^*}}\cos^2\frac{t}{2}\right]dt  \nonumber\\
    &=& \pi\frac{1297-23\sqrt{1297}}{192}   \label{Convex:A1'xx}
\end{eqnarray}
and
\begin{eqnarray}
    \<\frac{\partial}{\partial e}A_C(\bb^*,0)x_0,x_0\>
    &=& \<\frac{\partial}{\partial e}K_{\bb^*,0}z,z\>   \nonumber\\
    &=& -\int_0^{2\pi}\left[{9-3\sqrt{9-\bb^*}\over2}\tilde{a}_0^2\cos{t}\sin^2\frac{t}{2}
                +\sqrt{9-\bb^*}\cos{t}\cos^2\frac{t}{2}\right]dt  \nonumber\\
    &=&-\pi\frac{1201+41\sqrt{1297}}{2304}.    \label{Convex:A2'xx}
\end{eqnarray}
Therefore by (\ref{Convex:equality.of.bb'}) and (\ref{Convex:A1'xx})-(\ref{Convex:A2'xx}),
we obtain
\begin{equation}
    \bb'(0) = \frac{2525+67\sqrt{1297}}{288\sqrt{1297}}.
\end{equation}
The other tangent can be compute similarly.
Thus the theorem is proved.
\hb

\subsection{The region division and the symplectic normal forms of $\ga_{\bb,e}(2\pi)$}

For every $e\in[0,1)$, we recall
\be\label{bb_r}
    \bb_r(e)=\sup\{\bb'\in[0,1]|\sigma(\gamma_{\bb,e}(2\pi))\cap\U\ne\emptyset,\;\forall \bb\in[0,\bb']\},
\ee
and
$$
    \Gamma_r=\{(\bb_r(e),e)\in[0,1]\times[0,1)\}.
$$
By the similar arguments of Lemma 9.1 and Corollary 9.2 in \cite{HLS}, we have
\begin{lemma}\label{Convex:Lm:hyperbolic.region}
(i) If $0\le\bb_1<\bb_2\le5$ and $\gamma_{\bb_1,e}(2\pi)$ is hyperbolic, so does $\gamma_{\bb_2,e}(2\pi)$.
Consequently, the hyperbolic region of $\gamma_{\bb,e}(2\pi)$ in $[0,5]\times[0,1)$ is connected.

(ii) For any fixed $e\in[0,1)$, every matrix $\gamma_{\bb,e}(2\pi)$ is hyperbolic if $\bb_r(5)<\bb\le5$
for $\bb_r(e)$ defined by (\ref{bb_r}).

(iii) We have
\begin{equation}
    \sum_{\bb\in[0,\bb_r(e)]}\nu_\omega(\gamma_{\bb,e}(2\pi))=2,\quad \forall\omega\in\U\backslash\{1\}.
\end{equation}

(iv) For every $e\in[0,1)$, we have
\begin{equation}
    \sum_{\bb\in(0,\bb_m(e)]}\nu_{-1}(\gamma_{\bb,e}(2\pi))=2,\quad
    \sum_{\bb\in(\bb_m(e),5]}\nu_{-1}(\gamma_{\bb,e}(2\pi))=0.
\end{equation}
\end{lemma}

Now we can give the
\medskip

{\bf Proof of  Theorem \ref{main.theorem.convex}. }
(i) The starting point is follows by (\ref{Convex:-1-index.of.b0})
and (\ref{Convex:null.-1-index.of.b0}).
$\lim_{e\rightarrow1}\bb_i(1,e)=1,i=1,2$ can be proved similarly by
the proof of Theorem 1.7 in \cite{HLS}.

(ii) follows by Theorem \ref{Convex:Th:om.degenerate.curves} 
and Theorem \ref{Convex:Th:tangent.direction}.

(iii) and (iv) are follow by Lemma \ref{Convex:Lm:hyperbolic.region}.

(v) and (vi) can be proved by the similar arguments of Theorem 1.2 (vi)-(viii) in \cite{HLS}.

(vii) If $0<\beta<\beta_l(e)$, then by the definitions of the degenerate curves and
Lemma \ref{Lemma:decreasing.of.index} (iii), we have
\be\label{convex:region.I.1.index}
i_1(\ga_{\beta,e})=0,\quad \nu_1(\ga_{\beta,e})=0
\ee
and
\be\label{convex:region.I.-1.index}
i_{-1}(\ga_{\beta,e})=2,\quad \nu_{-1}(\ga_{\beta,e})=0.
\ee

Firstly, if $\gamma_{\beta,e}(2\pi)\approx N_2(e^{\sqrt{-1}\theta},b)$
	for some $\theta\in(0,\pi)\cup(\pi,2\pi)$, we have
	\begin{equation}
	i_{-1}(\gamma_{\beta,e})=i_1(\gamma_{\beta,e})-S_{N_2(e^{\sqrt{-1}\theta},b)}^-(e^{\sqrt{-1}\theta})
	+S_{N_2(e^{\sqrt{-1}\theta},b)}^+  (e^{\sqrt{-1}\theta})
	=i_1(\gamma_{\beta,e})
	\end{equation}
	or
	\begin{equation}
	i_{-1}(\gamma_{\beta,e})=i_1(\gamma_{\beta,e})-S_{N_2(e^{\sqrt{-1}\theta},b)}^-(e^{\sqrt{-1}(2\pi-\theta)})
	+S_{N_2(e^{\sqrt{-1}\theta},b)}^+(e^{\sqrt{-1}(2\pi-\theta)})
	=i_1(\gamma_{\beta,e}),
	\end{equation}
	which contradicts (\ref{convex:region.I.1.index}) and (\ref{convex:region.I.-1.index}).
	
Then we can suppose $\gamma_{\beta,e}(2\pi)\approx M_1\diamond M_2$ where $M_1$ and $M_2$ are two basic normal forms in $\Sp(2)$.
By Lemma 3.1 in \cite{ZL1}, there exist two paths $\gamma_1,\gamma_2\in\P_{2\pi}(2)$ such that $\gamma_1(2\pi)=M_1$,
$\gamma_2(2\pi)=M_2$ and $\gamma_{\beta,e}\sim\gamma_1\diamond\gamma_2$. Then
\begin{equation}
0=i_{1}(\gamma_{\beta,e})=i_{1}(\gamma_1)+i_{1}(\gamma_2).
\end{equation}
By the definition of $\beta_l$, $M_1$ and $M_2$ cannot be both hyperbolic,
and without loss of generality, we suppose $M_1=R(\theta_1)$.
Then $i_1(\gamma_1)$ is odd, and hence $i_1(\gamma_2)$ is also odd.
By Theorem 4 to Theorem 7 of Chapter 8 on pp.179-183
in \cite{Lon4} and using notation there,
we must have $M_2=D(-2)$ or $M_2=R(\theta_2)$ for some $\theta_2\in(0,\pi)\cup(\pi,2\pi)$.

If $M_2=D(-2)$, then we have $i_{-1}(\gamma_1)-i_1(\gamma_1)=\pm1$ and $i_{-1}(\gamma_2)-i_1(\gamma_2)=0$.
Therefore $i_{-1}(\gamma_{\beta,e}(2\pi))=i_{-1}(\gamma_1)+i_{-1}(\gamma_2)$
and $i_{1}(\gamma_{\beta,e}(2\pi))=i_{1}(\gamma_1)+i_{1}(\gamma_2)$ has the different odevity,
which contradicts (\ref{convex:region.I.1.index}) and (\ref{convex:region.I.-1.index}).
Then we have $M_2=R(\theta_2)$.

Moreover, if $\theta_1\in(0,\pi)$, we must have $\theta_2\in(0,\pi)$,
otherwise $i_{-1}(\gamma_1)-i_1(\gamma_1)=-1$ and $i_{-1}(\gamma_2)-i_1(\gamma_2)=1$ and hence
\begin{equation}
i_{-1}(\gamma_{\beta,e})=i_{-1}(\gamma_1)+i_{-1}(\gamma_2)
=i_1(\gamma_1)+i_1(\gamma_1)=0,
\end{equation}
which contradicts (\ref{convex:region.I.-1.index}).
Similarly, if if $\theta_1\in(\pi,2\pi)$, we must have $\theta_2\in(\pi,2\pi)$.

(viii) If $\beta_l(e)<\beta<\beta_m(e)$, then by the definitions of the degenerate curves and
Lemma \ref{Lemma:decreasing.of.index} (iii), we have
\be\label{convex:region.II.1.index}
i_1(\ga_{\beta,e})=0,\quad \nu_1(\ga_{\beta,e})=0
\ee
and
\be\label{convex:region.II.-1.index}
i_{-1}(\ga_{\beta,e})=1,\quad \nu_{-1}(\ga_{\beta,e})=0.
\ee
Similarly, $\gamma_{\beta,e}(2\pi)\approx N_2(e^{\sqrt{-1}\theta},b)$
is impossible for any $\theta\in(0,\pi)\cup(\pi,2\pi)$ ,
and we can suppose $\gamma_{\beta,e}(2\pi)\approx M_1\diamond M_2$ where $M_1$ and $M_2$ are two basic normal forms in $\Sp(2)$.
By Lemma 3.1 in \cite{ZL1}, there exist two paths $\gamma_1,\gamma_2\in\P_{2\pi}(2)$ such that $\gamma_1(2\pi)=M_1$,
$\gamma_2(2\pi)=M_2$ and $\gamma_{\beta,e}\sim\gamma_1\diamond\gamma_2$. Then
\begin{equation}
0=i_{1}(\gamma_{\beta,e})=i_{1}(\gamma_1)+i_{1}(\gamma_2).
\end{equation}
By the definition of $\beta_l$, $M_1$ and $M_2$ cannot be both hyperbolic,
and without loss of generality, we suppose $M_1=R(\theta_1)$.
Then $i_1(\gamma_1)$ is odd, and hence $i_1(\gamma_2)$ is also odd.
By Theorem 4 to Theorem 7 of Chapter 8 on pp.179-183
in \cite{Lon4} and using notation there,
we must have $M_2=D(-2)$ or $M_2=R(\theta_2)$ for some $\theta_2\in(0,\pi)\cup(\pi,2\pi)$.

If $M_2=R(\theta_2)$, then we have $i_{-1}(\gamma_1)-i_1(\gamma_1)=\pm1$ and $i_{-1}(\gamma_2)-i_1(\gamma_2)=\pm1$.
Therefore $i_{-1}(\gamma_{\beta,e}(2\pi))=i_{-1}(\gamma_1)+i_{-1}(\gamma_2)$
and $i_{1}(\gamma_{\beta,e}(2\pi))=i_{1}(\gamma_1)+i_{1}(\gamma_2)$ has the same odevity,
which contradicts (\ref{convex:region.II.1.index}) and (\ref{convex:region.II.-1.index}).
Then we have $M_2=D(-2)$.

Moreover, if $\theta_1\in(0,\pi)$, we have
\begin{eqnarray}
i_{-1}(\gamma_{\beta,e})&=&i_{-1}(\gamma_1)+i_{-1}(\gamma_2)  \nonumber\\
&=&i_1(\gamma_1)-S_{R(\theta_1)}^-(e^{\sqrt{-1}\theta})+S_{R(\theta_2)}^+(e^{\sqrt{-1}\theta})+i_{-1}(\gamma_2) \nonumber\\
&=&i_1(\gamma_1)-1+i_1(\gamma_1)  \nonumber\\
&=&-1,
\end{eqnarray}
which contradicts (\ref{convex:region.II.-1.index}).
Thus (viii) is proved.

(ix) If $\beta_m(e)<\beta<\beta_r(e)$, then by the definitions of the degenerate curves and
Lemma \ref{Lemma:decreasing.of.index} (iii), we have
\be\label{convex:region.III.1.index}
i_1(\ga_{\beta,e})=0,\quad \nu_1(\ga_{\beta,e})=0
\ee
and
\be\label{convex:region.III.-1.index}
i_{-1}(\ga_{\beta,e})=0,\quad \nu_{-1}(\ga_{\beta,e})=0.
\ee

Assume $\gamma_{\beta,e}(2\pi)\approx N_2(e^{\sqrt{-1}\theta},b)$ for some $\theta\in(0,\pi)\cup(\pi,2\pi)$.
Without loss of generality, we suppose $\theta\in(0,\pi)$.
Let $\omega_0=e^{\sqrt{-1}\theta}$, we have $\nu_{\omega_0}(\gamma_{\beta,e}(2\pi))\ge1$.
Then for any $\omega\in\U,\omega\ne\omega_0$, we have
\begin{equation}
i_{\omega}(\gamma_{\beta,e})=i_1(\gamma_{\beta,e})=0
\end{equation}
or
\begin{equation}
i_{\omega}(\gamma_{\beta,e})=i_1(\gamma_{\beta,e})-S_{N_2(e^{\sqrt{-1}\theta},b)}^-(e^{\sqrt{-1}\theta})
+S_{N_2(e^{\sqrt{-1}\theta},b)}^+(e^{\sqrt{-1}\theta})
=0.
\end{equation}
Then by the sub-continuous of $i_\omega(\gamma_{\beta,e})$ with respect to $\omega$,
we have $i_{\omega}(\gamma_{\beta,e})=0,\; \forall\omega\in\U$.
Moreover, by Corollary \ref{Convex:C5.4}, we have
\begin{equation}
i_{\omega}(\gamma_{\mu,e})=0,\qquad \forall\omega\in\U, \mu\in[\beta,{27\over4}).
\end{equation}
Therefore, by the definition of $\beta_r(e)$ of (\ref{beta_r}), we have $\beta_r(e)\le\beta$.
It contradicts $\beta_m(e)<\beta<\beta_r(e)$.

Now we suppose $\gamma_{\beta,e}(2\pi)\approx M_1\diamond M_2$
where $M_1$ and $M_2$ are two basic normal forms in $\Sp(2)$.
By Lemma 3.1 in \cite{ZL1}, there exist two paths $\gamma_1,\gamma_2\in\P_{2\pi}(2)$ such that $\gamma_1(2\pi)=M_1$,
$\gamma_2(2\pi)=M_2$ and $\gamma_{\beta,e}\sim\gamma_1\diamond\gamma_2$. Then
\begin{equation}
0=i_1(\gamma_{\beta,e})=i_1(\gamma_1)+i_1(\gamma_2).
\end{equation}
By the definition of $\beta_r$, $M_1$ and $M_2$ cannot be both hyperbolic,
and without loss of generality, we suppose $M_1=R(\theta_1)$.
Then $i_1(\gamma_1)$ is odd, and hence $i_1(\gamma_2)$ is also odd.
By Theorem 4 to Theorem 7 of Chapter 8 on pp.179-183
in \cite{Lon4} and using notation there,
we must have $M_2=D(-2)$ or $M_2=R(\theta_2)$ for some $\theta_2\in(0,\pi)\cup(\pi,2\pi)$.

If $M_2=D(-2)$, then we have $i_{-1}(\gamma_1)-i_1(\gamma_1)=\pm1$ and $i_{-1}(\gamma_2)-i_1(\gamma_2)=0$.
Therefore $i_{-1}(\gamma_{\beta,e}(2\pi))=i_{-1}(\gamma_1)+i_{-1}(\gamma_2)$
and $i_{1}(\gamma_{\beta,e}(2\pi))=i_{1}(\gamma_1)+i_{1}(\gamma_2)$ has the different odevity,
which contradicts (\ref{convex:region.III.1.index}) and (\ref{convex:region.III.-1.index}).
Then we have $M_2=R(\theta_2)$.

Moreover, if $\theta_1\in(\pi,2\pi)$, we must have $\theta_2\in(0,\pi)$,
otherwise $i_{-1}(\gamma_1)-i_1(\gamma_1)=1$ and $i_{-1}(\gamma_2)-i_1(\gamma_2)=1$ and hence
\begin{equation}
i_{-1}(\gamma_{\beta,e})=i_{-1}(\gamma_1)+i_{-1}(\gamma_2)
=i_1(\gamma_1)+i_1(\gamma_1)+2=2,
\end{equation}
which contradicts (\ref{convex:region.III.-1.index}).
Similarly, if if $\theta_1\in(0,\pi)$, we must have $\theta_2\in(\pi,2\pi)$.

(x) is follows from (\ref{beta_r}).
\hb

\medskip

\noindent {\bf Acknowledgments.}
The authors thank sincerely Professor Yiming Long for his
precious help and useful suggestions.

\end{document}